\documentclass[a4paper,nohyper]{JHEP3}
\usepackage{cite}
\usepackage{epsfig}
\usepackage{amsmath}

\newcommand{\mycomm}[1]{\hfill\break $\phantom{a}$\kern-3.5em{\tt===$>$ \bf #1}\hfill\break}
\newcommand{\mycommA}[1]{\hfill\break $\phantom{a}$\kern-3.5em{\tt   $>$ \bf #1}\hfill\break}

\newcommand{\be}{\begin{equation}}
\newcommand{\ee}{\end{equation}}
\newcommand{\ba}{\begin{eqnarray}}
\newcommand{\ea}{\end{eqnarray}}

\def\tHooft{\hbox{\tiny 't Hooft}}
\def\eq#1{Eq.~(\ref{#1})}

\def\MSbar{\hbox{\tiny ${\overline{\rm MS}}$}}

\def\leading{\hbox{\tiny leading}}
\def\integrable{\hbox{\tiny integrable}}
\def\subleading{\hbox{\tiny subleading}}
\def\tot{\hbox{\tiny total}}

\def\eff{\hbox{\tiny eff}}
\def\sing{\hbox{\tiny sing.}}
\def\reg{\hbox{\tiny reg.}}

\def\PT{\hbox{\tiny PT}}
\def\PV{\hbox{\tiny PV}}

\catcode`\@=11 
\def\lsim{\mathrel{\mathpalette\@versim<}}
\def\gsim{\mathrel{\mathpalette\@versim>}}
\def\@versim#1#2{\vcenter{\offinterlineskip
        \ialign{$\m@th#1\hfil##\hfil$\crcr#2\crcr\sim\crcr } }}
\catcode`\@=12 

\title{Radiative B decay spectrum: DGE at NNLO}

\author{Jeppe R. Andersen$^{(a)}$ and Einan Gardi$^{(a,b)}$ \\
$^{(a)}$ Cavendish Laboratory, University of Cambridge, J J Thomson
Avenue,
Cambridge, CB3 0HE, UK\\
$^{(b)}$ Department of Applied Mathematics \& Theoretical Physics,
Wilberforce Road, Cambridge CB3 0WA,~UK}

\abstract{We compute the differential $\bar{B}\to X_s\gamma$
 decay width in the Standard Model as a function of the photon
 energy using Dressed Gluon Exponentiation (DGE). The resummed spectrum is
matched with the fixed--order expansion, making use of the
next--to--next--to--leading order (NNLO) results for the matrix
element of the magnetic dipole interaction $O_7$ and NLO ones for
other operators in the effective Weak Hamiltonian. We develop a new
technique to implement constraints on the analytic structure of the
Sudakov factor in moment space. This improves the behavior of the
resummed spectrum away from the Sudakov region. We also derive an
analytic expression for the Borel transform of the perturbative
series for the $O_7$ spectrum in the large--$\beta_0$ limit. Using
this example we demonstrate that exponentiation in moment space is
necessary for the calculation of the spectrum for $E_{\gamma}\gsim
2$ GeV. Finally, we investigate numerically the relation between
renormalons, power corrections and support properties. We present
predictions for the branching fraction and the first few spectral
moments as a function of a cut $E_{\gamma}>E_0$ and estimate the
theoretical uncertainty.}


\preprint{Cavendish-HEP-06/23}

\begin{document}

\section{Introduction}

Inclusive radiative B decays, $\bar{B}\longrightarrow X_s\gamma$, have become an essential
ingredient in precision tests of the Standard Model. The Standard Model $b\to s \gamma$ decay
 occurs only
through loops (penguin diagrams) involving the W Boson, whose mass is significantly larger than the available
energy, $m_W\gg m_b$. This makes the $\bar{B}\longrightarrow X_s\gamma$ width
a sensitive probe of any potential flavor--changing short--distance interaction
beyond the Standard Model, see e.g.~\cite{Ciuchini:1997xe,Gambino:2005dp,Allanach:2006jc}.

The Standard Model Branching Fraction (BF) is
known~\cite{Greub:1996tg,Chetyrkin:1996vx,Kagan:1998ym,Gambino:2001ew,Buras:2002tp}
since a few years to next--to--leading order (NLO) in
renormalization--group improved perturbation theory. The summation
of large logarithms of $m_b/m_W$ is conveniently formulated in the
framework of the effective Weak Hamiltonian, where virtualities of
order of the Weak scale are integrated out to obtain a set of local
operators $O_i$ of dimension 6. The result for the inclusive
decay width with a minimal photon--energy cut, $E_{\gamma}>E_0$,
takes the form:
\begin{eqnarray}
\label{total_width_E0_}
\Gamma(\bar{B}\longrightarrow X_s\gamma,E>E_0)&
=& \frac{\alpha_{\rm em} G_F^2}{32\pi^4}
\left|V_{\rm tb}V_{\rm ts}^*\right|^2  \,\left(m_b^{\MSbar}(m_b)\right)^2 \,m_b^3\,
\\ \nonumber
&&\hspace*{60pt}\times\,\sum_{i,j, \,i\leq j}C^{\eff}_i(\mu)C_j^{\eff}(\mu)\,
{G_{ij}(E_{0},\mu)},
\end{eqnarray}
where $m_b$ is the quark pole mass, $G_{ij}(E_0,\mu)$ are matrix elements of operators in the
effective Weak Hamiltonian,
$G_{ij}(E_0,\mu)=\sum_{X_s}\left<\bar{B}\right|O_i^\dagger(\mu)\left|X_s \gamma\right>
\left<X_s\gamma\right|O_j(\mu)\left|\bar{B}\right>$ where $E_{\gamma}>E_0$,
and $C^{\eff}_i(\mu)$ are the corresponding Wilson coefficients.
The matrix elements can be computed in perturbation theory, replacing the $\bar{B}$
meson by an on-shell b quark and the hadronic system $X_s$ by a partonic one,
owing to the inclusive sum over the final states. This replacement can be
 justified using the Operator Product Expansion (OPE) so long as the photon--energy
cut is insignificant. Upon removing the cut, power corrections to
the partonic calculation can be formally shown to be of ${\cal
O}(\Lambda^2/m_b^2)$, and they are numerically small, approximately
$+2.5\%$ of the total BF~\cite{Gambino:2001ew}.
The Standard Model BF, which has been determined with about~$\pm 10\%$
uncertainty~\cite{Gambino:2001ew,Buras:2002tp}, is found to be in
good agreement with experimental measurements by CLEO, Belle and
BaBar~\cite{Abe:2001hk,Chen:2001fj,Koppenburg:2004fz,Aubert:2005cb,Aubert:2005cu,Abe:2005cv}.

The current world average of all experimental data, prepared by the
Heavy Flavor Averaging Group~\cite{unknown:2006bi}, is ${\cal
B}(\bar{B}\longrightarrow X_s\gamma, E_\gamma>1.6 \,{\rm GeV})
=(355\pm24\pm10\pm 3)\times 10^{-6}$, where the errors are: combined
statistical and systematic uncertainty, ``shape--function''
uncertainty owing to the extrapolation from the region of
measurement $E_{\gamma}>E_0$, where $E_0\geq 1.8\,{\rm GeV}$, to the
reference range of $E_\gamma>1.6\,{\rm GeV}$, and uncertainty owing
to the $b\to d \gamma$ fraction. The ``shape--function'' uncertainty
varies significantly between different theoretical approaches. For
example, Ref.~\cite{Neubert:2004dd} assigns an error of about $\pm
8\%$ to the extrapolation below $E_0=1.8$ GeV, i.e. three times the
size of the error quoted here.

One of the essential ingredients for improving the precision of this
comparison in the future
is the theoretical calculation of the photon energy spectrum.
Owing to irreducible background,
experimental measurements are limited to the range $E_{\gamma}> 1.8\,{\rm GeV}$ and they can be
significantly improved if the requirement on the range of measurement is relaxed, for example, to
$E_{\gamma}> 2.0\,{\rm GeV}$. This, however, requires larger extrapolation that relies on the
theoretical description of the spectrum. Fortunately,
this extrapolation presents very little sensitivity to short--distance physics,
in sharp contrast with the total width. On the other hand, it requires detailed
understanding of the QCD dynamics.

The QCD calculation of inclusive decay spectra is essential for
other aspects of flavor physics. An important example is the
determination of $|V_{\rm ub}|$ from inclusive charmless
semileptonic decays, $\bar{B}\longrightarrow X_u l \bar{\nu}$, where
the background due to the $50$ times more abundant semileptonic
decay into charm restricts the region of measurement to $M_X<1.7 $
GeV. The dynamics there is similar to the one governing the
$\bar{B}\longrightarrow X_s\gamma$ spectrum to the extent that
spectral measurements of the photon--energy spectrum in
$\bar{B}\longrightarrow X_s\gamma$ are used for the determination of
$|V_{\rm ub}|$. In recent years the $\bar{B}\longrightarrow
X_s\gamma$ spectrum has become the prime testing--grounds for
theoretical approaches to inclusive distributions, see
e.g.~\cite{Andersen:2005bj,Gardi:2004ia,Gardi:2006gt,Becher:2006qw,Becher:2005pd,Bauer:1997fe,Bauer:2000ew,Bauer:2003pi,Bosch:2004th,Neubert:2004sp,Neubert:2004dd,Neubert:2005nt,Aglietti:2003sn,Benson:2004sg}.

The main challenge in computing inclusive decay spectra in QCD is
the complex dynamics of the threshold region. In this region the
decaying b quark is just slightly off its mass shell, owing to its
``primordial'' Fermi motion and to soft gluon
radiation~\cite{Neubert:1993um,Bigi:1993ex,Falk:1993vb,Korchemsky:1994jb}.
Specifically, the $\bar{B}\longrightarrow X_s\gamma$ spectrum peaks
near the partonic threshold, $E_\gamma\longrightarrow m_b/2$, or
\hbox{$x\equiv 2E_{\gamma}/m_b\longrightarrow 1$}. This region is
characterized by parametrically--large higher--order perturbative
corrections (Sudakov logs)~\cite{Korchemsky:1994jb} as well as
non-perturbative effects, predominantly ones related to the Fermi
motion of the b quark in the B meson.

\begin{figure}[t!]
\begin{center}
\epsfig{file=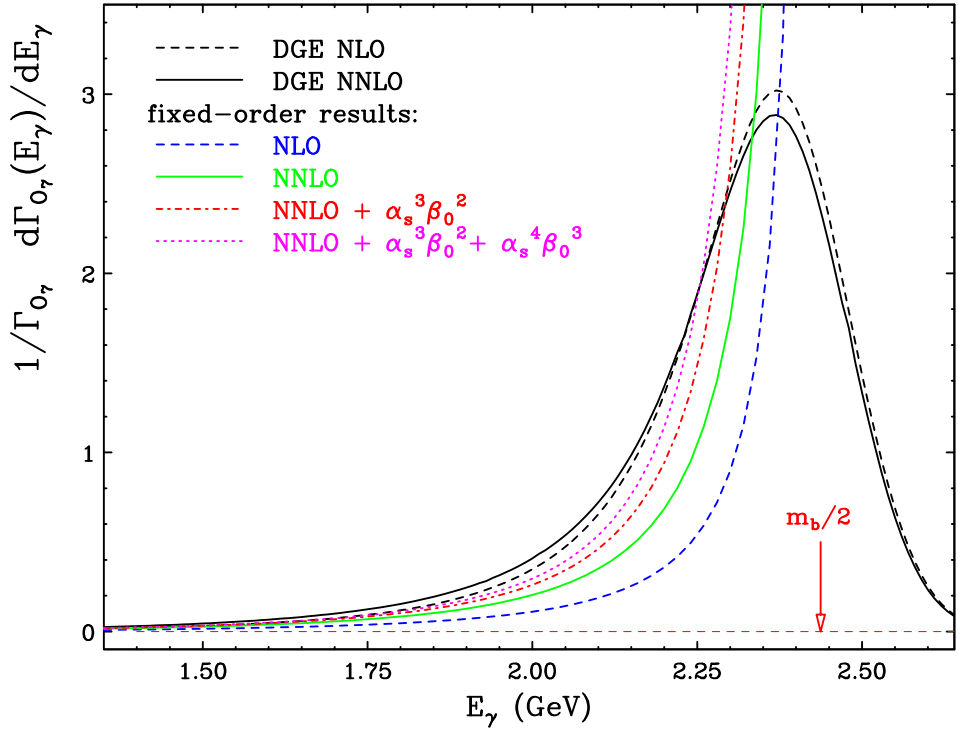,angle=90,width=12.7cm}
\caption{\label{fig:DGE_vs_FO} Comparison between the differential
spectra for $\bar{B}\longrightarrow X_s\gamma$ decay (through $O_7$
only) obtained by DGE, matched to NLO or NNLO, with fixed--order
results in $E_{\gamma}$ space ($x$ space). All curves correspond to
the same value of the b-quark mass, the PV pole mass $m_b=4.89$ GeV.
The end of partonic phase space, $E_{\gamma}=m_b/2$, is denoted by
an arrow. }
\end{center}
\end{figure}

It is universally acknowledged that
\emph{fixed--order} perturbative results cannot be directly used for
comparison with spectral data, not even the first few moments of the
photon energy with experimentally--relevant cuts, such as
$E_{\gamma}>1.8$ GeV.
Fixed--order results for the $\bar{B}\longrightarrow X_s\gamma$
spectrum are characterized by
\begin{itemize}
\item{} Sudakov logarithms, namely singular real-emission
corrections to the differential spectrum ${d\Gamma(x)}/{dx}$,
of the form~$\ln^{k}(1-x)/(1-x)$ with $k\leq 2n-1$ at order
${\alpha_s}^n$, owing to multiple soft and collinear radiation. The
perturbative spectrum is nevertheless integrable as there are also
infrared--singular virtual corrections proportional
to $\delta(1-x)$ --- the spectral moments,
\begin{equation}
\label{mom_def}
\Gamma_N^{\PT} \equiv \int_0^1 dx\,\frac{1}{\Gamma_{\tot}^{\PT}}\frac{d\Gamma^{\PT}(x)}{dx} \,x^{N-1}
\end{equation}
\emph{are} infrared safe.
\item{} Support for $E_{\gamma}<m_b/2$, where $m_b$ is the quark pole mass,
setting the upper limit of integration in \eq{mom_def} as $x=1$. The
perturbative support at any order is different from the physical
one, $E_{\gamma}<M_B/2$, where $M_B$ is the meson mass. Importantly,
the pole mass itself has a linear infrared renormalon
ambiguity~\cite{Bigi:1994em,Beneke:1994sw,Beneke:1994bc}, $m_b\to
m_b\pm {\cal O} (\Lambda)$, and therefore it cannot be assigned a
precise value without specifying an additional regularization
prescription. In a fixed--order framework one computes the pole mass
order-by-order from a given short--distance mass, such as
$m_b^{\MSbar}$, but the result strongly depends on the order, as the
series is badly divergent. The use of alternative mass
schemes~\cite{Neubert:2004sp,Neubert:2005nt,Bigi:1996si,Czarnecki:1997wy}
amounts to introducing an infrared cutoff, which hinders the
possibility of using of the inherent infrared safety of the on-shell
decay spectrum.
\item{} Large running--coupling effects, which completely dominate
the NNLO correction to the spectrum~\cite{Melnikov:2005bx} if the
coupling at NLO is renormalized at $m_b$.
Large running--coupling effects reflect the fact that
the typical gluon virtuality is much smaller than $m_b$.
Naturally, large running--coupling  corrections appear also at higher orders.
Real-emission corrections proportional to $C_F
\beta_0^{n-1}{\alpha_s}^{n}$, where $\beta_0$ (\ref{beta0}) is the leading coefficient of
 the beta function, were computed in \cite{Gardi:2004ia}
to all orders\footnote{Another calculation of these corrections is reported in
Ref.~\cite{Benson:2004sg}. There, an infrared cutoff was applied.}.
In Sec.~\ref{sec:large_beta0_Borel}  below we will
 show that the series composed of
these terms alone (i.e. with no exponentiation, no effect of real--virtual cancellation)
cannot be considered a
viable prediction for the
spectrum anywhere in the peak region. This series is Borel summable up to
$E_{\gamma}\sim 2$ GeV, but ceases to be so above this scale, where
the Borel integral diverges for $u\longrightarrow \infty$.
\end{itemize}
The numerical results for the spectrum obtained by fixed--order
calculations at NLO and NNLO, as well as running--coupling
corrections beyond this order, are shown in
Fig.~\ref{fig:DGE_vs_FO}, with the pole mass is set\footnote{The pole--mass value $m_b=4.89$ GeV
is obtained by Principal Value Borel summation starting from $m_b^{\MSbar}=4.20$ GeV, as explained in
\cite{Andersen:2005bj}. } to $m_b=4.89$
GeV. The lack of convergence in the peak region is apparent.

These characteristics of the fixed--order result point towards the
necessity for (1) resummation of all large perturbative corrections
(2) systematic separation between perturbative and non-perturbative
contributions and appropriate parametrization of the latter. Dressed
Gluon Exponentiation (DGE) offers a framework to do so, utilizing
directly the on-shell
scheme~\cite{Gardi:2004ia,Gardi:2006jc,Andersen:2005bj}. Owing to
its inherent infrared safety, the on-shell decay spectrum itself
provides then a first approximation to the physical B--meson decay
spectrum. Fig.~\ref{fig:DGE_vs_FO} shows these results as well.
Other approaches to compute the
spectrum~\cite{Benson:2004sg,Neubert:2004dd,Neubert:2005nt} have
been based on introducing an infrared cutoff as means of separation
between perturbative and non-perturbative corrections. With a cutoff
in place, a first approximation to the B--meson decay spectrum is
obtained only upon convoluting the perturbative result with a
leading--power non-perturbative quark distribution function, or
``shape
function''~\cite{Neubert:1993um,Bigi:1993ex,Falk:1993vb,Korchemsky:1994jb,Bosch:2004th,Neubert:2004sp,Neubert:2005nt,Bigi:1996si,Czarnecki:1997wy}.
Ref.~\cite{Andersen:2005bj} has shown that a cutoff and a
leading--power ``shape function'' can be avoided, and
\emph{replaced} by a resummation of the perturbative expansion, a
prescription for the renormalon singularities (Principal Value of
the Borel sum) and power corrections. This involves of course making
certain assumptions on the perturbative expansion and on
non-perurbative corrections. Aside from the separation issue, the
calculation in Ref.~\cite{Andersen:2005bj} applies a novel approach
to resummation per se. Different higher--order corrections are
considered important in different approaches:
\begin{itemize}
\item{} Refs.~\cite{Neubert:2004dd,Neubert:2005nt} emphasize the significance
of Sudakov logarithms and utilize a logarithmic accuracy criterion
and an infrared cuoff.
\item{} Ref.~\cite{Benson:2004sg} emphasizes the significance of
running--coupling corrections (dismissing Sudakov logarithms) and utilizes
large--$\beta_0$ resummation with a
Wilsonian cutoff.
\item{} Ref.~\cite{Andersen:2005bj} resums Sudakov logarithms as well as
running--coupling corrections in the Sudakov exponent and utilizes
additional information on the Borel transform of the exponent
to complement the logarithmic accuracy criterion.
\end{itemize}

Upon resumming the perturbative expansion for the
matrix elements $G_{ij}$ as a function of the cut $E_0$, normalized by $G_{ij}(E_{\min})$,
\eq{total_width_E0_} can be written as:
\begin{eqnarray}
\label{total_width_E0_resummed}
\Gamma(\bar{B}\longrightarrow X_s\gamma,E>E_0)&
=& \frac{\alpha_{\rm em} G_F^2}{32\pi^4}
\left|V_{\rm tb}V_{\rm ts}^*\right|^2  \,\left(m_b^{\MSbar}(m_b)\right)^2 \,m_b^3\,
\\ \nonumber
&&\hspace*{-20pt}\times\,\sum_{i,j, \,i\leq j}
C^{\eff}_i(\mu)C_j^{\eff}(\mu)\, {G_{ij}(E_{\min},\mu)}
\left[\frac{G_{ij}(E_{0})}{G_{ij}(E_{\min})}\right]_{\rm Resummed},
\end{eqnarray}
where it is assumed that $E_{\min} \ll m_b/2$ so
$G_{ij}(E_{\min},\mu)$ can be computed using fixed--order
perturbation theory (our choice will be $E_{\min}=m_b/20$) while
$E_0$ can take any value, in particular, experimentally--relevant
one: $E_0\geq 1.8$ GeV.

The essential element in the calculation of resummed spectra near
threshold, namely
$\left[{G_{ij}(E_{0})}/{G_{ij}(E_{\min})}\right]_{\rm Resummed}$ in
\eq{total_width_E0_resummed}, is the Sudakov factor, which sums up
the dominant corrections owing to multiple soft gluon emission that
exponentiate in moment space. In DGE the Sudakov exponent is
computed as Borel sum, combining Sudakov resummation with the
resummation of running--coupling effects. This immediately exposes
the power--like infrared sensitivity of the on-shell matrix elements
in the form of renormalons. This allows for
\begin{itemize}
\item{} Explicit cancellation of the
leading renormalon ambiguity, the one associated with the definition of the pole mass.
\item{} Definite regularization of all renormalons, using the Principal Value prescription.
\end{itemize}
This definition of the perturbative sum amounts to a systematic separation between
perturbative and non-perturbative corrections.
This procedure \emph{uniquely defines} the
\emph{non-perturbative Fermi motion effect}, distinguishing it from the \emph{radiation effect}
that is common to an on-shell heavy quark and to one that is part of a meson.
As already mentioned, in contrast with other formulations,
no infrared cutoff is needed here. Therefore, Fermi motion effects enter
exclusively through power corrections. These power corrections are power--enhanced at large~$N$:
they scale as powers of $N\Lambda/m_b$, so they can
modify the perturbative on-shell spectrum by ${\cal O}(1)$ corrections near threshold
while almost not affecting the first few moments and the distribution away from threshold.
According to the renormalon structure of the exponent, non-perturbative
corrections start at the \emph{third} power of $N\Lambda/m_b$,
making just a small correction to the spectrum.
The numerical significance of these power corrections will be
analyzed in Sec.~\ref{sec:numerical} below.

As shown in Fig.~\ref{fig:DGE_vs_FO} the DGE spectrum is
qualitatively different from the fixed--order results. It is
characterized by
\begin{itemize}
\item{} Approximate physical support: the spectrum smoothly extends beyond the
perturbative endpoint, $E_{\gamma}=m_b/2$, and tends to zero close to the physical one,
$E_{\gamma}=M_B/2$.
\item{} Mass--scheme independence, owing to the \emph{explicit} cancellation
of the pole--mass rerormalon ambiguity.
\item{} Stability in going from order to order, reflecting the fact
that the dominant higher--order corrections are indeed resummed.
\end{itemize}

The DGE $\bar{B}\longrightarrow X_s\gamma$ spectrum, and its first
few moments with experimentally relevant cuts, were computed in
Ref.~\cite{Andersen:2005bj}. Later, when experimental data for the
average energy the variance
appeared~\cite{Abe:2005cv,Aubert:2005cb,Aubert:2005cu}, these
predictions where found to be in good
agreement~\cite{Gardi:2005mf,Gardi:2006gt}.

Recently there has been significant progress in higher--order calculations.
The NNLO corrections to the spectrum associated with the matrix
element $G_{77}(E_0)$ of the magnetic dipole operator~
\begin{align}
\label{O7}
O_7\equiv \frac{\,e}{32\pi^2}\,m_{\MSbar}\,F^{\mu\nu}\,\,\bar{s}\,\sigma_{\mu\nu}(1+\gamma_5)\,b
\end{align}
have been computed in full~\cite{Melnikov:2005bx,Asatrian:2006sm}.
The resummed spectrum of Ref.~\cite{Andersen:2005bj} already
included NNLL corrections through the Sudakov
factor~\cite{Andersen:2005bj,Gardi:2005yi}; however, it included
non-logarithmic corrections to NLO only. One of the tasks of the
present paper is to match the resummed spectrum to full NNLO
accuracy in the $O_7$ sector.
We also consider other operators in the
Weak Hamiltonian, which have been so far computed to ${\cal O}(\alpha_s)$ only.
Knowing that independently of the nature of the short--distance interaction,
all important contributions in the peak region necessarily involve the same Sudakov factor,
we compute resummed spectra for individual matrix elements $G_{ij}(E_0)$, determining
the hard coefficient functions at ${\cal O}(\alpha_s)$ from known results.

In addition, significant progress towards NNLO calculation of the
\emph{total BF} was recently
made~\cite{Blokland:2005uk,Asatrian:2006ph}: two--loop matrix
element of the $O_7$ operator have been computed in full. This adds
to the already existing NNLO results in the framework of the
effective Weak Hamiltonian, which includes the matching coefficients
at the Weak scale~\cite{Misiak:2004ew}, partial information on the
evolution matrix~\cite{Gorbahn:2004my,Gorbahn:2005sa}, as well as
$\beta_0 {\alpha_s}^2$ contributions to $\bar{B}\longrightarrow
X_s\gamma$ of several matrix elements~\cite{Bieri:2003ue}. It is our
aim here to set a framework where the state-of-the-art calculation
of the spectrum can be used together with that of the total BF. This
would be particularly important once the NNLO calculation is
complete. A detailed knowledge of the partial BF as a function of
the photon--energy cut $E_{\gamma}>E_0$ can help making good use of
the data, since low cuts, such as $E_0=1.8$ GeV, are characterized
by large systematic experimental errors, in contrast with higher
cuts, such as $E_0=2.0$ GeV, that, in turn, requires larger
extrapolation.

An important new ingredient in the calculation of the spectrum that
we develop in this paper is the use of the analytic structure of the
perturbative result in moment space when writing the resummation
formula. It is a general problem in the application of Sudakov
resummation, that the region of interest may extend far beyond the asymptotic
Sudakov regime where the logs are large and the coupling is small. An
immediate implication is that the resummed result depends on the
(often implicit) assumptions made concerning non-logarithmic ${\cal O}(1/N)$
higher--order corrections. Non-logarithmic  corrections are of course
included to some fixed order in ${\alpha_s}$ in the process of
matching the resummed spectrum into the fixed--order expansion. This
procedure, however, may not be sufficient to avoid bias of the
result due to the resummation of logarithms in the region where the
logarithms are not at all dominant.

The $\bar{B}\longrightarrow X_s\gamma$ spectrum provides an
important motivation to address this problem, since the region where
the logarithms alone dominate is rather small, at least up to the
NNLO level~\cite{Melnikov:2005bx}. To make good use of perturbation
theory it is therefore important to impose additional constrains on
the resummation formula. Such constraints are indeed available: the
analytic structure of the perturbative result (at any order) in
moment space is known fairly well: perturbative coefficients are
composed of harmonic sums and rational functions whose singularities
appear on the negative real axis in $N$ space. The \emph{rightmost
singularity} appears at $N=-J$ where $J$, a non-negative integer,
corresponds to the power fall of the spectrum, $d\Gamma/dx\sim x^J$,
in the $x\to 0$ limit. In $\bar{B}\longrightarrow X_s\gamma$, for
most matrix elements, $J=3$. Thus, there is quite a strong
suppression of the spectrum at small $x$, which obviously would not
be respected by a generic large--$x$ resummation formula that
accounts only for $\ln^l (1-x)/(1-x)$ terms. A power fall is
generally expected when the $x\to 0$ behavior is
dominated\footnote{It does not apply when this limit is
characterized by singular matrix elements, as is the case of the
chromomagnetic operator contribution to $\bar{B}\longrightarrow
X_s\gamma$.} by phase space.

The remainder of this paper is organized as follows:
Sec.~\ref{sec:resummed_G77} is devoted entirely to the normalized
spectrum of the $G_{77}$ matrix element, corresponding to the
magnetic dipole operator, $O_7$. $G_{77}$ is the only matrix element
contributing at ${\cal O}(1)$, while other $G_{ij}$ start at ${\cal
O}({\alpha_s})$. Moreover, $G_{77}$ is the only matrix element for
which the full ${\cal O}({\alpha_s}^2)$ (NNLO) result is
available~\cite{Melnikov:2005bx,Asatrian:2006sm,Blokland:2005uk,Asatrian:2006ph}.
This facilitates matching of the resummed spectrum to NNLO as well
as performing an in-depth analysis of the perturbative expansion. In
Sec.~\ref{sec:DGE} we briefly summarize the main assumptions and the
necessary formulae in the application of DGE to the radiative B
decay spectrum. In Sec.~\ref{sec:small_x} we reformulate the
resummation formulae under constraints on the analytic structure in
moment space, in order to have a good description of the small
$E_{\gamma}$ tail. Next, in Sec.~\ref{sec:large_beta0_Borel} we
analyze the all--order result in the large--$\beta_0$ limit.
Sec.~\ref{sec:others} is devoted to computing the resummed spectra
for the matrix elements $G_{ij}$ other than $G_{77}$. In
Sec.~\ref{sec:numerical} we compute the total BF and incorporate the
resummed spectra of Secs.~\ref{sec:resummed_G77}
and~\ref{sec:others} into \eq{total_width_E0_resummed}, to obtain
the BF as a function of $E_0$. We also study there numerically the
relation between renormalons, power correction and support properties.
Finally, we present predictions for the first few moments as a function
of the cut and analyze the theoretical uncertainty. In Sec.~\ref{sec:conc}
we shortly summarize our conclusions.

\section{Resummed spectrum for the magnetic dipole operator $O_7$~\label{sec:resummed_G77}}

\subsection{Dressed Gluon Exponentiation~\label{sec:DGE}}

DGE is a general resummation formalism for
inclusive distributions that is designed to address kinematic
threshold problems in QCD where there is interplay between Sudakov
logarithms, running--coupling effects and parametrically--enhanced
power corrections. The formalism will not be described here in any
detail; we refer the reader to a recent review~\cite{Gardi:2006jc}.
Here we concentrate on the application to $\bar{B}\longrightarrow
X_s\gamma$~\cite{Andersen:2005bj}, briefly summarizing the main
assumptions and the necessary formulae.

For simplicity we consider in this section the normalized spectrum associated with the magnetic dipole
operator only,
postponing the calculation of the spectra of other matrix elements to
Sec.~\ref{sec:others}, and the calculation of the overall normalization of the
BF to Sec.~\ref{sec:numerical}, where the contributions of the different matrix elements are combined
with the appropriate Wilson coefficients.

\subsubsection*{Formulating Sudakov resummation in moment space}

The normalized differential spectrum associated with the magnetic dipole
operator takes the form
\begin{align}
\label{G77_diff_form}
\begin{split}
\frac{1}{\Gamma_{O_7}}\,\frac{d\Gamma_{O_7}(x)}{dx}&=V\left(\alpha_s(m_b)\right)\,\delta(1-x) \,+\, R\left(\alpha_s(m_b),x\right),
\\
&V\left(\alpha_s(m_b)\right)=1+C_F\sum_{n=1}^{\infty} k_n\, \left(\frac{\alpha_s(m_b)}{\pi}\right)^n
\\
\begin{split}
&R\left(\alpha_s(m_b),x\right)=R_{\sing}\left(\alpha_s(m_b),x\right)+
R_{\reg}\left(\alpha_s(m_b),x\right)\\
&\hspace*{70pt}R_{\sing}\left(\alpha_s(m_b),x\right)=C_F\,\sum_{n=1}^{\infty}\Big[
r_n^{\sing}(x)\Big]_{+}
\,\left(\frac{\alpha_s(m_b)}{\pi}\right)^n\\
&\hspace*{70pt}R_{\reg}\left(\alpha_s(m_b),x\right)
=C_F\,\sum_{n=1}^{\infty}r_n^{\reg}(x)\,\left(\frac{\alpha_s(m_b)}{\pi}\right)^n
\end{split}
\end{split}
\end{align}
where $R_{\sing}$ contains only plus distributions of the form
$\left[{\ln^l(1-x)}/{(1-x)}\right]_{+}$
where at order ${\alpha_s}^n$, \hbox{$0\leq l\leq 2n-1$}, while $R_{\reg}$ is integrable for $x\to 1$.
The constants $k_n$ are determined such that the integral of the normalized
spectrum would be exactly unity:
\begin{equation}
\label{cn_def}
k_n\equiv -\int_0^1 dx\, r_n^{\reg}(x),
\end{equation}
at every order in the perturbative expansion.

The partially--integrated matrix element defined with a $E_{\gamma}>E_0$ cut, normalized by
the fully--integrated one, is
\begin{eqnarray}
\label{integrated_G77}
\frac{G_{77}(E_0,m_b)}{G_{77}(0,m_b)}
=\int_{x=2E_0/m_b}^1\,dx\,\frac{1}{\Gamma_{O_7}}\,\frac{d\Gamma_{O_7}(x)}{dx}.
\end{eqnarray}
Defining the spectral moments of~\eq{G77_diff_form} as in \eq{mom_def} one may resum the
perturbative expansion corresponding to \eq{integrated_G77} as follows:
\begin{eqnarray}
\label{G_77_resummed}
\left[\frac{G_{77}(E_0,m_b)}{G_{77}(0,m_b)}\right]_{\rm Resummed}
&=&\frac{1}{2\pi i}
\int_{c-i\infty}^{c+i\infty} \frac{dN}{N-1} \,\left(\frac{2E_0}{m_b}\right)^{1-N}\,
H\left(\alpha_s(m_b),N\right)\,\times\,{\rm Sud}(N,m_b) \nonumber
\\&+&\,\,\int_{x=2E_0/m_b}^1 dx
\Delta R(\alpha_s(m_b),x),
\end{eqnarray}
where the first line, written as an inverse--Mellin
transform\footnote{The contour running parallel to the imaginary
axis is assumed to be to the right of all the singularities of the
integrand.}, is the dominant contribution, which contains in
particular the contributions originating in $V$ in
\eq{G77_diff_form} as well as all the log--enhanced contributions to
the matrix element, $R_{\sing}$ in \eq{G77_diff_form}, while the
second line contains some residual real--emission terms that are
integrable for $x\to1$.

\subsubsection*{The Sudakov factor}

The Sudakov factor, ${\rm Sud}(N,m_b)$, sums up the logarithmically--enhanced
corrections originating
$R_{\sing}$ to all orders. These corrections exponentiate:
\begin{equation}
\label{Sud_expand} {\rm Sud}(N,m_b)=\exp\left\{
C_F\left[E_1(N)\frac{\alpha_s(m_b)}{\pi}+E_2(N)\left(\frac{\alpha_s(m_b)}{\pi}\right)^2+\cdots\right]\right\}.
\end{equation}
In DGE the exponent is written as a Borel sum:
\begin{align}
\label{Sud_DGE} {\rm Sud}(N,m_b)&=\exp\Bigg\{\frac{C_F}{\beta_0}
\int_0^{\infty}\frac{du}{u}\,T(u)\,\left(\frac{\Lambda^2}{m_b^2}\right)^u\,
\bigg[B_{\cal
S}(u)\Gamma(-2u)\,\left(\frac{\Gamma(N)}{\Gamma(N-2u)}-\frac{1}{\Gamma(1-2u)}\right)
\nonumber \\  &
-B_{\cal J}(u)\Gamma(-u)
\left(\frac{\Gamma(N)}{\Gamma(N-u)}-\frac{1}{\Gamma(1-u)}\right)\bigg]\Bigg\},
\end{align}
where $\beta_0$ is defined in (\ref{beta0}) below.
$B_{\cal S}(u)$ and $B_{\cal J}(u)$ are the Borel transforms
\begin{eqnarray}
\label{Sud_anom}
{\cal S}(\alpha_s(\mu)) &=&\frac{C_F}{\beta_0}\int_0^{\infty} du \,T(u)
\left(\frac{\Lambda^2}{\mu^2}\right)^u B_{\cal S}(u)\nonumber \\
{\cal J}(\alpha_s(\mu)) &=&\frac{C_F}{\beta_0}\int_0^{\infty} du \,T(u)
\left(\frac{\Lambda^2}{\mu^2}\right)^u B_{\cal J}(u)
\end{eqnarray}
of the Sudakov anomalous dimensions of the quark distribution
function~\cite{Gardi:2005yi} and the jet
function~\cite{Gardi:2002xm}, respectively, and $T(u)$ is the
Laplace conjugate of the 't~Hooft coupling~\cite{Grunberg:1992hf}:
\begin{eqnarray}
\label{tHooft_coupling}
A(\mu)&=&\frac{\beta_0{\alpha_s}^{\tHooft}(\mu)}{\pi}=
\int_0^{\infty}{du} \,T(u)\, \left(\frac{\Lambda^2}{\mu^2}\right)^u;
\qquad \qquad \frac{dA}{d\ln \mu^2} =-A^2(1+\delta A),
\nonumber \\
T(u)&=&\frac{(u\delta)^{u\delta}{\rm
e}^{-u\delta}}{\Gamma(1+u\delta)};\qquad \qquad \ln
(\mu^2/\Lambda^2)=\frac{1}{A}-\delta\ln\left(1+\frac{1}{\delta
A}\right)
\end{eqnarray}
with $\delta\equiv \beta_1/\beta_0^2$.

Let us point out that the particular $N$--dependence in
(\ref{Sud_DGE}) is based on the Mellin transform of the plus
distributions of the log--enhanced terms,
$\left[{\ln^l(1-x)}/{(1-x)}\right]_{+}$, making no further
approximation. Eq.~(2.17) in Ref.~\cite{Andersen:2005bj} is based on
an approximation to this result, a minimal scheme where \emph{only}
$\ln N$ terms are exponentiated; it differs from (\ref{Sud_DGE})
above by ${\cal O}(1/N)$ terms. As discussed in Sec. 3.4
in~\cite{Andersen:2005mj}, Eq.~(\ref{Sud_DGE}) goes beyond this
minimal scheme in exponentiating a particular class of ${\cal
O}(1/N)$ terms. In the present paper we will take one further step
in this direction (see Sec.~\ref{sec:small_x} below) and modify the
exponent to conform with the analytic structure of the full perturbative result in
moment space. The Sudakov factor then takes the form
(\ref{Sud_with_cubic_small_x_general_color}). It should be clear
that these modification do not affect the large--$N$ behavior, and
their influence on the distribution in the peak region is small.

The calculation of the exponent proceeds as in
Refs.~\cite{Andersen:2005mj,Andersen:2005bj}. Formally, \eq{Sud_DGE} (or
Eq.~(\ref{Sud_with_cubic_small_x_general_color}) below)
is computed with NNLL accuracy, using the
known~\cite{Gardi:2005yi,Moch:2004pa,Andersen:2005bj,Melnikov:2005bx,Melnikov:2004bm,Korchemsky:1992xv}
${\cal O}({\alpha_s}^3)$ expansions of ${\cal S}(\alpha_s(\mu))$ and
${\cal J}(\alpha_s(\mu))$\footnote{The two-loop results for the
Sudakov anomalous dimensions have recently been checked by
additional, independent
calculations~\cite{Becher:2006qw,Becher:2005pd}.}. However, since
the Borel integral is evaluated, not expanded(!), large subleading
corrections, notably running--coupling effects are accounted for to
all
orders~\cite{Gardi:2001ny,Gardi:2002bg,Gardi:2001di,Cacciari:2002xb,Andersen:2005mj,Andersen:2005bj}.
Importantly, the Borel integrand presents singularities at integer
and half integer values of $u$. These induce ambiguities that scale
as integer powers of $N\Lambda/m_b$ and $N\Lambda^2/m_b^2$ that are
inherent to the perturbative quark distribution and the jet,
respectively. In DGE the perturbative exponent is \emph{defined} as
the Principal Value of the integral in \eq{Sud_DGE}, while
non-perturbative corrections are assumed to follow the ambiguity
structure of the exponent.

The details of the spectrum are dictated by the two functions:
$B_{\cal S}(u)$ and $B_{\cal J}(u)$ in \eq{Sud_DGE}. In this paper
we adopt the approximations to these two functions that were
developed and used in previous papers, namely Eq. (3.27) in
Ref.~\cite{Andersen:2005mj} (or Eq. (2.35) in
Ref.~\cite{Andersen:2005bj}). These approximations are based on the
known NNLO expansions of the anomalous dimensions and on additional
constraints on the behavior of these functions away from the origin.
These are particularly important in the case of the quark
distribution function $B_{\cal S}(u)$, since $N\Lambda/m_b$ is not
necessarily small; the contribution of $B_{\cal J}(u)$ away from
$u=0$ has a rather significant suppression, since
$N\Lambda^2/m_b^2\ll 1$ at any relevant $N$. The additional
constraints can be briefly summarized as
follows~\cite{Andersen:2005mj,Andersen:2005bj} (see also the recent
review~\cite{Gardi:2006jc}):
\begin{itemize}
\item{} \emph{Properties of the large--$\beta_0$
results~\cite{Gardi:2004ia} that are expected to hold in the full
theory:} the Sudakov anomalous dimensions in \eq{Sud_anom} have no
Borel singularities. Moreover, their Borel transform vanishes at
certain integer values of $u$, eliminating some potential renormalon
singularities in the exponent of \eq{Sud_DGE}. Specifically in the
quark distribution function there is one zero at $u=1$: $B_{\cal
S}(u=1)=0$, so ${\cal O}(N^2\Lambda^2/m_b^2)$ ambiguities are absent
while higher\footnote{The ${\cal O}(N\Lambda/m_b)$ ambiguity cancels
against that of the pole mass~\cite{Gardi:2004ia,Andersen:2005bj}.}
power ambiguities are present. This suggests that the width of the
spectrum is ``protected'' from non-perturbative power corrections,
while higher moments receive such corrections.
\item{} \emph{The computed value of $B_{\cal S}(u=1/2)$, corresponding to the
large--order asymptotic behavior of the Sudakov exponent} (which is
different from its large--$\beta_0$ limit). This calculation (see
Sec. 2.3 and Appendix B in Ref.~\cite{Andersen:2005bj}) is based on
the known cancellation mechanism~\cite{Gardi:2004ia} of the leading,
${\cal O}(N\Lambda/m_b)$, renormalon ambiguity in the exponent with
that of the pole mass, the known structure of this Borel
singularity~\cite{Beneke:1994rs,Andersen:2005bj} and the
perturbative expansion of the ratio between the pole mass
and~$m_b^{\MSbar}$.
\end{itemize}

It was further observed in
Refs.~\cite{Andersen:2005bj,Andersen:2005mj} that given the
constraints described above --- in particular, the expansion of
$B_{\cal S}(u)$ near the origin and its values at $u=1/2$ and at
$u=1$ --- and so long as $B_{\cal S}(u)$ does not get large at
intermediate values of $u$ (specifically for $u \sim 3/2$) the
support properties of the resummed spectrum are close to these of
physical spectrum. In this scenario power corrections are
expected to be small. When making predictions we shall not assume
that this is necessarily the case, but instead allow for variation
of $B_{\cal S}(u)$ and for power corrections as explained below.

\subsubsection*{Renormalons and power corrections in the exponent}

We base our analysis on the parametrization of $B_{\cal S}(u)$ in
Ref.~\cite{Andersen:2005mj} (see Eqs.~(3.27) to (3.29) there), where
\begin{equation}
\label{C32}
B_{\cal
S}(u=3/2)=-0.23366\, C_{3/2}.
\end{equation}
$C_{3/2}=1$ is the default value
used in Refs.~\cite{Andersen:2005bj,Andersen:2005mj}, and variation
of $C_{3/2}$ between $0.1$ and~$10$ is considered. Here, however, we
take one step further in the way power corrections are taken into
account. Introducing power corrections based on the renormalon
ambiguities in the quark--distribution part of the Sudakov factor
(\ref{Sud_with_cubic_small_x_general_color}) we
have~\cite{Andersen:2005mj}:
\begin{align}
\label{Sud_with_cubic_Renormalons} &\left.\widetilde{\rm
Sud}^{(J)}(N,m_b )\right\vert_{\PV} \,\longrightarrow \,\left.
\widetilde{\rm Sud}^{(J)}(N,m_b)\right\vert_{\PV}\times \exp\Bigg\{
\sum_{k=3}^{\infty}\epsilon^{\PV}_k
\left(\frac{\Lambda}{m_b}\right)^k \,R^{(J)}(N,k/2)\Bigg\},
\end{align}
where the $N$--dependence of each renormalon residue is carried by
\begin{equation}
R^{(J)}(N,k/2)\equiv {\rm Res}\left.
\Gamma(-2u)\,\left(\frac{\Gamma(N+J)}
{\Gamma(N+J-2u)}-\frac{\Gamma(J+1)}{\Gamma(J+1-2u)}
\right)\right\vert_{u=k/2},
\end{equation}
and where we defined
\begin{equation}
\label{epsilon_k}
\epsilon^{\PV}_k \equiv \frac{C_F}{\beta_0}\,\pi\,
f^{\PV}_k\,\frac{T(k/2)}{k/2}\, \,B_{\cal S}(k/2),
\end{equation}
where $f_k^{\PV}$ are dimensionless non-perturbative parameters.
Assuming that the power corrections are of order of the ambiguity itself,
\eq{epsilon_k} with $f_k^{\PV}\thickapprox 1$ provides an estimate
of their magnitude. This assumption will be eventually tested by data.

The application of (\ref{Sud_with_cubic_Renormalons}) is limited in
practice for several reasons: (1) going to large $E_{\gamma}$ an
increasing number of power terms become relevant. It is not clear a
priori how many would be needed in a given situation; (2)
$B_{\cal S}(k/2)$ with $k\geq3$ is not known beyond the
large--$\beta_0$ limit, and this limit most likely provides an
overestimate. If one varies $C_{3/2}$ in (\ref{C32}) over a large range,
also the magnitude of the power correction
$\epsilon^{\PV}_3$ in \eq{epsilon_k} varies a lot if $f_3^{\PV}$ is kept
at its natural size, $f_3^{\PV}\thickapprox 1$.

Fortunately, there is another selection criterion that can be used
to determine the allowed range in the parametrization of $B_{\cal
S}(u)$ as well as in the size of the power corrections~$\epsilon_k$:
the support properties. In the calculation itself the heavy--meson
mass is not used. However, since it sets the physical support, it
can be used to distinguish acceptable spectra from non-acceptable
ones. In Sec.~\ref{sec:mom} we perform a numerical analysis
examining the parameter space of $C_{3/2}$ and $f^{\PV}$ under constrains on the support
of the corresponding spectra, see
\eq{Sud_with_cubic_Renormalons_single_parameter} and Fig.~\ref{fig:map}
there.

\subsubsection*{Matching the resummed spectrum to fixed order}

Given the Sudakov factor, one can determine the hard coefficient function
$H\left(\alpha_s(m_b),N\right)$ as well as the residual terms $\Delta R(\alpha_s(m_b),x)$ in \eq{G_77_resummed}
 order-by-order in perturbation theory from the expansion of the differential
spectrum in \eq{G77_diff_form}. It should be noted that as far as
the terms that are regular for $x\to 1$ are concerned, the
separation between the contributions that are taken into account in
moment space and the residual terms that are included directly in
$x$ space is arbitrary. Moreover, the matching between the resummed
exponent and the fixed--order expansion can be done in a variety of
ways that differ by subleading corrections. In
Appendix~\ref{sec:matching} we derive explicit expressions to ${\cal
O}({\alpha_s}^2)$ for $H\left(\alpha_s(m_b),N\right)$ and $\Delta
R(\alpha_s(m_b),x)$ based on the available NNLO results of
Ref.~\cite{Melnikov:2005bx,Asatrian:2006sm}. In doing so we rely on
previous experience in the applications of Sudakov resummation to
inclusive distributions in QCD,
e.g.~\cite{Gardi:2001ny,Gardi:2002bg,Cacciari:2002xb,Andersen:2005bj,Andersen:2005mj},
in the following ways:
\begin{itemize}
\item{} we give preference
to moment space, so $\Delta R(\alpha_s(m_b),x)$ reduces to small
corrections that vanish at $x\to 1$. This guarantees, in particular,
smooth transition from the perturbative region $x\leq 1$ to the one
above $x=1$.
\item{} in moment space we use ``log-R'' matching~\cite{Catani:1992ua}, where the perturbative expansion of the
\emph{logarithm} of the spectral moments is constructed as a sum of the perturbative expansions of
$\ln {\rm Sud}(N,m_b)$ and of $\ln H\left(\alpha_s(m_b),N\right)$. Consequently
$H\left(\alpha_s(m_b),N\right)$ itself is constructed as an exponential function.
\end{itemize}
Following these considerations we arrive at the NNLO matching
formula of \eq{after_full_exponentiation_most_moment_space_mu}.

\subsection{Analytic structure in moment space and the
small--$E_{\gamma}$ asymptotics\label{sec:small_x}}

In general, Sudakov resummation is a parametrically--controlled
approximation in a specific kinematic region where the logarithms
are large and the coupling is small. The logarithmic--accuracy
criterion applies only if the product of the coupling times the
logarithm is small enough. The application to b decay, where the
coupling at the hard scale is $\alpha_s(m_b)\simeq 0.2$, is a
borderline case at the outset: near threshold, where the logarithms
are really large, the logarithmic--accuracy criterion does not hold
and non-perturbative corrections become important, while away from
threshold the logarithms are no more the dominant perturbative
corrections.

As discussed above, DGE is designed to deal with the first problem,
extending the applicability of the resummation deeper into the
threshold region, $x\simeq 1$. In this section we would like to address
the second, namely the application of the resummed spectrum in the
region where the logarithms are not necessarily dominant. Usually,
Sudakov--resummed spectra suffer from artifacts when evaluated away
from the Sudakov region. We will show that by imposing constraints
on the analytic structure of the Sudakov factor in moment space, one
can extend the applicability of the resummed spectrum away from the
Sudakov region, and even use it for $x\ll 1$, where in the absence
of such constraints artifacts from the resummed $\ln^k(1-x)/(1-x)$
terms are significant.

Much like the problem, the solution we suggest is general. We
nevertheless consider here the concrete case of the photon--energy
spectrum in $b\to s\gamma$. As we shall see, it provides a
particularly good example because of the strong suppression of the
differential spectrum at small~$x$, which makes potential
resummation--artifacts more pronounced.

As was observed by Melnikov and Mitov~\cite{Melnikov:2005bx}, at
small $x$, the differential $b\to s\gamma$ spectrum
$d\Gamma_{O_7}/dx$ falls as $x^3$:
\begin{equation}
\label{small_x_power_falloff}
\frac{1}{\Gamma_{O_7}}
\frac{d\Gamma_{O_7}}{dx}\,\,\simeq\,\, x^3\times
\left(\frac{C_F\alpha_s}{2\pi} +\cdots\right)
+{\cal O}(x^4).
\end{equation}
This is a general property that
holds to all orders in the perturbative expansion. In particular, the known
two--loop results~\cite{Melnikov:2005bx,Asatrian:2006sm} as well as all-order
results in the large--$\beta_0$ limit~\cite{Gardi:2004ia}
--- see Appendix \ref{sec:large_beta0_small_x} below --- all share
this cubic suppression.
This suppression is partially owing to the fact that the available phase space shrinks as $E_{\gamma} dE_{\gamma}$ for $E_{\gamma}\to 0$, and partially a consequence of the dynamics: the coupling of the photon to the flavor--changing current in (\ref{O7}) involves $F^{\mu\nu}=\partial^\mu A^\nu-\partial^\nu A^\mu$, which translates into a power of the photon momentum compared to the usual $A^{\mu}$ coupling. In the \emph{squared} matrix element this results in \emph{two} powers of $E_{\gamma}$ on top of the phase space suppression, namely $d\Gamma_{O_7}\sim  E_{\gamma}^3dE_{\gamma}$.

It is obvious that any resummation procedure that
considers only the terms that are singular at $x\to 1$, namely,
${\alpha_s}^n \ln^k(1-x)/(1-x)$, is bound to generate artifacts in the region
where these terms no longer dominate. The largest artifacts are associated with
the most subleading logarithms, $1/(1-x)$, which behave as a constant for
$x\to 0$. Such resummation artifacts are limited in size by virtue of
matching the resummed Sudakov factor to the fixed--order expansion.
However, matching to a fixed order
may not always be sufficient to guarantee a good approximation
away from the Sudakov region, especially in cases where the coupling is large,
so subleading perturbative corrections are not negligible.

The strong suppression of the differential spectrum in
\eq{small_x_power_falloff} makes such artifacts particularly important:
if \emph{only} ${\alpha_s}^n \ln^k(1-x)/(1-x)$ terms are resummed,
resummation artifacts that behave as a constant at small $x$
are expected to appear at any order. NNLO matching guarantees that
such terms only appear at ${\cal
O}({\alpha_s}^3)$ and beyond.
Nevertheless, at sufficiently small $x$ these can easily compete with the
terms in \eq{small_x_power_falloff}, and eventually dominate.
Consequently, the resummed spectrum
would develop a (small, ${\cal O}({\alpha_s}^3)$) constant $x\to 0$ tail
instead of the correct cubic fall-off. Obviously, this should be avoided.
In the following we develop a formalism
where such artifacts are systematically avoided to any order by
imposing the correct analytic structure on the moment space Sudakov
factor.

The basic observation underlying our method is that there is
a direct correspondence between the small--$x$ asymptotic behavior
and the analytic structure in moment space (where moments are defined according to
\eq{mom_def}): a pole at $N=-J$ corresponds to a small--$x$ fall-off of the form:
$d\Gamma/dx\sim x^J$.

Motivated by \eq{small_x_power_falloff}, we assume
the asymptotic behavior $d\Gamma/dx\sim x^J$ (in our case $J=3$)
and construct the Sudakov--resummed
spectrum to accommodate this behavior. In order to capture the $x^J$
suppression at $x\to 0$ at the level of the Sudakov exponent
we modify \eq{Sud_DGE} into:
\begin{align}
\label{Sud_with_cubic_small_x_general_color} &\,\,\widetilde{{\rm
Sud}}^{(J)}(N,m_b)=\exp\Bigg\{\frac{C_F}{\beta_0}
\int_0^{\infty}\frac{du}{u}\,T(u)\,\left(\frac{\Lambda^2}{m_b^2}\right)^u\,
\bigg[B_{\cal S}(u)\Gamma(-2u)\,\times\\ \nonumber &
\hspace*{-5pt}
\left(\frac{\Gamma(N+J)}{\Gamma(N+J-2u)}-\frac{\Gamma(J+1)}{\Gamma(J+1-2u)}\right)
-B_{\cal J}(u)\Gamma(-u)\,
\left(\frac{\Gamma(N+J)}{\Gamma(N+J-u)}-\frac{\Gamma(J+1)}{\Gamma(J+1-u)}\right)\bigg]\Bigg\}.
\end{align}
This guarantees that \emph{no poles are generated} for $N>-J$ and therefore already before any matching is done
the Sudakov factor would not give rise to a tail that falls slower than $x^J$.

\begin{figure}[t]
\begin{center}
\epsfig{file=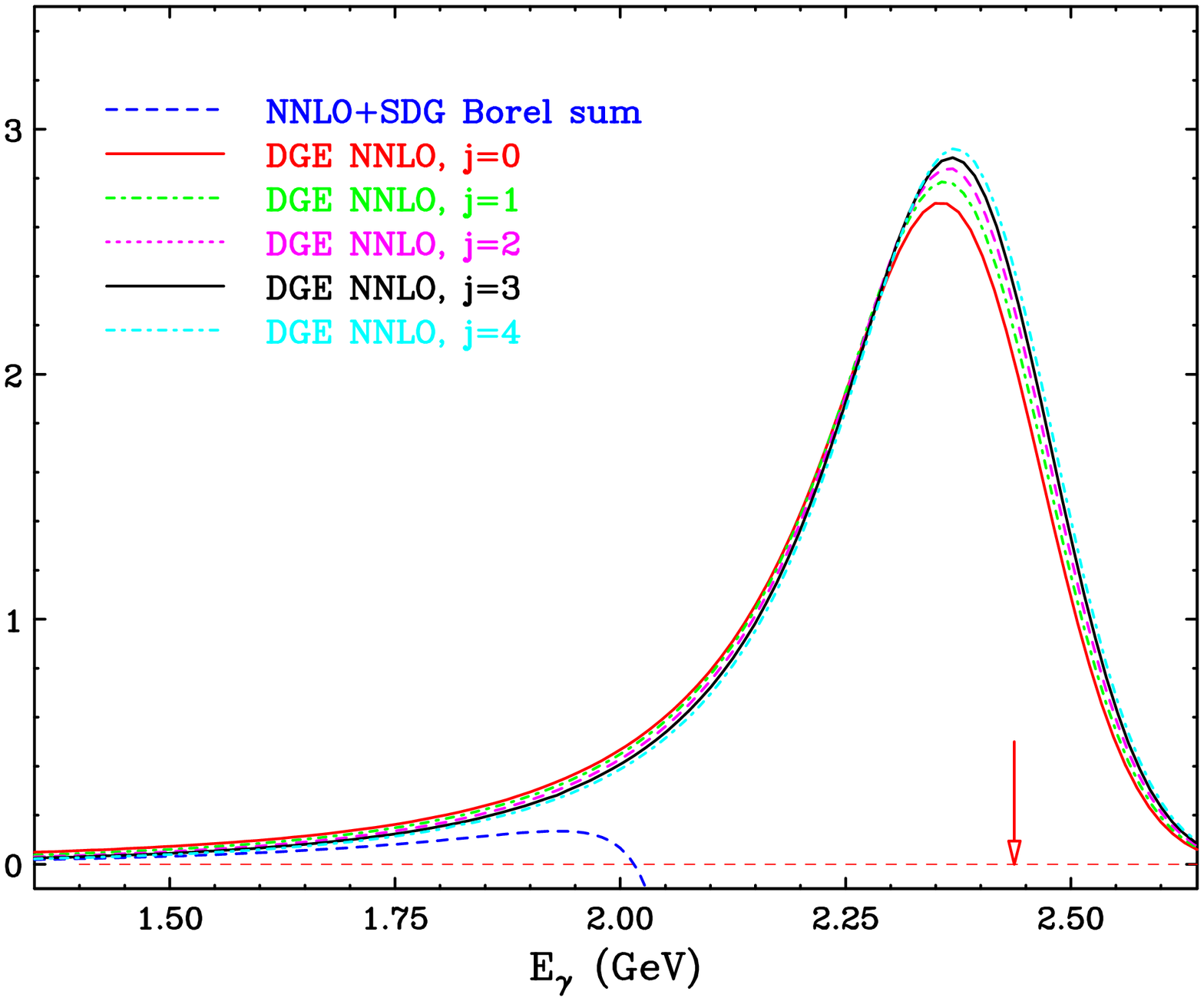,angle=90,height=6cm,width=0.49\textwidth}\hfill
\epsfig{file=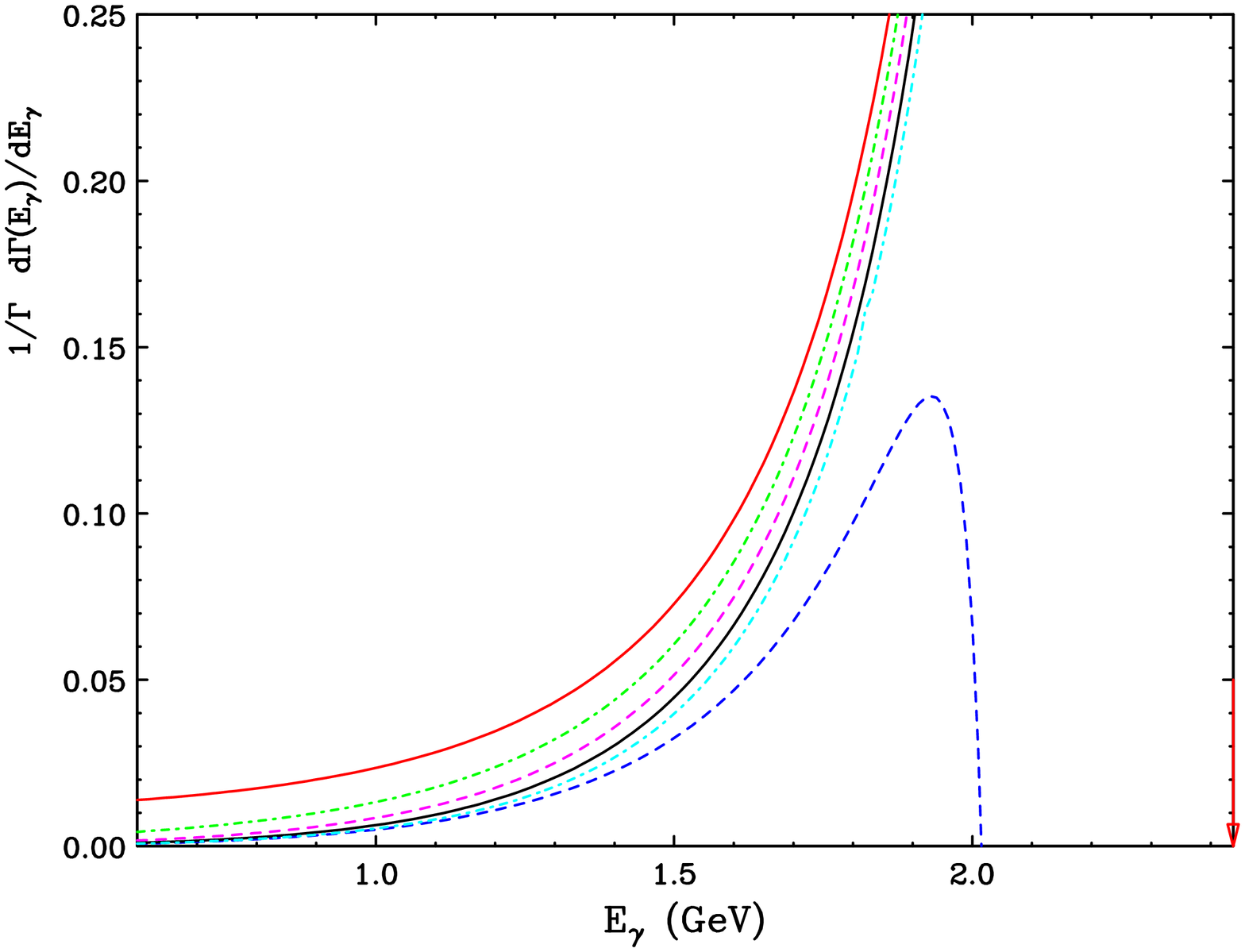,angle=90,height=6cm,width=0.49\textwidth}
\caption{\label{fig:J_dep} Comparison between the differential
spectra for $\bar{B}\longrightarrow X_s\gamma$ decay (through $O_7$
only) obtained by DGE using the exponent (\ref{Sud_with_cubic_small_x_general_color}),
matched to NNLO according
to~(\ref{tilde_after_full_exponentiation_most_moment_space_mu}) with
$J=0$ through $4$ (and $\mu=m_b$), and results obtained in $x$ space
with NNLO using Borel summation of running--coupling effects in the
SDG approximation according to Eq.~(\ref{BXg_full_with_full_NNLO}).
In the plot on the left the horizontal axis ends at the physical
endpoint $E_{\gamma}=M_B/2$ while the partonic endpoint is denoted
by an arrow. The plot on the right hand side enlarges the tail of the
distribution; it ends at the partonic end of phase space. }
\end{center}
\end{figure}

In Appendix~\ref{sec:mathcing_Jneq0} we develop the matching
procedure of \eq{Sud_with_cubic_small_x_general_color} with the NNLO results, in
analogy with what was done in Appendices~\ref{sec:basic_matching}
and~\ref{sec:more_in_mom} for the $J=0$ case.
Similarly to
\eq{Sud_with_cubic_small_x_general_color}, the matching coefficients are
constructed under a constraint on the analytic structure in moment space:
no poles should appear for $N>-J$, and so the small--$x$ asymptotic behavior
would coincide with that of the fixed--order result, $d\Gamma/dx\sim x^J$.
The final matching formula, for a general $J$, appears in
\eq{tilde_after_full_exponentiation_most_moment_space_mu}.
Note that while both the exponent (\ref{Sud_with_cubic_small_x_general_color})
and the matching coefficients entering
(\ref{tilde_after_full_exponentiation_most_moment_space_mu}) vary with $J$,
the NNLO expansion does not; only higher orders do. Moreover, the
log--enhanced terms are $J$ independent, only ${\cal O}(1/N)$ terms
depend on $J$.

\begin{figure}[t!]
\begin{center}
\epsfig{file=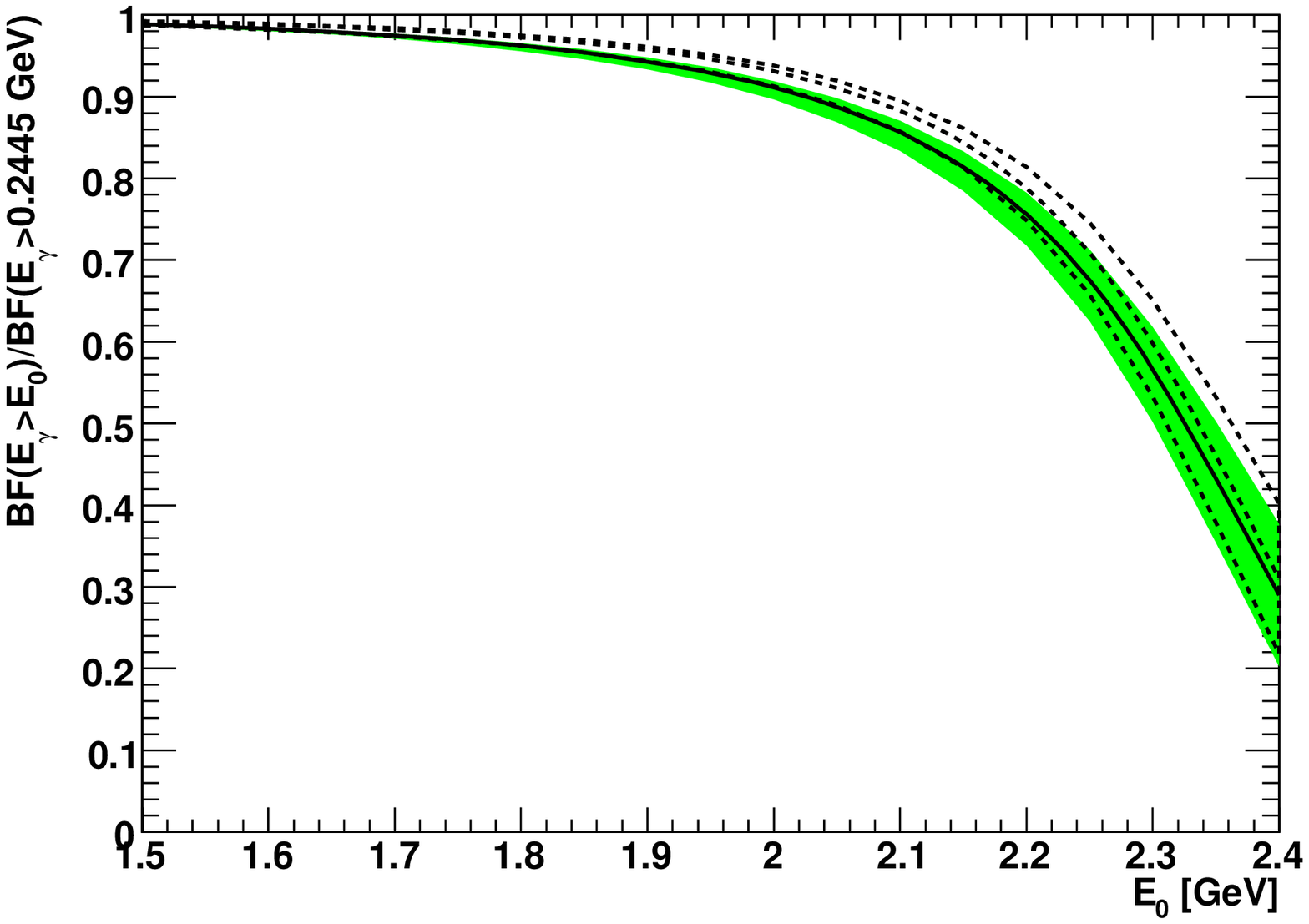,width=12.7cm}
\caption{The partial BF ratio, ${\rm BF}(E_{\gamma}>E_0)/{\rm BF}(E_{\gamma}>0)$,
corresponding to magnetic dipole operator, as a function of the cut $E_0$,
computed by DGE (\ref{Sud_with_cubic_small_x_general_color}) with $J=3$,
matched to NLO (dashed) and NNLO (full line) according to
\eq{tilde_after_full_exponentiation_most_moment_space_mu}. In both cases
the renormalization scale in the matching coefficient is $\mu=m_b$. The shaded
area represents the uncertainty band at NNLO while the two external dashed lines
represent the NLO one. The lower NLO uncertainty band trails the central line
for the NNLO matching prediction for $E_0\lesssim 2.15$ GeV. The width of the
uncertainty band is reduced roughly by a factor of two by including the NNLO corrections.
The calculation of the theoretical errors is
 explained in Sec.~\ref{sec:partial_BF} below.
\label{fig:J_eq_3_BF_NLO_NNLO}
}
\end{center}
\end{figure}
The effect of varying $J$  is shown in Fig.~\ref{fig:J_dep}. As
expected, variations in the peak region as a function of $J$ are
small, while variations in the tail are rather significant. In
particular, as shown in the plot on the right, the $J=0$ curve
approaches a constant at small $x$, the $J=1$ one falls as $x^1$,
etc. These are artifacts of the resummation. Such artifacts are
absent for $J=3$ (or larger) where the power fall-off is determined
by the perturbative expansion itself, \eq{small_x_power_falloff}.
Obviously, the natural choice in our case is $J=3$: singularities at
$N=-3$ do appear in the perturbative expansion, so they should be
allowed in the resummed spectrum. We will use the $J=3$ spectrum as
the default choice below.

Having set the final resummation and matching formulae, let us examine the
prediction for the partial BF as a function of the cut. The result is shown
in Fig.~\ref{fig:J_eq_3_BF_NLO_NNLO}. The figure compares the results obtained
at NLO and at NNLO. Note that the Sudakov factor, which is computed as a Principal
Value Borel sum, is the same
(formally it has NNLL accuracy) ---
only the matching coefficient (\ref{tilde_after_full_exponentiation_most_moment_space_mu}) is
truncated at different orders. The error is significantly reduced at small $E_0$ cuts,
where the corrections in the matching coefficient are essential.

\subsection{All--order results in the large--$\beta_0$ limit~\label{sec:large_beta0_Borel}}

Running--coupling effects often constitute a major part of the radiative
corrections in QCD~\cite{Brodsky:1982gc,Brodsky:2000cr,Beneke:1998ui}. The typical situation
is that the average gluon virtuality is lower, sometimes quite
significantly lower, than the hard scale in the process, $m$.
Thus, when the latter is used as the default renormalization scale for
coupling ($\mu=m$), one encounters large corrections involving powers of
$(\beta_0\alpha_s(m)/\pi)$, to any order in perturbation theory.
In simple cases perturbative predictions can therefore be significantly
improved by using the BLM scale instead.
Differential spectra involve other parametrically--large corrections,
such as Sudakov logarithms. Since they inherently
depend on several scales, a straightforward scale--setting procedure is
excluded. In such cases resummation is necessary.

\begin{figure}[th]
\begin{center}
\epsfig{file=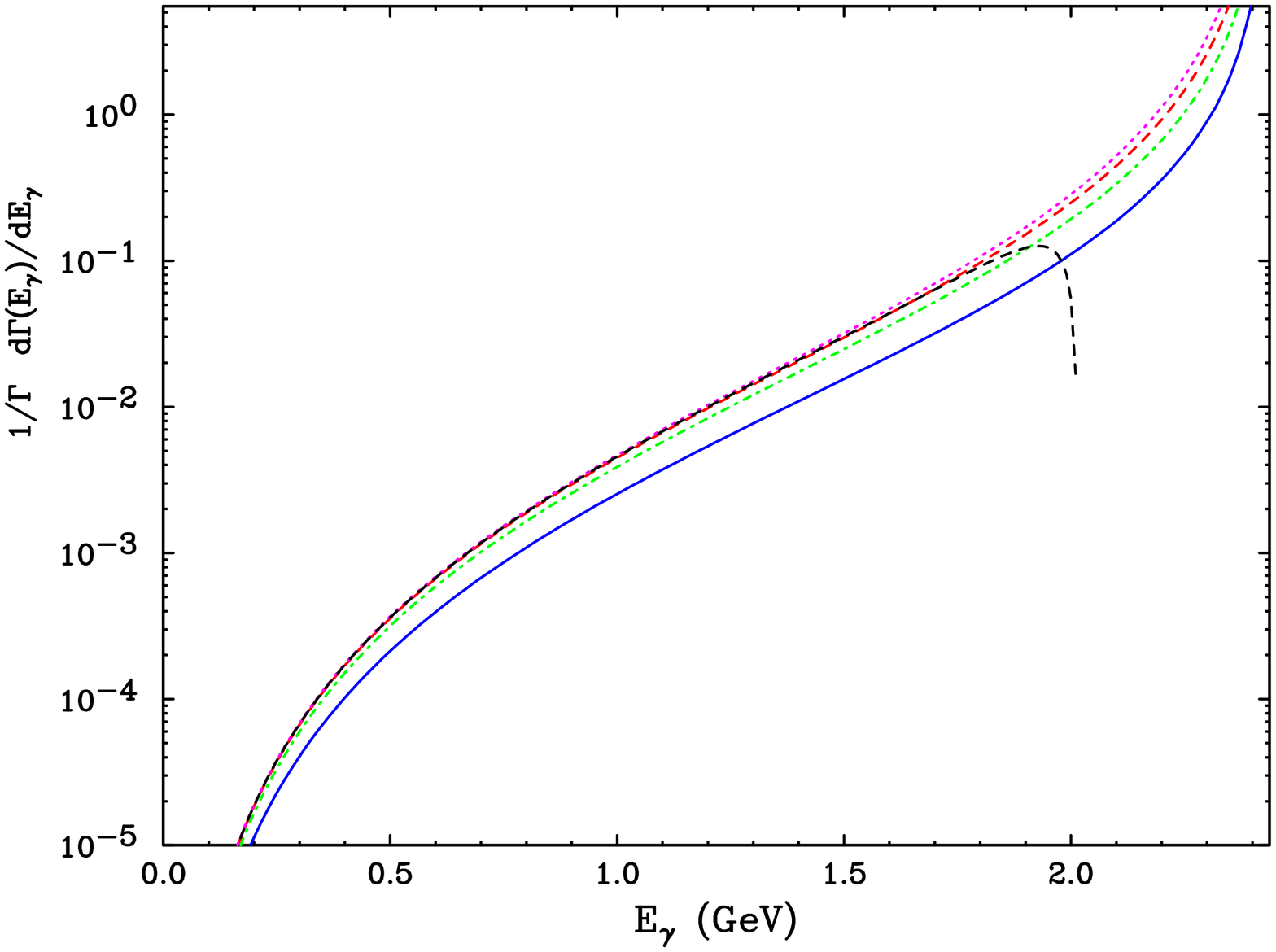,angle=90,height=6cm,
width=.49\textwidth} \hfill
\epsfig{file=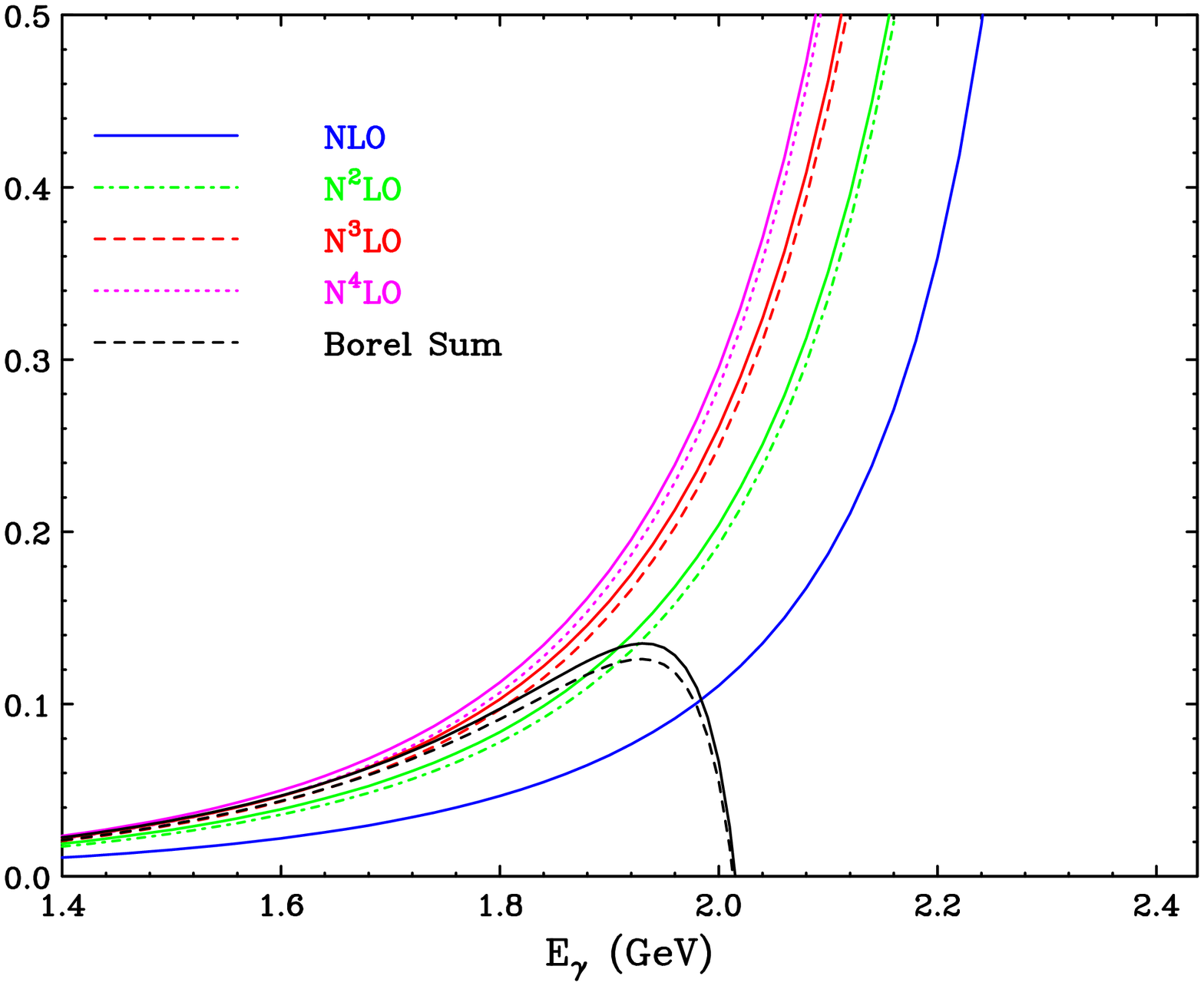,angle=90,height=6cm,
width=.49\textwidth} \caption{\label{fig:Fixed_order_x_space}
Convergence of the perturbative expansion in
\eq{large_beta_0_partial_sum}, generated by running--coupling
effects only, for the $O_7$ decay spectrum: partial sums of order
${\cal O}({\alpha_s}^n)$ with $n=1$ through 4 are shown together with
the Borel sum of~\eq{BXg_full}. Left: the whole range of energies on
a logarithmic vertical axis. Right: a restricted range on a linear
scale. In both plots the horizontal axis ends at $E_\gamma=m_b/2$,
as dictated by the support of these functions.  The
$\delta(E_{\gamma}-m_b/2)$ is not shown. In the
 plot on the right hand side the additional full lines at NNLO and beyond include the
full ${\cal O}({\alpha_s}^2)$ corrections based on the calculation of
Ref.~\cite{Melnikov:2005bx}, see \eq{full_NNLO} and (\ref{BXg_full_with_full_NNLO}) below.}
\end{center}
\end{figure}

In case of the $b\to X_s \gamma$ spectrum
${\cal O} (C_F\beta_0{\alpha_s}^2)$ running--coupling (BLM)
corrections have been computed long ago~\cite{Ligeti:1999ea}. When
using $\mu=m_b$, these corrections are very significant.
A couple of years ago running--coupling terms in the $G_{77}$
spectrum, $C_F\beta_0^{n-1}{\alpha_s}^n$,
were computed to all orders~\cite{Gardi:2004ia,Benson:2004sg}. Recently,
the full ${\cal O} ({\alpha_s}^2)$ calculation of $G_{77}$ was
performed~\cite{Melnikov:2005bx,Asatrian:2006sm}, finding that the additional,
non-BLM corrections are, instead, moderate.
Figs.~\ref{fig:Fixed_order_x_space} and \ref{fig:DGE_vs_FO}
present these results for the
normalized differential $b\to X_s \gamma$ spectrum.

The plots clearly show that the $C_F\beta_0^{n-1}{\alpha_s}^n$ corrections are
very large. In particular, if the expansion is performed
in terms of $\alpha_s(m_b)$, the ${\cal O} (C_F\beta_0{\alpha_s}^2)$ term completely
dominate the ${\cal O} ({\alpha_s}^2)$ correction. This is true even at fairly large $E_{\gamma}$,
despite the presence of formally leading Sudakov logarithms with other color factors.
With this numerical evidence it is easy to reach the conclusion that Sudakov resummation
is not needed, or cannot significantly improve the description of the spectrum (see e.g.
the discussion in Ref.~\cite{Benson:2004sg}). The purpose of this section is to demonstrate
that this conclusion is wrong, and that multiple soft and collinear
gluon emission does in fact play an important r\^ole in shaping the spectrum through
higher--order corrections. To this end we compare the fixed--order results with (1) DGE,
where Sudakov logarithms are taken into account through exponentiation in moment space;
and (2) the single--dressed--gluon (SDG) approximation, which accounts for
real--emission running--coupling $C_F \beta_0^{n-1}{\alpha_s}^{n}$ corrections,
to all orders, but neglects Sudakov logarithms. We will show that
while the SDG series is Borel--summable for \hbox{$E_{\gamma}\lsim 2$}
GeV, it does not provide a viable description of the spectrum anywhere in the peak region,
in sharp contrast with the DGE result.

\subsubsection*{The single--dressed--gluon approximation}

The all--order calculation of the $G_{77}$ decay spectrum in the large--$\beta_0$ limit
was performed in Ref.~\cite{Gardi:2004ia}. Using the scheme--invariant Borel
transform with $\Lambda$ defined in the ${\overline{\rm MS}}$ scheme the result
takes the form:
\begin{eqnarray}
\label{BXg_full}
\frac{1}{\Gamma_{O_7}}\frac{d\Gamma_{O_7}}{dx}&=&\delta(1-x)
+\left[\frac{C_F}{2\beta_0}
\int_0^{\infty}du\,T(u)\,\left(\frac{\Lambda^2}{m_b^2}\right)^u\,B(x,u)\right]_{+}
\end{eqnarray}
with $m_b$ the bottom pole mass,
\begin{equation}
\label{beta0}
\beta_0=\frac{11}{12}N_c-\frac16N_f
\end{equation}
and
\begin{eqnarray}
\label{B_BXSG}
 B(x,u)&\equiv&{\rm e}^{\frac53 u}\,\frac{\sin\pi u}{\pi u}
\,x^3\,(1-x)^{-u} \int_0^1 d\alpha\,
\alpha\,(1-\alpha)^{-u}\,\times \\ \nonumber &&\hspace*{-30pt}
\bigg[\frac{1}{(1-x\alpha)^2}\left(1-4\alpha+\alpha^2
-\frac{(1-\alpha)^2}{1-u}\right)
+\frac{1-\alpha}{(1-x\alpha)(1-u)} +\frac
2{1-x}\frac{1}{1-x\alpha} +\frac 1{1-x}\bigg]\\ \nonumber
&=&
{\rm e}^{\frac53 u}\,\frac{\sin\pi u}{\pi u}
\,x^3\,(1-x)^{-u}\Bigg\{\frac{1}{1-x}\frac{1}{(1-u)(2-u)}+\\ \nonumber&&
\left[-\frac{(1-4x+x^2)}{x^3}\left(\frac{1}{1-x}+\frac{1}{1-u}\right)+\frac{2(1-x)^2}{(1-u)^2x^3}\right]
\,\, _2F_1\Big([1, 1],[2-u],x\Big)\\ \nonumber&&
+\frac{(1-4x+x^2)}{x^3}\frac{1}{1-x}+\frac{(x+1)(x^2-3x+1)}{(1-u)x^3(1-x)}-\frac{(2-x)}{(2-u)x^2}-
\frac{2(1-x)}{(1-u)^2x^3}
\Bigg\}.
\end{eqnarray}
In Ref.~\cite{Gardi:2004ia} the Borel function was expressed as an
integral over a single Feynman parameter (the first expression
above). Here we performed this integral and wrote the Borel function
in terms of a hypergeometric function.

Strictly within the large--$\beta_0$ limit $T(u)=1$. It is
straightforward to take into account running--coupling effects
beyond this level by an appropriate choice of the function $T(u)$ in
the Borel integral\footnote{Of course, this can be done in a
renormalization--scheme invariant way only as far as two--loop
effects ($\beta_1$) are concerned.}. To account for $\beta_1$ terms
to all orders it is sufficient to use the $T(u)$ of
\eq{tHooft_coupling}, see Ref.~\cite{Grunberg:1992hf}. This will be done
in the perturbative expansion, \eq{large_beta_0_partial_sum}, and in
the numerical analysis that follows.

Let us now examine the result of \eq{B_BXSG}. The first observation
is that this Borel function has \emph{no renormalon singularities}.
To see this note first that (despite appearance) there is no double
pole at $u=1$ inside the curly brackets: for $u\longrightarrow 1$
the hypergeometric function reduces to $_2F_1\Big([1,
1],[2-u],x\Big)\longrightarrow 1/(1-x)$, so the double pole cancels.
Now, since the curly brackets contain just simple poles at integer
values of $u$, upon taking the $\sin \pi u$ factor into account, one
concludes that there in no renormalon in $B(x,u)$.

The absence of Borel singularities may be interpreted as an
indication that the perturbative expansion in $x$ space should be
good. As we shall see this is not the case. First of all, at large
$x$ the effective scales are parametrically lower than
$m_b$~\cite{Gardi:2004ia}: the dynamics is dominated by momenta of
order $m_b\sqrt{1-x}$, the jet mass, and $m_b(1-x)$, the soft scale
associated with the momentum distribution of the b quark. The
analytic expression in (\ref{B_BXSG}) immediately indicates the
presence of the former through its $(1-x)^{-u}$ dependence; to
isolate the latter it is convenient to first apply the identity:
\begin{equation}
\label{2F1_identity} _2F_1\Big([1,1],[2-u],x\Big)
=(1-u)\left[-\frac{1}{u}\, _2F_1\Big([1,
1],[1+u],1-x\Big)+\frac{\pi}{\sin\pi u}(1-x)^{-u}x^{u-1}\right]
\end{equation}
in (\ref{B_BXSG}), revealing the dependence on $(1-x)^{-2u}$.
Moreover, the absence of renormalons \emph{does not} imply the
existence of the Borel sum: the Borel integral (\ref{BXg_full}) may
not converge at $u\to \infty$. It turns out that the Borel integral
exists for sufficiently small $x$ values, but it does not exist at
large $x$. We shall return to discuss this issue below, after
considering the perturbative expansion order by order.

As usual, renormalons \emph{do appear} upon taking
moments~\cite{Gardi:2004ia}: when integrating over all values of
$x$, \emph{including the endpoint region} $x\to 1$, soft gluons
necessarily contribute and consequently, infrared renormalon
singularities at integer and half integer values of $u$ are
generated. We emphasize that this is not a disadvantage of the
moment space formulation, but rather its strength: renormalon
singularities are a useful tool to understand the r\^ole of
non-perturbative power corrections.

\subsubsection*{Perturbative expansion in $x$ space in the large--$\beta_0$ limit}

The perturbative expansion of \eq{BXg_full} is:
\begin{equation}
\label{large_beta_0_partial_sum}
\frac{1}{\Gamma_{O_7}}\frac{d\Gamma_{O_7}}{dx}=\delta(1-x)+
\frac{C_F}{\beta_0}\,\sum_{n=1}^{\infty}a_n\,
\left[r^{\beta_0}_n(x)\right]_{+}\times \Big(1+{\cal
O}(1/\beta_0)\Big),
\end{equation}
where
\begin{align}
a_n &\equiv \frac{1}{n!}
\int_0^{\infty}du\,T(u)\,\left(\frac{\Lambda^2}{m_b^2}\right)^u\,
u^n= \left(-\frac{d}{d\ln m_b^2}\right)^n\left(
\frac{\alpha_s(m_b)\beta_0}{\pi}\right) =
\left(
\frac{\alpha_s(m_b)\beta_0}{\pi}\right)^n+\cdots,
\end{align}
where the dots stand for higher--order terms, ${\cal
O}({\alpha_s}^{n+1})$ and so on, containing powers of
$\beta_1/\beta_0^2$. Up to ${\cal O}({\alpha_s}^2)$ such terms do not
appear, and \eq{large_beta_0_partial_sum} reduces to
\begin{equation}
\frac{1}{\Gamma_{O_7}}\frac{d\Gamma_{O_7}}{dx}=\delta(1-x)+\frac{\alpha_s(m_b)}{\pi}
C_F \left[r^{\beta_0}_1(x)\right]_{+}
+\left(\frac{\alpha_s(m_b)}{\pi}\right)^2C_F \beta_0
\left[r^{\beta_0}_2(x)\right]_{+}+\cdots.
\end{equation}
The coefficients $r^{\beta_0}_n$ can be obtained order by order from
the expansion of
 $\frac12 B(x,u)=\sum_{n=0}^{\infty} r^{\beta_0}_n(x)\, u^n/n!$ in \eq{B_BXSG}.
To this end we need the expansion of the hypergeometric function,
which is available based on Ref.~\cite{Kalmykov:2006pu}.
We present this expansion and the resulting coefficients up to
${\cal O}({\alpha_s}^4)$ in Appendix~\ref{sec:bar_r_reg_expression}.

It is straightforward to include the remaining $C_F^2$ and $C_AC_F$
terms at ${\cal O}({\alpha_s}^2)$ based on the results of
Ref.~\cite{Melnikov:2005bx}\footnote{We wish to thank Melnikov and
Mitov for providing us with their final expressions in the form of a
Maple file.}:
\begin{eqnarray}
\label{full_NNLO}
\frac{1}{\Gamma_{O_7}}\frac{d\Gamma_{O_7}}{dx}&=&\delta(1-x)+\frac{\alpha_s(m_b)}{\pi}
C_F \left[r_1(x)\right]_{+}
\\\nonumber
&+&C_F \left[N_f r_2^{N_f}(x)+C_F r_2^{C_F}(x) +C_A
r_2^{C_A}(x)\right]_{+}\left(\frac{\alpha_s(m_b)}{\pi}\right)^2
+\cdots\\\nonumber &=&\delta(1-x)+\frac{\alpha_s(m_b)}{\pi} C_F
\left[r^{\beta_0}_1(x)\right]_{+}
\\\nonumber
&+&C_F \left[\beta_0 r^{\beta_0}_2(x)+C_F r_2^{C_F}(x) +C_A
\left(r_2^{C_A}(x)
+\frac{11}{2}r_2^{N_f}(x)\right)\right]_{+}\left(\frac{\alpha_s(m_b)}{\pi}\right)^2
+\cdots.
\end{eqnarray}
The partial sums of \eq{large_beta_0_partial_sum} as well as the
Borel sum of \eq{BXg_full} include the large--$\beta_0$,
$r^{\beta_0}_n(x)$ terms at each order. Thus, to have a complete
result to ${\cal O}({\alpha_s}^2)$ we now include the non-BLM ($C_F$
and $C_A$) terms in the square brackets of \eq{full_NNLO}. When
using Borel summation for the running--coupling contributions the
normalized spectrum at NNLO can be expressed as:
\begin{eqnarray}
\label{BXg_full_with_full_NNLO}
\frac{1}{\Gamma_{O_7}}\frac{d\Gamma_{O_7}}{dx}&=&\delta(1-x)
+\left[\frac{C_F}{2\beta_0}
\int_0^{\infty}du\,T(u)\,\left(\frac{\Lambda^2}{m_b^2}\right)^u\,B(x,u)\right]_{+}
\\ \nonumber &+&C_F \left[C_F r_2^{C_F}(x) +C_A \left(r_2^{C_A}(x)
+\frac{11}{2}r_2^{N_f}(x)\right)\right]_{+}\left(\frac{\alpha_s(m_b)}{\pi}\right)^2
+\cdots.
\end{eqnarray}
The results are shown together with the pure running--coupling
contributions in the plot on the right hand side in
Fig.~\ref{fig:Fixed_order_x_space}. As already concluded in
Ref.~\cite{Melnikov:2005bx}, these additional contributions are
moderate, so at least at this order, the running--coupling
contributions dominate.

Let us now examine the convergence of the expansion and the
possibility to sum up the series \`a la Borel.
Fig.~\ref{fig:Fixed_order_x_space} summarizes the numerical results
using $m_b=4.875$ GeV and $\Lambda=0.332$ GeV with $N_f=4$ and
two--loop running coupling. The partial sums obtained by truncation
of \eq{large_beta_0_partial_sum} at increasing orders converge well
at small energies $E_{\gamma}\lsim 1.5$ GeV and less so as the
energy increases. Nevertheless, the increasing--order contributions
do decrease monotonically. This decrease is  related of course to
the absence of renormalons in \eq{B_BXSG}. This usually means that
the series is Borel summable. However, since it is a divergent
series there is no straightforward relation between partial sums and
the Borel sum.

As anticipated, the series is in fact Borel summable only below a
certain energy, \hbox{$x<x_{\max}$}:
 the convergence of the Borel integral at large $u$ is guaranteed only
owing to the suppression by $(\Lambda^2/m_b^2)^u$; as already
observed in Ref.~\cite{Gardi:2004ia}, because of the presence of
$(1-x)^{-2u}$ contributions in $B(x,u)$ (see \eq{2F1_identity}), it
is predominantly the soft scale $m_b(1-x)$ which sets the argument
of the coupling at large $x$. Based on these considerations we can
deduce from \eq{B_BXSG} an estimate of where the Borel sum will
cease to exist:
\begin{equation}
\label{E_gamma_Borel_x_breakdown}
E_{\gamma}^{\max}\simeq \frac{m_b}{2}\left(1-{\rm e}^{5/6}\Lambda/m_b\right)\simeq 2.06\,\, {\rm GeV}.
\end{equation}

Indeed, as shown in Fig.~\ref{fig:Fixed_order_x_space} the Borel sum
is close to the high--order partial sums at small energies and to
lower--order ones at higher energies\footnote{For very low
$E_{\gamma}\lsim 0.3$ GeV the Borel sum is still larger than the
${\cal O}({\alpha_s}^4)$ partial sum; as the energy increases it
gradually approaches the ${\cal O}({\alpha_s}^3)$ partial sum,
crossing it around $E_{\gamma}\simeq 1.65$ GeV; around
$E_{\gamma}\simeq 1.9$ GeV it crosses the ${\cal O}({\alpha_s}^2)$
partial sum and soon after turns over and starts decreasing; finally
close to 2 GeV it crosses the leading--order result and continues to
decrease.}, until it eventually reaches a peak, bends downwards and
becomes negative. Soon after it breaks down completely owing to
non-convergence of the $u$ integral at $u\longrightarrow \infty$, in
accordance with \eq{E_gamma_Borel_x_breakdown}.

We note that increasing--order partial sums become quite different
from each other at $E_{\gamma}\gsim 2$ GeV; this obviously means
they cannot be thought of as an approximation to the spectrum. Also
the corresponding Borel sum, although unique where it exists, does
not look anything like a physical spectrum would. Its complete breakdown
according to (\ref{E_gamma_Borel_x_breakdown}) is indicative of the
low scales involved. But even if the bottom mass were much higher,
this SDG approximation cannot be expected to describe the Sudakov region.
Running--coupling corrections are not the most important corrections for
$E_{\gamma}\longrightarrow m_b/2$. In the large--$\beta_0$ limit
the most singular terms at large $x$ are ${\alpha_s}^n\, C_F \,{\beta_0}^{n-1}
\ln^{n}(1-x)/(1-x)$, while the full perturbative expansion contains
double logarithms of form ${\alpha_s}^n\, {C_F}^n\,\ln^{2n-1}(1-x)/(1-x)$, as
well as other subleading logarithms which are associated with
multiple soft and collinear emission. With increasing order, the
latter become more important at large $x$ compared to the former.
Finally, returning to Fig.~\ref{fig:DGE_vs_FO}, we observe the
remarkable difference between increasing--order partial sums and the
DGE result in the peak region. The gap between them should be
understood as the contribution of multiple soft and collinear gluon
emission, with a particular regularization of the divergent sum. In
contrast with the fixed--order partial sums and the SDG Borel sum, the
DGE result provides a useful approximation to the spectrum in the peak region.

\section{Resummed spectra for individual matrix elements other than $G_{77}$~\label{sec:others}}

In Sec.~\ref{sec:resummed_G77} we considered in some detail the
calculation of the normalized spectrum of the magnetic dipole
operator. It is well known that in contrast with the total width,
the spectrum is not so sensitive to the details of the
short--distance interaction, and therefore the $G_{77}$ spectrum
computed above can be thought of as an approximation to the physical
spectrum. However, to make precision estimates of the partial BF as
a function of the cut, it is important to compute the spectra of
other matrix elements as well. Since fixed--order results for all
matrix elements but $G_{77}$ are available to ${\cal O}(\alpha_s)$
only, the accuracy we can hope to achieve in the description of the
spectra of individual $G_{ij}$ matrix elements is not as high.

We will make the approximation where the matrix elements $G_{ij}$ (for any $i$ and $j$ other than 77)
are written as ${\cal O}(\alpha_s)$ hard coefficient functions, which are ${\cal
O}(N^0)$ for $G_{i7}$ interference terms and ${\cal O}(N^{-1})$ for other
terms, times the \emph{same} Sudakov factor discussed in the context of $G_{77}$.
This approximation is motivated by the following observations:
\begin{itemize}
\item{} Independently of the nature of the short--distance interaction,
all non--integrable real--emission corrections at any order, namely corrections to the moments
that scale as $N^0$ times logarithms at large $N$, are necessarily associated with soft and collinear
radiation around the same Born--level configuration involving a b quark in the initial state and
an unresolved quark jet in the final state.
\item{} All operators mix into $O_7$ under
renormalization. Moreover, the ${\cal O}(\alpha_s)$ result of $G_{ij}$
is dominated by the virtual diagrams, which are proportional to the tree level $G_{77}$.
This observation was already used in Ref.~\cite{Bieri:2003ue}
arguing that ${\cal O}(N_f \alpha_s^2)$ corrections can be well approximated by computing
both real and virtual diagrams for~$O_{7}$  and only virtual diagrams for $O_{1,2,8}$.
\end{itemize}
It should be noted that while the ${\cal O}(N^{0})$ interference terms do indeed involve the same
Sudakov factor, ${\cal O}(N^{-1})$ contributions associated with
integrable bremsstrahlung corrections, do not share this property, and in general
involve a new jet function. In this respect, our calculation of the resummed spectra of $G_{ij}$
that do not involve $O_7$ is a cruder approximation; as we shall see the
relative weight of these contributions, especially in the Sudakov region, is small.

The structure of this section is as follows: in Sec.~\ref{sec:G_ij_matching}
we discuss the small
$E_{\gamma}$ behavior of the different matrix elements and then
extend the matching procedure of Sec.~\ref{sec:resummed_G77}
accordingly, using the known ${\cal O}(\alpha_s)$ coefficients. In
Sec.~\ref{sec:G_ij_spectra} we present numerical results for the
resummed spectra of individual matrix elements, based on this ${\cal
O}(\alpha_s)$ matching procedure.

\subsection{The small $E_{\gamma}$ asymptotic behavior and matching at
${\cal O}(\alpha_s)$~\label{sec:G_ij_matching}}

Looking at the small photon--energy limit, $E_\gamma\to 0$, we find
the following asymptotic behavior:
\begin{eqnarray}
\label{small_E_asympt} &&\frac{dG_{ij}(E_{0},m_b)}{dE_{0}}={\cal
O}(E_{0}^3); \\ \nonumber &&{\mathrm{except}}
\,\,{\mathrm{for}}\qquad\left\{
\begin{array}{l}{\displaystyle
\frac{dG_{88}(E_{0},m_b)}{dE_{0}}=\frac{1}{E_{0}}}\left[
C_F\frac{\alpha_s}{\pi}\frac29
\left(1-\ln\frac{m_b}{m_s}\right)\,+\,{\cal O}({\alpha_s}^2)\right];
\\\nonumber\\\nonumber {\displaystyle
\frac{dG_{78}(E_{0},m_b)}{dE_{0}}=\frac{4E_{0}}{m_b^2}}\left[C_F\frac{\alpha_s}{\pi}
\frac13\,+\,{\cal O}({\alpha_s}^2)\right].
\end{array}
\right.
\end{eqnarray}
As discussed in Sec. 2.2, the phase--space suppression for $E_\gamma\to 0$ is $E_{\gamma}dE_{\gamma}$, and any additional suppression or enhancement of the $G_{ij}$ spectrum in this limit depends on the dynamics, and therefore on the operators $O_i$ and $O_j$. To understand the difference between $O_8$ and $O_7$ in this respect, note that in the $O_8$ amplitude the gluon couples to the flavor--changing current through $G^{\mu\nu}$ while the photon is emitted as bremsstrahlung, whereas in the $O_7$ case it is the other way around. While the $O_7$ coupling (\ref{O7}) gives rise to a linear dynamical \emph{suppression} with the photon energy, photon bremsstrahlung involves a soft singularity, namely a $1/E_{\gamma}$ \emph{enhancement}. Combining this behavior of the amplitudes with the phase--space factor one immediately obtains the different $E_\gamma\to 0$ limits in (\ref{small_E_asympt}) for $G_{77}$, $G_{88}$ and the interference~$G_{78}$.
At ${\cal O}(\alpha_s)$, the behavior of matrix elements containing $O_{1,2}$ is similar to that of $O_7$. The reason for this is that the photon cannot be emitted as bremsstrahlung, but must instead couple to the virtual charm propagating in the loop (the diagram with just gluon coupling to the charm loop vanishes).

Thus in most cases there is a cubic power suppression of the
spectrum for extremely soft photons, $E_{\gamma}\to 0$.
When matching Sudakov resummation for the high photon energy
endpoint ($E_\gamma=m_b/2$) into the fixed--order results, it is
therefore useful to take $J=3$ as for the $O_7$ contribution. As
explained in detail in Sec.~\ref{sec:small_x}, in this way artifacts
of resummation away from the hard photon endpoint are avoided (the
more pronounced soft--energy tail of $O_8$ and its interference with
$O_7$ will be accounted for at fixed order).

In general, the matching involves a separation of the real--emission
contributions between momentum and moment space (see Sec.~3 and
Appendix C in Ref.~\cite{Andersen:2005bj}). The separation we use is
defined such that the leading contributions at small $\Delta$ are
always treated in moment space:
\begin{equation}
\label{separation} \phi_{ij}(\Delta) = \tilde{\eta}_{ij}(\Delta) +
\tilde{\xi}_{ij}(\Delta) \quad {\mathrm{with}}\quad
\tilde{\xi}_{ij}(\Delta) = O(\Delta^2).
\end{equation}
Introducing the Mellin transform of $\tilde{\eta}_{ij}(\Delta)$ as
in \cite{Andersen:2005bj}:
\begin{equation}
\label{mu_def} \tilde{\mu}_{ij}(N)\equiv
\int_0^1dx\,x^{N-1}\left.\frac{d\tilde{\eta}_{ij}(\Delta)}{d\Delta}
\right\vert_{\Delta=1-x}; \qquad
\tilde{\eta}_{ij}(\Delta)=\int_{c-i\infty}^{c+i\infty}
\frac{dN}{2\pi\,
i}\,\frac{(1-\Delta)^{1-N}}{N-1}\,\tilde{\mu}_{ij}(N),
\end{equation}
we get:
\begin{eqnarray}
\label{Gij_separated}
G_{ij}(E_{\min},m_b)&=&C_F\frac{\alpha_s(m_b)}{\pi}\left[\frac38\delta_{j=7}{\rm
Re} \left\{r_i+\gamma_{i7}^{(0)}\ln\frac{m_b}{\mu}\right\}
+{\tilde{\eta}}_{ij}(\Delta)\right]\nonumber \\\hspace*{-80pt}&&+\,
C_F\frac{\alpha_s(m_b)}{\pi}{\tilde{\xi}}_{ij}(\Delta)
+{\cal O}({\alpha_s}^2)\nonumber\\
&&\hspace*{-85pt}= \int_{c-i\infty}^{c+i\infty} \frac{dN}{2\pi\,
i}\,\frac{(1-\Delta)^{1-N}}{N-1} \left[C_F\frac{\alpha_s(m_b)}{\pi}
V_{ij}^{(1)}(N) +{\cal O}({\alpha_s}^2)
 \right]\times\widetilde{\rm Sud}^{(J)}(N,m_b)\nonumber \\\hspace*{-80pt}&&
+\,C_F\frac{\alpha_s(m_b)}{\pi}{\tilde{\xi}}_{ij}(\Delta) +{\cal
O}({\alpha_s}^2),
\end{eqnarray}
where in the second expression used \eq{mu_def}, included the
Sudakov factor resumming large logarithms at ${\cal O}({\alpha_s}^2)$
and beyond, and defined
\begin{equation}
V_{ij}^{(1)}(N)\equiv \frac38\delta_{j=7}{\rm Re}
\left\{r_i+\gamma_{i7}^{(0)}\ln\frac{m_b}{\mu}\right\}
+{\tilde{\mu}}_{ij}(N).
\end{equation}
Note that
 $\tilde{\mu}_{ij}(N)$ are all ${\cal O}(1/N)$ (Appendix C in~\cite{Andersen:2005bj}) so
$V_{ij}^{(1)}(N)$ is of ${\cal O}(1)$ at large $N$ for $j=7$ (the
interference terms for any $i$ are  ${\cal O}(1)$ at large $N$)
while it is ${\cal O}(1/N)$ otherwise.

For $77$, $78$ and $88$ matrix elements the NLO contribution is
treated \emph{entirely} in moment space:
$\tilde{\eta}_{ij}(\Delta)=\tilde{\phi}_{ij}(\Delta)$ and
$\tilde{\xi}_{ij}(\Delta)=0$, so
\begin{eqnarray}
\label{mu78}
\tilde{\mu}_{78}(N)&=&\frac{2}{3}\left[-\frac{1}{(N-1)N}(\Psi(N)+\gamma_E)
+\frac{1}{N^2}+\frac{1}{4}\frac{1}{N+2} +\frac{1}{N}\right].
\end{eqnarray}
and
\begin{eqnarray}
\label{mu88}
\tilde{\mu}_{88}(N)&=&-\left[\frac{2}{9}\frac{1}{N}-\frac{2}{9}\frac{1}{N-1}-\frac{1}{9}
\frac{1}{N+1} \right] \left(\ln\frac{m_b}{m_s}-\frac12
\Psi(N)-\frac{1}{2}\gamma_E\right)\\ \nonumber &&
-\frac{1}{18}\frac{1}{N+2}+\frac{1}{6}\frac{1}{N}-\frac{2}{9}\frac{1}{N-1}+
\frac{1}{36}\frac{1}{N+1}+\frac{1}{9N^2}-\frac{1}{18}\frac{1}{(1+N)^2}.
\end{eqnarray}
Note that the latter expression has a pole at $N=1$, owing to the
soft--photon singularity. However, this should not affect the
calculation of $G_{ij}(E_{\min},m_b)$ with non-zero  $E_{\min}$ so
long as the contour in \eq{Gij_separated} is to the right of $N=1$.

For all the other contributions, namely the ones associated with the
operators $O_1$ and $O_2$ and their interference with $O_7$ and
$O_8$ (this includes $\phi_{22}$, $\phi_{27}$, $\phi_{11}$,
$\phi_{12}$ $\phi_{17}$, $\phi_{18}$ and $\phi_{28}$) we introduce a
separation according to \eq{separation} with a non-trivial
$\tilde{\xi}_{ij}(\Delta)$. The specific separation used in Appendix
C in~\cite{Andersen:2005bj} was based on defining
$\tilde{\eta}_{ij}(\Delta)$ as the leading--order term in the
expansion of $\phi_{ij}(\Delta)$ at small $\Delta$. This simple
procedure is not adopted here, since it would introduce a constant
differential spectrum at very small $E_\gamma$, in contradiction
with the power suppressed behavior of~\eq{small_E_asympt}. Instead
we will define $\tilde{\eta}_{ij}(\Delta)$ such that it would
capture both the small $\Delta$ limit and certain features of the
$\Delta\to 1$ limit. As shown in Appendix C
in~\cite{Andersen:2005bj} for small $\Delta$, ${\phi}_{ij}(\Delta)$
is ${\cal O}(\Delta)$, so according to \eq{separation} we require
\begin{equation}
\label{small_Delta} \mathrm{for}\,\,\mathrm{small}\,\,
\Delta:\hspace*{110pt} \tilde{\eta}_{ij}(\Delta)\simeq
{\phi}_{ij}^{\prime}(\Delta=0)\times \Delta +{\cal O}(\Delta^2).
\end{equation}
As explained above (see \eq{small_E_asympt}) for $\Delta\to 1$ the
differential spectrum falls as the \emph{third} power $1-\Delta$ so
the integrated spectrum approaches a constant as the fourth power of
$(1-\Delta)$. We therefore require:
\begin{equation}
\label{large_Delta} \mathrm{for}\,\,\mathrm{large}\,\, \Delta:
\hspace*{90pt} \tilde{\eta}_{ij}(\Delta)\simeq
{\phi}_{ij}(\Delta=1)+{\cal O}\Big((1-\Delta)^{J+1}\Big),
\end{equation}
where $J=3$, such that $(1-\Delta)^4$ contributions appear through
both the moment--space expression ($\tilde{\eta}_{ij}$) and the
momentum--space residual ($\tilde{\xi}_{ij}$), while ones with
$(1-\Delta)^h$ for $h=1,2,3$, which are absent in the physical
spectrum, do not appear in either. Both
${\phi}_{ij}^{\prime}(\Delta=0)$ and ${\phi}_{ij}(\Delta=1)$ can be
explicitly computed from the expressions in Eq.~(C.1) in
\cite{Andersen:2005bj}. To accommodate (\ref{small_Delta}) and
(\ref{large_Delta}) we define:
\begin{equation}
\label{tilde_eta_def} \tilde{\eta}_{ij}(\Delta)\equiv
\phi_{ij}(\Delta=1)\bigg[1-(1-\Delta)^{J+1}\,\Big(1+(J+1)\Delta\Big)\bigg]
+\phi^{\prime}_{ij}(\Delta=0)\,(1-\Delta)^{J+1}\,\Delta.
\end{equation}
In computing $G_{ij}(E_0,m_b)/G_{ij}(E_{\min},m_b)$ using
\eq{Gij_separated}, $\tilde{\xi}_{ij}(\Delta)$ is readily obtained
using $\tilde{\xi}_{ij}(\Delta)= \phi_{ij}(\Delta) -
\tilde{\eta}_{ij}(\Delta)$, while $\tilde{\mu}_{ij}(N)$ is computed
according to \eq{mu_def} by the Mellin conjugate of the derivative
of \eq{tilde_eta_def}, namely:
\begin{align}
\label{mu_def_explicit} \tilde{\mu}_{ij}(N) &\equiv
\int_0^1d\Delta\,(1-\Delta)^{N+J-1}\Big[
\phi_{ij}(\Delta=1)\,(J+1)(J+2)\Delta\,+\,
\phi^{\prime}_{ij}(\Delta=0)\,\big(1-(J+2)\Delta\big)
\Big]\nonumber \\
&= \frac{(J+1)(J+2)}{(N+J)(N+J+1)}\phi_{ij}(\Delta=1)+
\frac{N-1}{(N+J)(N+J+1)}\phi^{\prime}_{ij}(\Delta=0) .
\end{align}
As required, the first moment $\tilde{\mu}_{ij}(N=1)$, corresponding
to the total width,
 is given by $\phi_{ij}(\Delta=1)$, while the large--$N$ asymptotic behavior is
 determined by $\phi^{\prime}_{ij}(\Delta=0)$. Importantly, when using the matching
 procedure
 of \eq{Gij_separated},
 where $\widetilde{\rm Sud}^{(J)}(N,m_b)$ is given by
 \eq{Sud_with_cubic_small_x_general_color} and $\tilde{\mu}_{ij}(N)$ by \eq{mu_def_explicit},
  we are guaranteed that no spurious singularities at $N>-J$ would appear
  (singularities do appear for $N=-J$, $N=-J-1$ etc.), which could
 alter the small--$E_{\gamma}$ behavior.

\subsection{Resummed spectra for individual matrix elements~\label{sec:G_ij_spectra}}

In the previous sections we established a procedure for matching the
Sudakov--resummed spectrum with the known fixed--order expansion.
For $G_{77}$ this is done to NNLO, ${\cal O}({\alpha_s}^2)$, while for
the matrix elements of other operators and their interference with
$O_7$, to ${\cal O}(\alpha_s)$. This allows us to study the shape of
each contribution and eventually have better control on the cut
dependence of the total BF.
\begin{figure}[t!]
\begin{center}
\epsfig{file=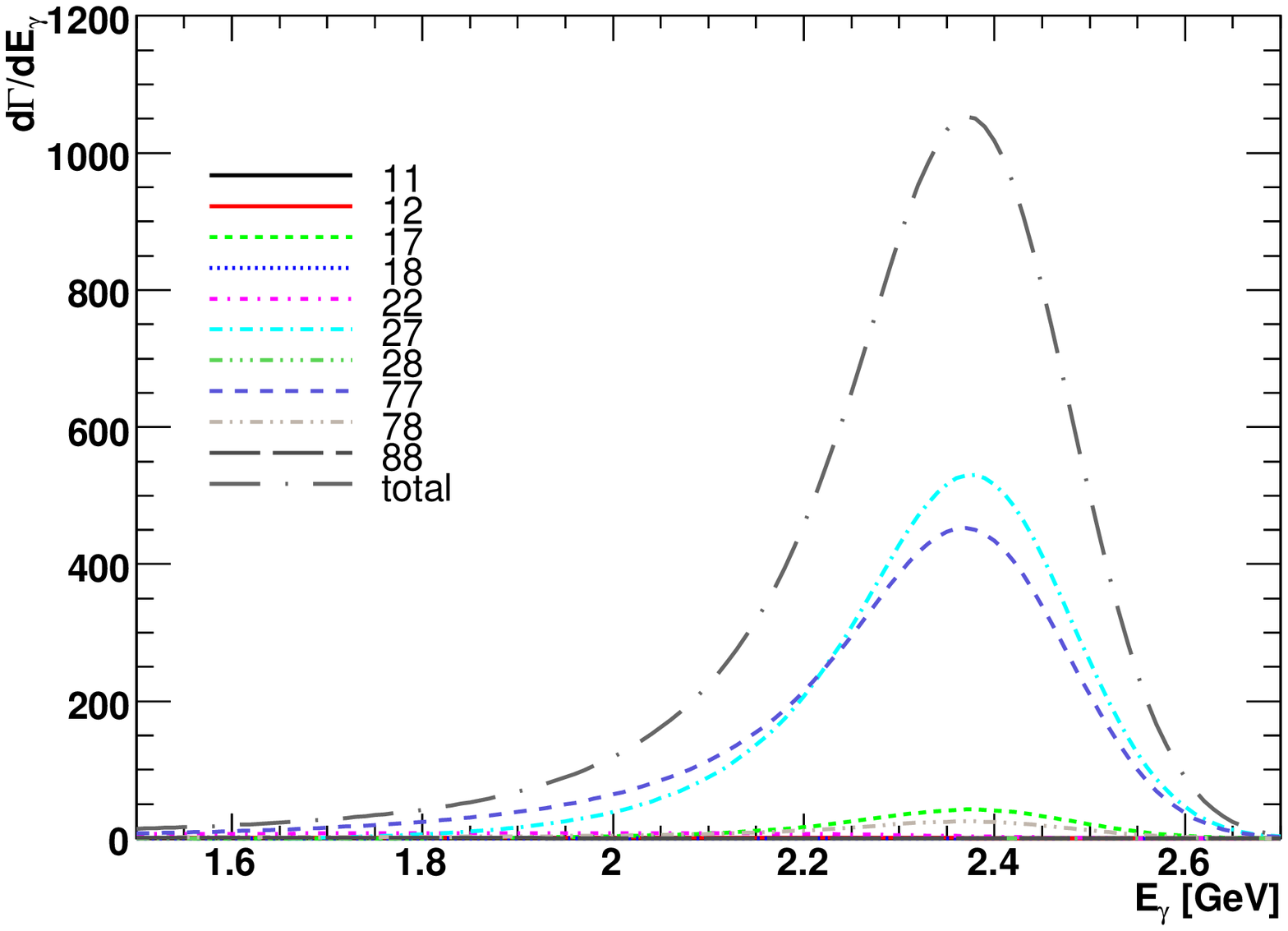,angle=0,width=.49\textwidth}\hfill
\epsfig{file=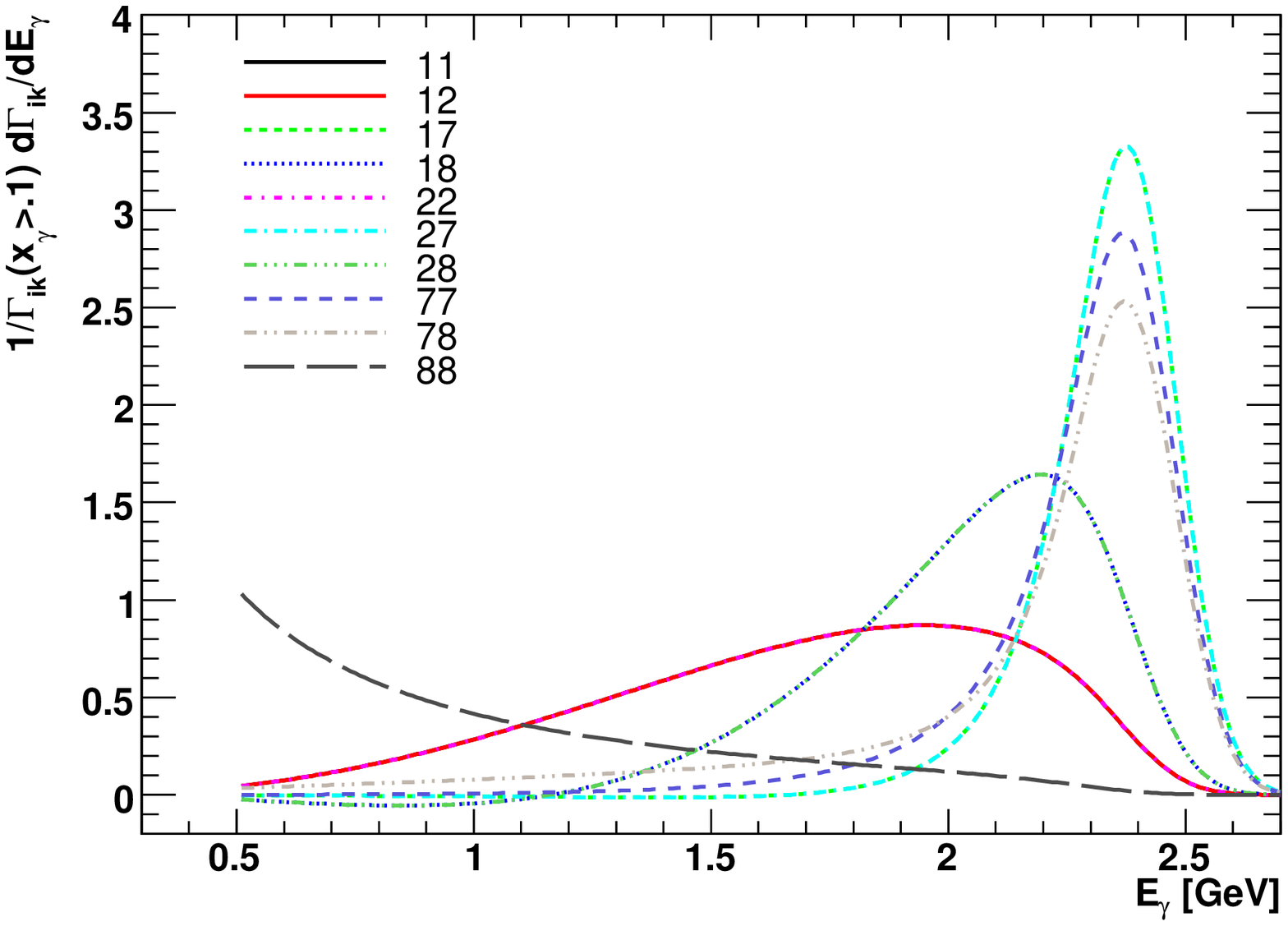,angle=0,width=.49\textwidth}
\caption{\label{fig:spec_ind_contr} Left: the contribution to the
spectrum from the separate operators and their mixing, together with
their sum, the total width, based on
Eq.~\ref{total_width_E0_using_normalized}. We notice that the
dominant contributions (from 77 and 27) have similar shapes in the
Sudakov region, although their tails for small $E_\gamma$ are
significantly different. Right: the spectrum of each matrix element
$\Gamma_{ij}$ (see Eq.~(\ref{total_width_E0_using_normalized}))
has been normalised to better display the various
shapes. Since effectively $O_1=\frac 1 6 O_2$, the curves for 17 and
27 are identical. The same is true for the groups (18, 28) and (11,
12, 22).}
\end{center}
\end{figure}

Figure \ref{fig:spec_ind_contr} shows the contributions to the
differential BF from each matrix element $G_{ij}$ according to
Eq.~\ref{total_width_E0_using_normalized} below, as well as a
comparison between the shapes (normalized spectra) of the different
matrix elements. In general, it confirms the lore that the shapes of the spectra
of the most contributing operators do not vary much, and the shape of the
total spectrum is roughly the same as the one for $O_7$.
However, the shapes are not completely identical and the details do depend on
the relative contributions of the different operators. These depend
of course also on the coefficient functions. Numerically, the most
important effect in the experimentally accessible energy range is
that of $G_{27}$ which is comparable in magnitude to $G_{77}$, but
is somewhat narrower. The effect it has on the differential
distribution along the tail, e.g. between $1.8$ and $2$ GeV, is
significant. The $G_{78}$ spectrum, on the other hand, is somewhat
wider and it has a higher small-$x$ tail. A rather different, and
significantly wider shape is presented by $G_{28}$ and even wider
still by $G_{22}$ (and likewise those corresponding to $O_1$ instead
of $O_2$). These affect the differential spectrum at smaller
energies, below $1.8$ GeV.

\section{Partially--integrated BF and moments with a photon--energy cut\label{sec:numerical}}

In the previous sections we computed the ${\bar B} \longrightarrow
X_s\gamma$ spectra using DGE. We improved the calculation in
Ref.~\cite{Andersen:2005bj} in various ways: we included the NNLO
corrections in full in the $G_{77}$ sector (Sec.~\ref{sec:DGE} and
Appendix~\ref{sec:matching}), introduced a new method to constrain
the behavior of the resummed spectrum away from the Sudakov region
(Sec.~\ref{sec:small_x} and Appendix~\ref{sec:mathcing_Jneq0}), and
matched the resummed spectrum with the fixed--order expansion,
separately for each matrix element $G_{ij}$ (Sec.~\ref{sec:others}).
In this section we will make use of these advancements to make
predictions for measurable quantities: the partial BF and the first
few moments as a function of the cut on the photon energy. We begin
in Sec.~\ref{sec:total_BF} by computing the total BF and discussing
its theoretical uncertainty, then in Sec.~\ref{sec:partial_BF}, we
consider the partial BF with a cut and discuss cut--dependent
uncertainties, and finally in Sec.~\ref{sec:mom} we present
results for the first few moments and analyze numerically the
relation between renormalon contributions, power corrections and
support properties.

\subsection{The total BF\label{sec:total_BF}}

The calculation of the total BF presents a challenge in its
own~\cite{Gambino:2001ew,Buras:2002tp}. First, this calculation has
to accommodate a large hierarchy of scales $M_W,m_t\gg m_b$. This is
dealt
with~\cite{Greub:1996tg,Chetyrkin:1996vx,Kagan:1998ym,Gambino:2001ew,Buras:2002tp}
by the well--established machinery of the effective Weak
Hamiltonian, which facilitates the resummation of large logarithms,
$\ln m_W/m_b$, to all orders in the strong coupling. Estimating the
BF therefore involves the perturbative calculation of the matching
coefficients at the high scales, evolution of the different
operators from $M_W$ to $m_b$ and finally, the contribution of the
matrix elements of the different operators to ${\bar
B}\longrightarrow X_s \gamma$. The main challenge here is at the
perturbative level: even the matrix elements present little
sensitivity\footnote{A possible exception is the large distance
sensitivity associated with $c\bar{c}$ pairs.} to non-perturbative
effects.

The NLO calculation was completed 5 years ago (see
\cite{Buras:2002tp} and Refs. therein) and some of the ingredients
have been recently brought to the NNLO level. This includes, in
particular, the matching coefficients at the Weak
scale~\cite{Misiak:2004ew}, partial calculation of the evolution
matrix~\cite{Gorbahn:2004my,Gorbahn:2005sa} and, most importantly,
the two--loop matrix element of the $O_7$
operator~\cite{Blokland:2005uk,Asatrian:2006ph}. However, other
essential ingredients are still missing. Amongst these are the
matrix elements of other operators, especially $G_{27}$, which
contributes to the BF almost as much as $G_{77}$, despite the fact
that its perturbative expansion starts at ${\cal O}(\alpha_s)$,
rather than at ${\cal O}(1)$. The renormalization scale of $G_{ij}$
other than $G_{77}$ is obviously not fixed at this level.

Beyond $G_{77}$, which is known to NNLO in
full~\cite{Blokland:2005uk,Asatrian:2006ph}, the virtual
$\beta_0 {\alpha_s}^2$ corrections to $G_{ij}$ have been computed in
Ref.~\cite{Bieri:2003ue}. It is likely that these running--coupling
corrections constitute a major part of the NNLO contribution --- this
has been supported by the finding of Ref.~\cite{Blokland:2005uk}.
Here we shall make use of this available, partial NNLO information on the matrix
elements, which are the most important ingredient, neglecting NNLO
effects in the Wilson coefficients at the $M_W$ scale and in the
evolution. A more complete NNLO BF estimate is due when some of the
missing ingredients --- in particular $G_{27}$  at ${\cal
O}({\alpha_s}^2)$ --- become available.

Another difficulty in the evaluation of the BF is its sensitivity to
the bottom mass, which is raised to the fifth power
in~(\ref{total_width_E0_}). Amongst these five powers, two are
associated with the operator itself, and therefore correspond to the
short--distance mass $m_b^{\MSbar}$, whereas the additional three
result from phase--space integration, and therefore correspond to
the pole mass, $m_b$. As discussed in the introduction, the pole
mass has a leading renormalon ambiguity at $u=1/2$, implying that
the perturbative expansion of the total width in the on-shell scheme
is badly divergent. Writing~(\ref{total_width_E0_}) order by order,
\begin{eqnarray}
\label{total_width_E0_with_F} \Gamma(\bar{B}\longrightarrow
X_s\gamma,E>E_0)& =& \frac{\alpha_{\rm em} G_F^2}{32\pi^4}
\left|V_{\rm tb}V_{\rm ts}^*\right|^2  \,\left(m_b^{\MSbar}(m_b)\right)^2
\,m_b^3\,
\\ \nonumber
&&\hspace*{60pt}\times\,\sum_{i,j, \,i\leq
j}C^{\eff}_i(\mu)C_j^{\eff}(\mu)\, {G_{ij}(E_{0},\mu)},\\ \nonumber
& =& \frac{\alpha_{\rm em} G_F^2}{32\pi^4}
\left|V_{\rm tb}V_{\rm ts}^*\right|^2  \,\left(m_b^{\MSbar}(m_b)\right)^2
\,m_b^3\,
\\ \nonumber
&&\hspace*{60pt}\times\,\underbrace{\bigg[f_0(\mu)+f_1(\mu)\frac{\alpha_s(\mu)}{\pi}+
f_2(\mu)\left(\frac{\alpha_s(\mu)}{\pi}\right)^2+\cdots\bigg]}_{F},
\end{eqnarray}
the expansion coefficients $f_n(\mu)$ are therefore expected to grow
factorially with the order~$n$ owing to running--coupling
corrections. Judging from what we know of the pole mass
--- see Appendix B in Ref.~\cite{Andersen:2005bj} --- this divergence
sets in already at the first few orders. A similar situation occurs
for other inclusive decays, for example for the charmless
semileptonic decay ${\bar{B}}\longrightarrow X_u
l\bar{\nu}$~\cite{Bigi:1994em,Beneke:1994sw,Beneke:1994bc}:
\begin{eqnarray}
\label{SL_total} \Gamma_{\tot}\left({\bar B}\longrightarrow X_u
l\bar{\nu}\right)=
\frac{G_F^2|V_{\rm ub}|^2m_b^5}{192\pi^3}\underbrace{\bigg[1+s_1\frac{\alpha_s(\mu)}{\pi}+
s_2(\mu)\left(\frac{\alpha_s(\mu)}{\pi}\right)^2+\cdots\bigg]}_{G_u}.
\end{eqnarray}
In this case the coefficients have been computed to NNLO in
Ref.~\cite{vanRitbergen:1999gs}, where large running--coupling
corrections already appear. Moreover, in this case it is possible to
interpolate reliably~\cite{Andersen:2005mj} (see
also Ref.~\cite{Lee:2002px,Cvetic:2001sn,Pineda:2001zq}) between the
first few orders and the known asymptotic behavior of the series,
which is set by the $u=1/2$ renormalon of the pole mass. In this way
a precise evaluation of the total charmless semileptonic BF can be
made directly using the on-shell scheme, where the $u=1/2$
renormalon ambiguities of the pole mass and the series in
$G_u$ in \eq{SL_total} are regularized using the PV prescription.
According to Sec.~2 in Ref.~\cite{Andersen:2005mj}, this results in
the following values\footnote{It is crucial to defined both $m_b$ and $G_u^{1/5}$ using the
\emph{same} prescription; the product $m_b \,G_u^{1/5}$ is prescription independent.
Throughout this paper we use the PV prescription when refering to these parameters,
although this will not be explicitly reflected in the notation.}:
\begin{equation}
\label{G_u} \left.m_b\right\vert_{\PV}=4.89\,\pm 0.05\, {\rm GeV} \qquad \qquad
\left.G_u^{1/5}\right\vert_{\PV}=0.928\,\pm 0.002,
\end{equation}
corresponding to $N_f=4$. Here the error in the pole mass $m_b$ is dominated by the error on the short
distance mass $m_b^{\MSbar}=4.20\pm 0.04$ (as quoted in Table \ref{table:param}).
The uncertainty in $\Gamma_{\tot}\left({\bar B}\longrightarrow X_u
l\bar{\nu}\right)/|V_{\rm ub}|^2$ is largely determined by this error.

\begin{table}
\begin{center}
  \begin{tabular}{l}\\
  $G_F=1.16637\cdot10^{-5}$ GeV$^{-2}$; \quad $\alpha_{\rm em}=1/(130.3\pm 2.3)$ \\
  $|V_{\rm ts}|=(41.13\pm 0.63)\cdot 10^{-3}$;\quad
  $|V_{\rm tb}|=0.999119 \pm 0.000026$~\cite{Charles:2004jd}\\
  $M_W=80.388\pm 0.035$  GeV\\
  $m_t^{\MSbar}(m_t^{\MSbar})=165\pm 5$ GeV\\
  $\alpha_s(M_Z=91.19)=0.1176\pm 0.0020$ \quad $\Longrightarrow$
  \quad $\eta\equiv \alpha_s(M_W)/\alpha_s(\mu)\simeq 0.561$\\
  $m_s^{\MSbar}\simeq 0.15$ GeV\\
  $m_c^{\MSbar}(m_c^{\MSbar})=1.295 \pm 0.015$ GeV~\cite{Boughezal:2006px}\\
  $m_b^{\MSbar}(m_b^{\MSbar})=4.20 \pm 0.04$ GeV~\cite{Boughezal:2006px,Buchmuller:2005zv}
\end{tabular}
 \caption{\label{table:param}Values of input parameters in
   the calculation of the BF and their uncertainties.}
\end{center}
\vspace*{-10pt}
\end{table}

A useful tool in the calculation of the total ${\bar B}
\longrightarrow X_s\gamma$ width is to normalize it by the
semileptonic width; both $b\to c$ and $b\to u$ have been used in the
past.  In this way one exploits the better control one has on
higher--order corrections in the semileptonic case, and reduces the
sensitivity to the pole mass. There is however an important
difference between the radiative decay width
(\ref{total_width_E0_with_F}) and the semileptonic one
(\ref{SL_total}) in the power of the pole mass. Thus, while the
ratio
\begin{equation}
\label{simply_ratio}
\left.\Gamma(\bar{B}\longrightarrow
X_s\gamma,E>E_0)\right/\Gamma_{\tot}\left({\bar B}\longrightarrow
X_u l\bar{\nu}\right)
\end{equation}
itself is an observable, and therefore renormalon free, it still
contains an explicit dependence on the pole mass through
$(m_{\MSbar}/m_b^{\rm pole})^2$. This has been dealt with in the
past by resorting to some alternative renormalon--free mass scheme,
a step that unavoidably introduces some uncertainty that is hard to
quantify. Here we suggest an alternative procedure that utilizes the
charmless semileptonic width to eliminate the explicit power
dependence on the pole mass and yet does not require any additional
mass scheme. This procedure is explained below.

\subsubsection*{Using the semileptonic width}

In order to use the semileptonic result (\ref{SL_total}) to
normalize the radiative one (\ref{total_width_E0_with_F})
 in a renormalon free manner, we first write the series for $F$
over $G_u^{3/5}$. By virtue of the cancallation of renormalon
ambiguities in Eq.~(\ref{total_width_E0_with_F}) and in
Eq.~(\ref{SL_total}), this ratio is renormalon free, so it can be
evaluated at fixed order. To recover the result for the radiative
decay width in Eq.~(\ref{total_width_E0_with_F}) we then multiply the
series for $F/G_u^{3/5}$ by the PV result~\cite{Andersen:2005mj} for
$G_u$ quoted in Eq.~(\ref{G_u}), namely $G_u^{3/5}=(0.928)^{3/5}$ and
use the corresponding PV pole mass in the overall $m_b^3$ power (the
product $m_b^3 G_u^{3/5}$ is prescription--independent, see
Eq.~(\ref{SL_total})):
\begin{align}
\begin{split} \label{total_width_E0_F_over_G} \Gamma(\bar{B}\longrightarrow
X_s\gamma,E>E_{\min})& = \frac{\alpha_{\rm em} G_F^2}{32\pi^4}
\left|V_{\rm tb}V_{\rm ts}^*\right|^2
\,\left(m_b^{\MSbar}(m_b)\right)^2\,\times \\&\hspace*{70pt}
\,m_b^3\, G_u^{3/5} \bigg[F(E_{\min})/G_u^{3/5}\bigg]_{\rm Fixed\,\,
Order}
\end{split}
\end{align}
with
\begin{eqnarray}
\label{F_over_G35} \bigg[F(E_{\min})/G_u^{3/5}\bigg]_{\rm Fixed\,\,
Order} &\equiv& f_0(\mu) +\left(f_1(\mu)-\frac35 s_1 f_0(\mu)
\right)\frac{\alpha_s(\mu)}{\pi} \\ \nonumber &&\hspace*{-100pt}
+\left( f_2(\mu)-\frac35 s_1 f_1(\mu) -\frac35s_2(\mu)f_0(\mu)
+\frac{12}{25}s_1^2\,f_0(\mu)
\right)\left(\frac{\alpha_s(\mu)}{\pi}\right)^2+\cdots.
\end{eqnarray}

Similarly, in order to use the resummed calculation of the
normalized integrated spectra for individual matrix elements, namely
${G_{ij}(E_{0})}/{G_{ij}(E_{\min})}$, we define
\begin{eqnarray}
F_{ij}(E_{\min})&=&C^{\eff}_i(\mu)C_j^{\eff}(\mu)\, {G_{ij}(E_{\min},\mu)}\\ \nonumber
&=& f_0^{ij}(\mu)+f_1^{ij}(\mu)\frac{\alpha_s(\mu)}{\pi}+
f_2^{ij}(\mu)\left(\frac{\alpha_s(\mu)}{\pi}\right)^2+\cdots
\end{eqnarray}
and write
\begin{eqnarray}
\label{total_width_E0_using_normalized}
\Gamma(\bar{B}\longrightarrow X_s\gamma,E>E_0)&
=& \frac{\alpha_{\rm em} G_F^2}{32\pi^4}
\left|V_{\rm tb}V_{\rm ts}^*\right|^2  \,\left(m_b^{\MSbar}(m_b)\right)^2 \,m_b^3\,
\\ \nonumber
&&\hspace*{-20pt}\times\,\sum_{i,j, \,i\leq j} G_u^{3/5}
\bigg[F_{ij}(E_{\min})/G_u^{3/5}\bigg]_{\rm Fixed\,\, Order}
\left[\frac{G_{ij}(E_{0})}{G_{ij}(E_{\min})}\right]_{\rm Resummed}.
\end{eqnarray}
This formula will be used for the numerical analysis that follows. In
Fig.~\ref{fig:spec_ind_contr} above we show the contribution from each of the terms
$\Gamma_{ij}$. \eq{total_width_E0_using_normalized} will allow us to study
the theoretical uncertainties associated with different ingredients as a function of the cut.

\subsubsection*{The total BF and the theoretical uncertainty}

Using \eq{total_width_E0_F_over_G} we obtain the following result for the total BF:
\begin{equation}
\label{total_BF_result}
{\rm BF}(E_{\gamma}>m_b/20)=\Big(357 \,\,\pm 40_{(\mu)} \,\pm 19_{ (\mu_c)}\,
\pm {21}_{({\rm param.})}\Big) \,\cdot 10^{-6} ,
\end{equation}
where, as explained above, we used (\ref{F_over_G35}) to NNLO where $f_1(\mu)$ is the full NLO
coefficient while $f_2(\mu)$ is the $\beta_0$ part of the NNLO corrections to the
matrix elements, with the parameters of Table~\ref{table:param}, which yield (\ref{G_u}).
The central value is based on
$\mu=m_b$ and $\mu_c=m_b/2$, and the three errors represent
the variation of (1) $\mu$ and (2) $\mu_c$ as explained below, and (3) the parametric uncertainty in
$m_b^{\MSbar}$ and ${\alpha_s}^{\MSbar}$ according to Table~\ref{table:param}, respectively.

Let us discuss now the theoretical uncertainty in some more detail.
Having separated the calculation into that of Wilson
coefficients on the one hand, and matrix elements on the other,
 the stability of the result with respect to the
 renormalization point of the operators becomes an
essential measure of the accuracy of the calculation. It should be
emphasized that by modifying $\mu$ in (\ref{total_width_E0_with_F})
or in (\ref{total_width_E0_using_normalized}) both the
renormalization scale of the coupling and the factorization scale
(i.e. the separation between the coefficient functions and the
matrix elements) is changed. This necessarily implies reshuffling of
contributions between different matrix elements according to the
anomalous dimension matrix.

It should be noted that there are various possibilities for choosing
the renormalization point of the short distance mass in
(\ref{total_width_E0_using_normalized}), which are of course
reflected in the anomalous dimension matrix. We have chosen to fix
this scale at the pole mass $m_b$, rather than to vary it with
$\mu$.
\begin{figure}[t]
\begin{center}
\epsfig{file=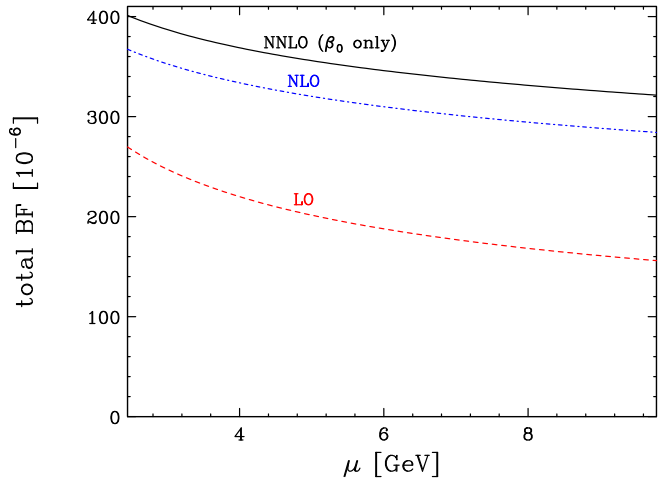,angle=90,width=8.7cm}
\caption{\label{fig:total_BF_scale_dep} The scale dependence of the
total BF (the BF with $E_{\gamma}>m_b/20$) at LO, NLO and NNLO, according to
Eqs.~(\ref{total_width_E0_F_over_G}) and (\ref{F_over_G35}).
At NNLO only $\beta_0{\alpha_s}^2$ terms are included in $f_2(\mu)$. }
\end{center}
\end{figure}

Figure~\ref{fig:total_BF_scale_dep} presents the variation of the
total BF as a function of $\mu$, with the default parameters of
Table~\ref{table:param} and with $\sqrt{z}\equiv
m_c(\mu_c)/m_b=0.2215$ (see below). We see that at leading order
there is a large scale dependence (that of $(C_7^{\eff}(\mu))^2$)
which get significantly smaller at NLO and NNLO. Note that at NNLO
only $\beta_0{\alpha_s}^2$ corrections are included here. The reason
for suppressing the known ${\cal O}({\alpha_s}^2)$ terms for $G_{77}$
beyond this approximation, is that there is a large cancellation, at
each order in the expansion, between $G_{77}$ and $G_{27}$. It is
therefore essential to include the same type of corrections for
both\footnote{For the default values of the parameters and with
$\mu=m_b$, using the full $G_{77}$ NNLO correction instead of its
$\beta_0{\alpha_s}^2$ component (while keeping other matrix elements
with the $\beta_0{\alpha_s}^2$ according to~\cite{Bieri:2003ue}) the
BF increases by $\sim 2.5\%$.}. Note that there is no significant
reduction of the scale dependence by including the
$\beta_0{\alpha_s}^2$ NNLO corrections. This is a reflection of the
fact that important NNLO corrections are still missing: the leading
scale dependence at each order, ${\cal O}({\alpha_s}^n\ln^n m_b/\mu)$,
is associated with the evolution of the operators and their mixing.

Another important source of uncertainty in the calculation of the
total BF~\cite{Gambino:2001ew} is the renormalization point $\mu_c$
of the charm mass $m_c$ entering through the ratio $z\equiv
m_c^2(\mu_c)/m_b^2$ into the expressions for the matrix element
involving $O_2$ (and $O_1$), notably $G_{27}$. We have chosen the
central value for $z$ as $\sqrt{z}=0.2215$, corresponding to an
intermediate renormalization point, $\mu_c\simeq m_b/2$ GeV, which
is in between the b mass and the c mass. This is close to the
central value chosen in \cite{Gambino:2001ew}. For the error
estimate we vary $z$ in the range suggested in that paper, namely
\hbox{$0.18\,<\, \sqrt{z}\,<\,0.26$}. This uncertainty will reduce
once the calculation of the matrix elements involving $O_2$ is
complete.

Let us also note that the results of Ref.~\cite{Bieri:2003ue} for the $\beta_0 {\alpha_s}^2$
corrections involving charm loops were computed as an expansion in $z=m_c(\mu_c)/m_b$.
When using this expansion we assume that it is valid at the physical value of $z$, an issue that was checked in Ref.~\cite{Bieri:2003ue}. Given that $z$ is already varied over a large range, we do not consider the residual error associate with using a finite--order expansion in $z$ as an independent source of uncertainty\footnote{See Note Added concerning a new publication
on the subject that appeared upon completion of this work.}.

We note that the central value of Eq.~(\ref{total_BF_result}) is
somewhat lower than that of previous estimates. This is despite the fact that NNLO
corrections increase the BF: using the same procedure at NLO we get
${\rm BF}(E_{\gamma}>m_b/20)=322 \cdot 10^{-6}$. These results can be compared with
the result by Gambino and Misiak~\cite{Gambino:2001ew}:
${\rm BF}(E_{\gamma}>m_b/20)=373 \cdot 10^{-6}$, who used the NLO calculation. The main change
is due to the different procedure for normalizing the BF by the semileptonic width.
Note, in particular, that Ref.~\cite{Gambino:2001ew} evaluated the factor
$m_b^{\MSbar}/m_b^{\rm pole}$ appearing in (\ref{simply_ratio}) by replacing it with
$m_b^{\MSbar}/m_b^{1S}$. At NLO this replacement has no effect on the expansion since the expansion
of $m_b^{1S}/m_b^{\rm pole}$ starts at ${\cal O}({\alpha_s}^2)$, however, numerically, this ratio
is $\sim 0.95$. This explains much of the difference in the central values between
the two estimates. It should also be noted that the uncertainty we find from varying the
renormalization scale~$\mu$, $\pm 11\%$, is larger than those
in Refs.~\cite{Gambino:2001ew,Buras:2002tp},
and consequently the combined uncertainty on the total BF
is about $\pm 13.7\%$, somewhat larger than previous estimates.

Finally, we point out that certain small corrections have been neglected in our calculation.
The largest amongst these is the electroweak corrections whose effect has been
estimated~\cite{Gambino:2001ew} to be
$-3.8\%$ of the BF. Another correction is the power--suppressed $\Lambda^2/m_b^2$ that was
estimated as $+2.5 \%$ of the BF~\cite{Gambino:2001ew}. As soon as more complete NNLO results
become available these corrections should be taken into account as well.

\subsection{BF for $E_{\gamma}>E_0$\label{sec:partial_BF}}

\begin{figure}[t]
\begin{center}
\epsfig{file=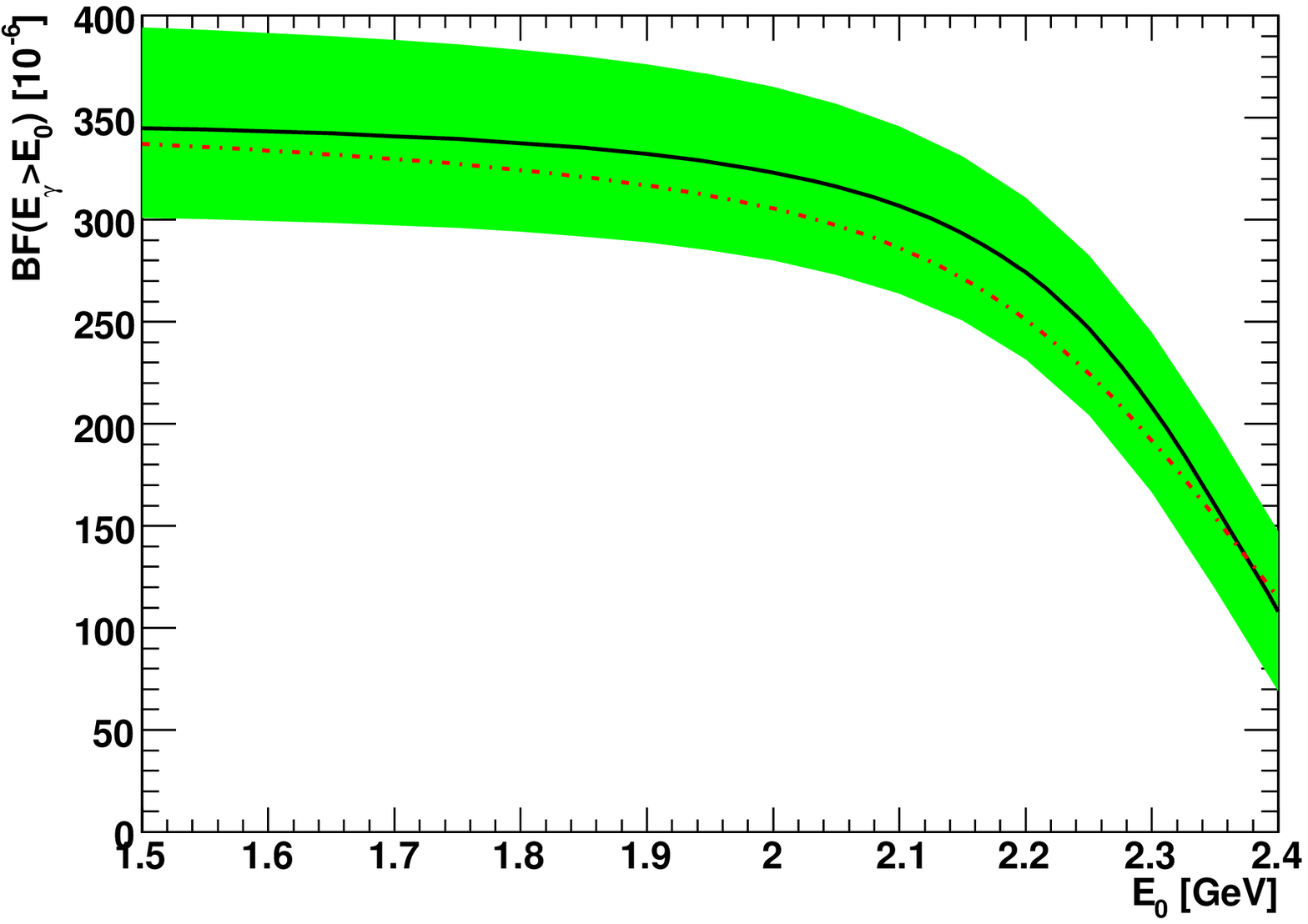,width=.49\textwidth} \hfill
\epsfig{file=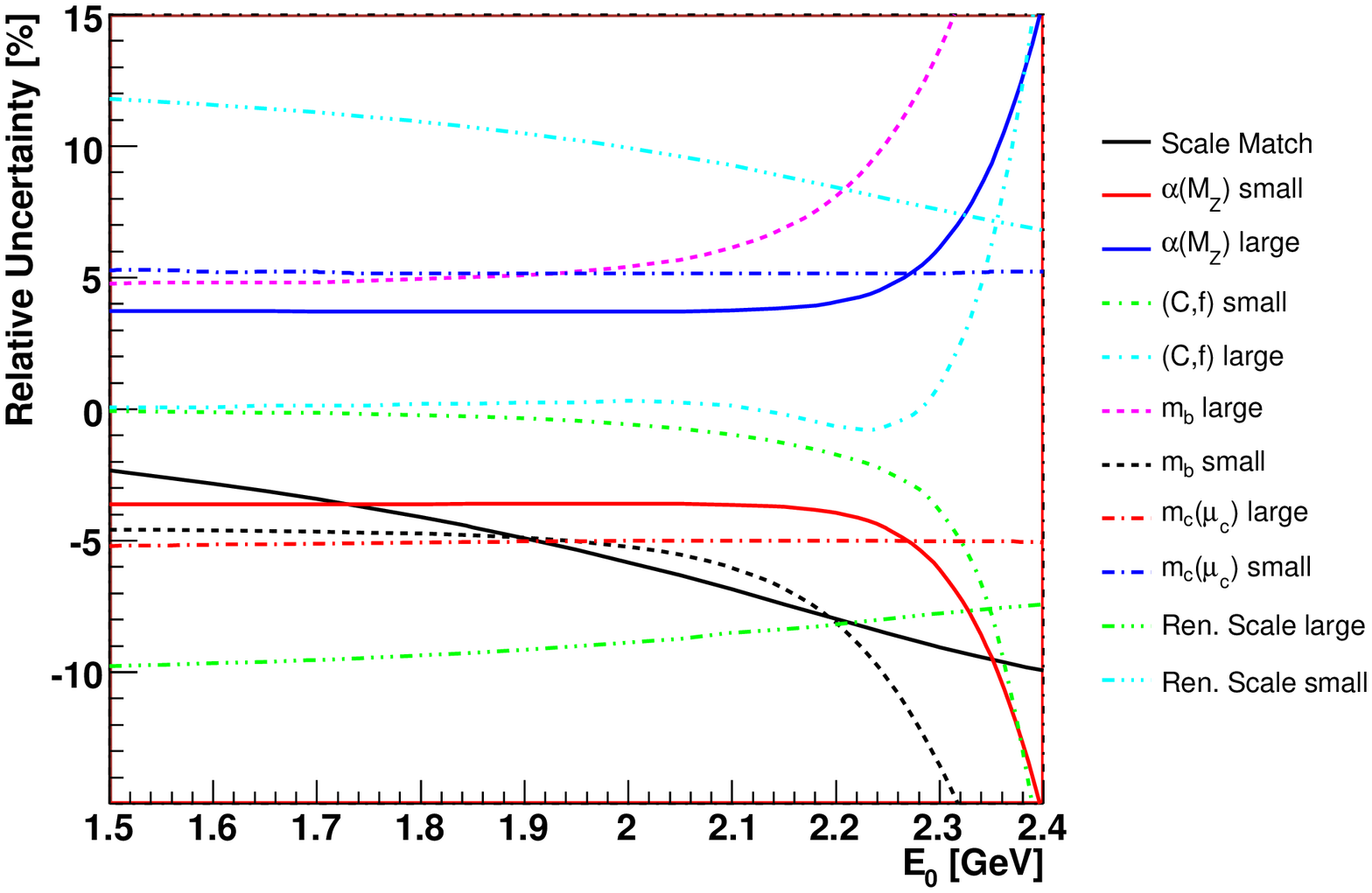,width=.49\textwidth}
\caption{\label{fig:BF} Left: The central value and uncertainty on the
  branching fraction ratio ${\rm BF}(E_\gamma>E_0)$
  as a function of the cut on the photon energy $E_0$. The black central line
  is obtained with the standard values of the parameters, while the (red)
  dot-dashed line is obtained by the same choice of parameters
  as for the dot-dashed line in
  Figs.~\ref{fig:AvgEnergy} and~\ref{fig:Variance}. Right: The
  breakdown on sources to the uncertainty in the branching fraction.}
\end{center}
\end{figure}

\begin{figure}[t]
\begin{center}
\epsfig{file=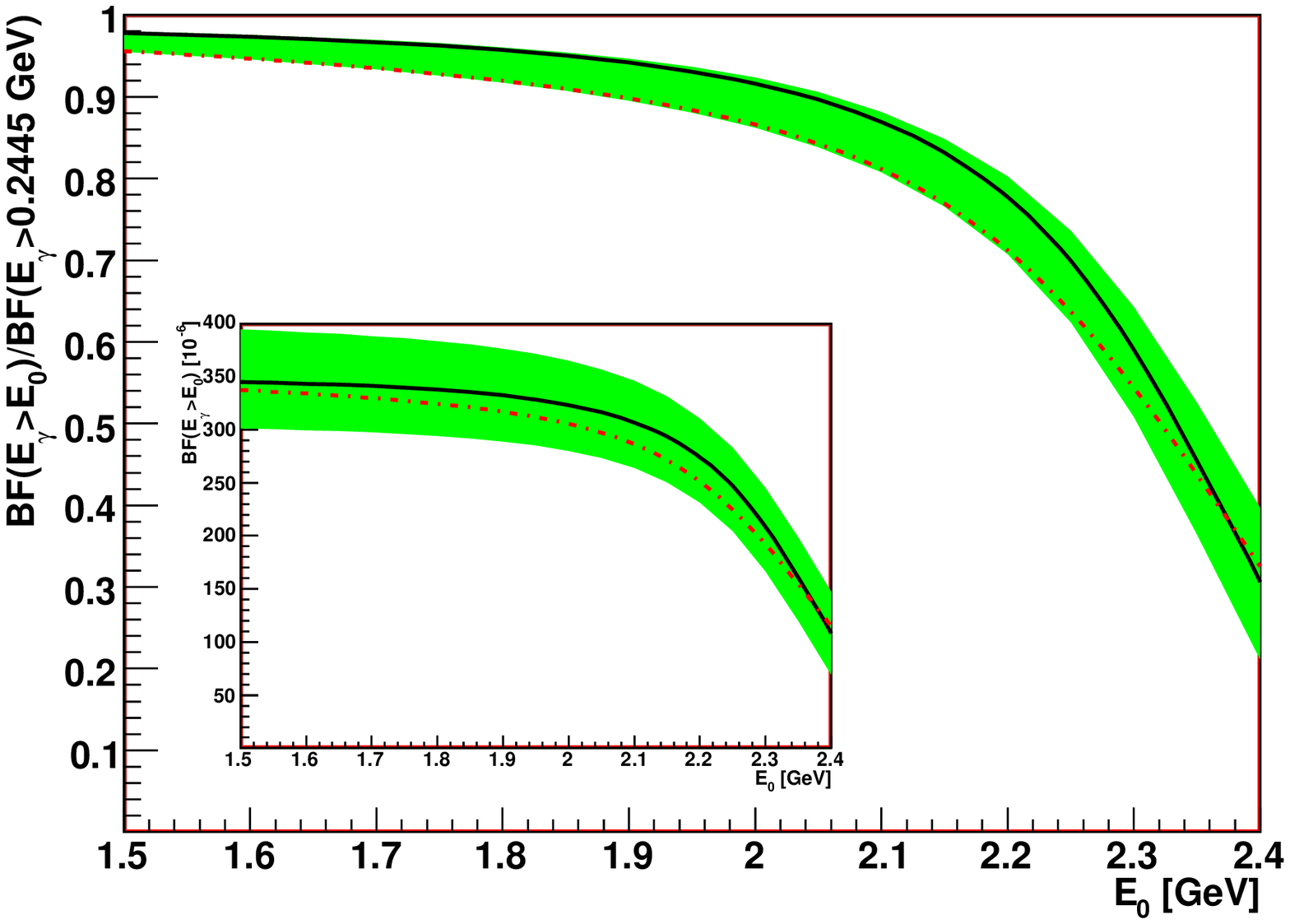,angle=0,width=.49\textwidth} \hfill
\epsfig{file=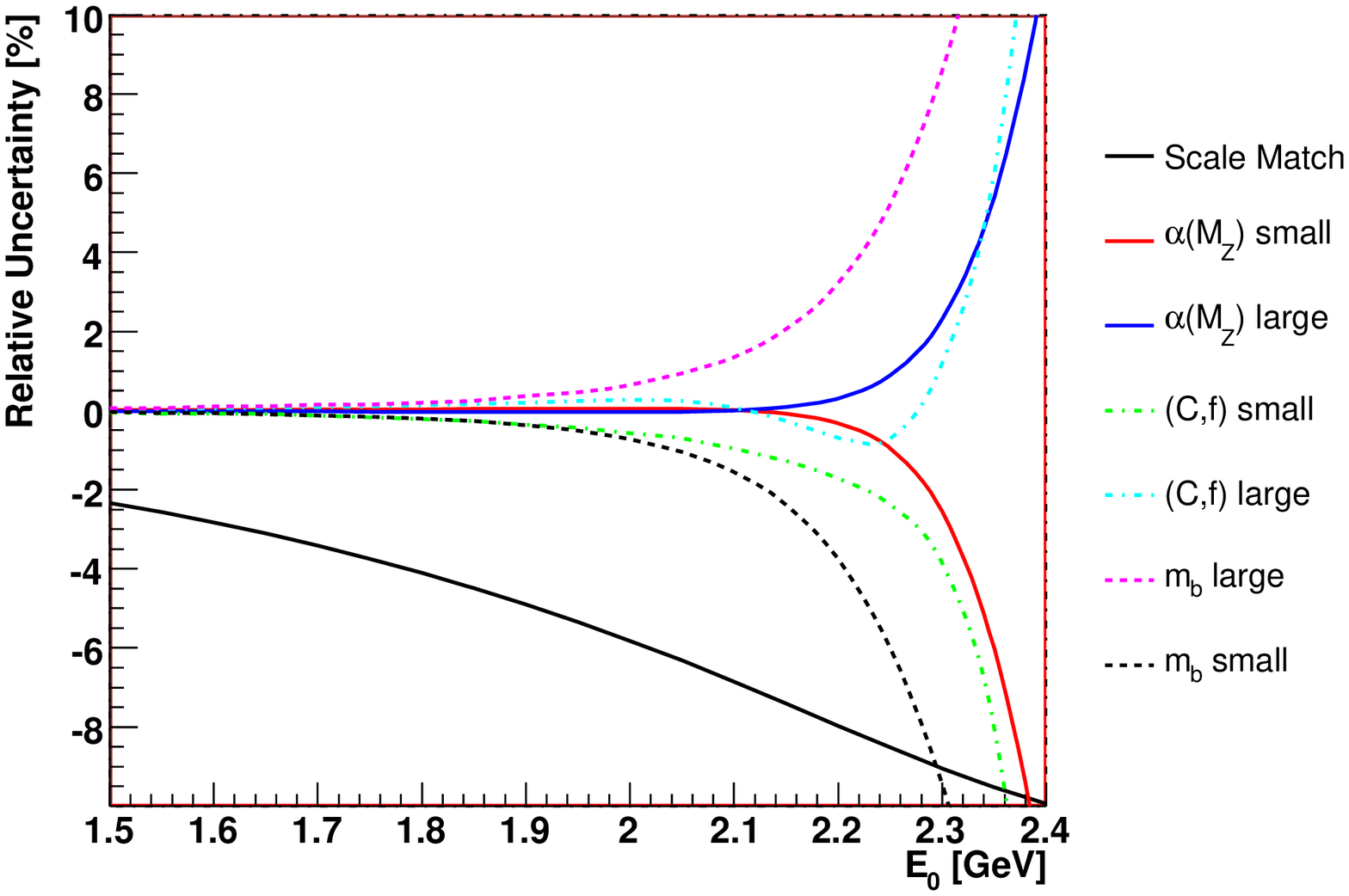,width=.49\textwidth}
\caption{\label{fig:BF_ratios} Left: The central value and uncertainty on the
  branching fraction ratio ${\rm BF}(E_\gamma>E_0)/{\rm
    BF}(E_\gamma>.2445\mathrm{GeV})$ and the branching fraction itself
  (insert) as a function of the cut on the photon energy $E_0$. Right: The
  breakdown on sources to the uncertainty in the branching fraction ratio.
  For low values of the cut, the uncertainty is dominated by the contribution
  from a variation in the renormalization scale in the matching
  coefficients~(\ref{tilde_after_full_exponentiation_most_moment_space_mu}).
  This relatively large effect is originating from the matrix elements
  other than $G_{77}$ that are known to NLO accuracy only.
  For $E_0\lesssim2.1\,\mathrm{GeV}$, the
  parameters controlling the behavior in the Sudakov region contributes only
  little to the overall uncertainty. The (red) dot-dashed lines are obtained
  by the same choice of parameters as for the dot-dashed line in
  Figs.~\ref{fig:AvgEnergy} and~\ref{fig:Variance}.}
\end{center}
\end{figure}

In Fig.~\ref{fig:BF} we present the BF as a function of the cut, ${\rm
  BF}(E_{\gamma}>E_0)$. The uncertainty band in this figure as well as in Figs.
\ref{fig:J_eq_3_BF_NLO_NNLO}, \ref{fig:BF_ratios} and \ref{fig:AvgEnergy}
to \ref{fig:third_central_moment},
indicates the theoretical uncertainty obtained by varying separately the following parameters,
and summing the respective uncertainties in quadrature:
\begin{itemize}
\item{} The renormalization scale of the operators (and the coupling) in the calculation of
the total width according to Eqs. (\ref{total_width_E0_F_over_G}) and (\ref{F_over_G35})  between
$m_b/2$ and $2m_b$ where $\mu=m_b$ is the default.
\item{} The renormalization point $\mu_c$ of the
charm mass $m_c(\mu_c)$ entering the matrix elements involving $O_2$ and $O_1$.
As mentioned above we use: $0.18\,m_b<\, m_c(\mu_c) \,<\,0.26\,m_b$.
\item{} The renormalization scale in the matching
coefficients of the resummed spectra
(\ref{tilde_after_full_exponentiation_most_moment_space_mu}), between $m_b$ and $m_b/2$,
where $\mu=m_b$ is the default.
\item{} The parameters controlling the details of the quark distribution
function (corresponding to the leading ``shape function'') $B_{\cal S}(u)$,
which enter the Sudakov factor in
Eq.~(\ref{Sud_with_cubic_small_x_general_color}),
with the corresponding power corrections, namely $(N\Lambda/m_b)^k$ for any $k\geq 3$.
Specifically,  as discussed in
Sec.~\ref{sec:mom} below, we vary $C_{3/2}$ in \eq{C32} (see Eqs. (3.27)
to (3.29) in Ref.~\cite{Andersen:2005mj}) and $f^{\mathrm{PV}}$ in
\eq{Sud_with_cubic_Renormalons_single_parameter} below within a range that is determined based on
the support properties of the spectra, according to
$\mathrm{CutA,\,CutB}<0.01$, where the functions $\mathrm{CutA}$ and $\mathrm{CutB}$ are defined in
(\ref{eq:Cepscuts}).
\item{} The short distance parameters, $\alpha_s^{\MSbar}$ and $m_b^{\MSbar}$,
within their error ranges as indicated in Table \ref{table:param}.
\end{itemize}
Let us point out that the effect of the charm mass on the running of the coupling
is neglected throughout the calculation of the spectra. We use $N_f=4$, and ignore
the charm threshold. The consequences of this approximation have not yet been studied. They
certainly worthwhile investigating since at large $N$ the coupling in the Sudakov factor
is effectively evaluated at scales below the charm mass. We further neglect \emph{non-perturbative}
effects that go beyond the summation of the leading powers,
$(N\Lambda/m_b)^k$. This includes
the so-called ``subleading shape functions'', namely power corrections that
are suppressed by~$1/N$ compared to the
ones we include, as well as the non-perturbative structure of the final
state, which can be viewed as power corrections in $(N\Lambda^2/m_b^2)$, or directly in momentum space
as exclusive states.

Excluding very high cut values, the largest uncertainties shown in Fig.~\ref{fig:BF}
have very little dependence on the cut. This includes, in particular, the
renormalization scale of the operators and the value of
$m_c(\mu_c)$. The cut dependence of these contributions, is entirely due to the change their
variation induces on the relative weight of the different matrix elements
 in Eq.~(\ref{total_width_E0_using_normalized}). Furthermore, Fig.~\ref{fig:BF} shows that
 the uncertainty associated with the values of the short distance parameters,
 is also insensitive to the cut for sufficiently mild cuts.
It is therefore useful to analyze separately the cut--dependent uncertainty.
This is done in Fig.~\ref{fig:BF_ratios}, where we display the partial BF ratio
${\rm BF}(E_\gamma>E_0)/{\rm
    BF}(E_\gamma>m_b/20)$
as a function of $E_0$. This ratio is needed, and can be used
for extrapolating experimental measurements from the region $E_{\gamma}>E_0$ to
the whole of phase space. Some numerical values for the extrapolation factor
are presented in Table \ref{table:ext_factors}.
\begin{table}[t]
\begin{center}
\begin{tabular}{|l|l|l|l|}
\hline
$E_0$  (GeV) & default & $\min$ & $\max$\\
\hline
1.6          & 1.028& 1.025& 1.058\\
1.7          & 1.034& 1.031& 1.071\\
1.8          & 1.045& 1.041& 1.090\\
1.9          & 1.062& 1.056& 1.117\\
2.0          & 1.092& 1.083& 1.160\\
2.1          & 1.150& 1.134& 1.238\\
2.2          & 1.287& 1.247& 1.414\\
2.3          & 1.692& 1.556& 1.965\\
\hline
\end{tabular}
\vspace*{10pt}
\caption{Extrapolation factors for the ${\bar B}\longrightarrow X_s \gamma$ BF,\,\,
  ${\rm BF}(E_{\gamma}>m_b/20) / {\rm BF}(E_\gamma>E_0)$}
  \label{table:ext_factors}
\end{center}
\vspace*{-10pt}
\end{table}
As expected, the uncertainty of the extrapolation increases as the cut is increased. However,
the rate of increase is not too high: precise experimental measurements with fairly
stringent cuts, such as $E_{\gamma}>2.0$ GeV, can still be useful.
As shown in Fig.~\ref{fig:BF_ratios}, up to cuts as high as $E_0=2.3$ GeV, the largest source
of uncertainty is still the dependence on the renormalization scale in the
matching coefficients of Eq.~\eqref{tilde_after_full_exponentiation_most_moment_space_mu}.
At higher cuts the $m_b$ dependence and
the details of the quark distribution function, $B_{\cal S}(u)$ and the associated power corrections
become dominant. For the higher moments of the photon energy these ingredients become important
already at milder cuts. We therefore discuss these issues in more detail in the next section.

\subsection{Spectral moments for $E_{\gamma}>E_0$\label{sec:mom}}

\begin{figure}[t]
\begin{center}
\epsfig{file=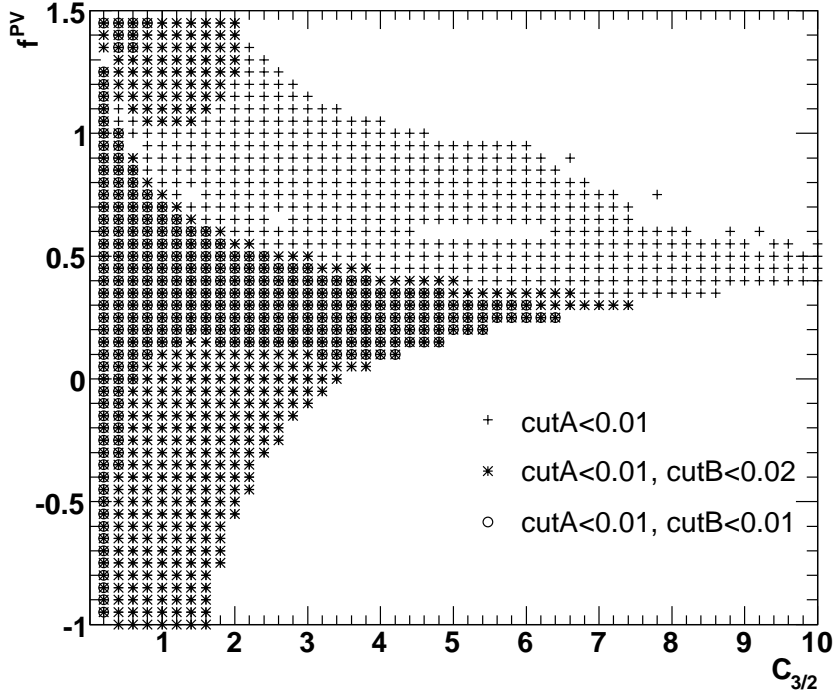,width=.8\textwidth}
\caption{\label{fig:map} The region in the $u=3/2$ renormalon residue and power corrections plain,
  $(C_{3/2},f^{\mathrm{PV}})$, where the spectra conforms with the physical support
properties. For example, in the most restricted region, marked by circles, both the requirement
on the BF $\mathrm{CutA}(C_{3/2},f^{\mathrm{PV}})<0.01$ and the one on the average energy
$\mathrm{CutB}(C_{3/2},f^{\mathrm{PV}})<0.01$ are satisfied ($\mathrm{CutA}$ and
  $\mathrm{CutB}$ are defined in Eq.~\eqref{eq:Cepscuts}). }
\end{center}
\end{figure}

\begin{figure}[t]
\begin{center}
\epsfig{file=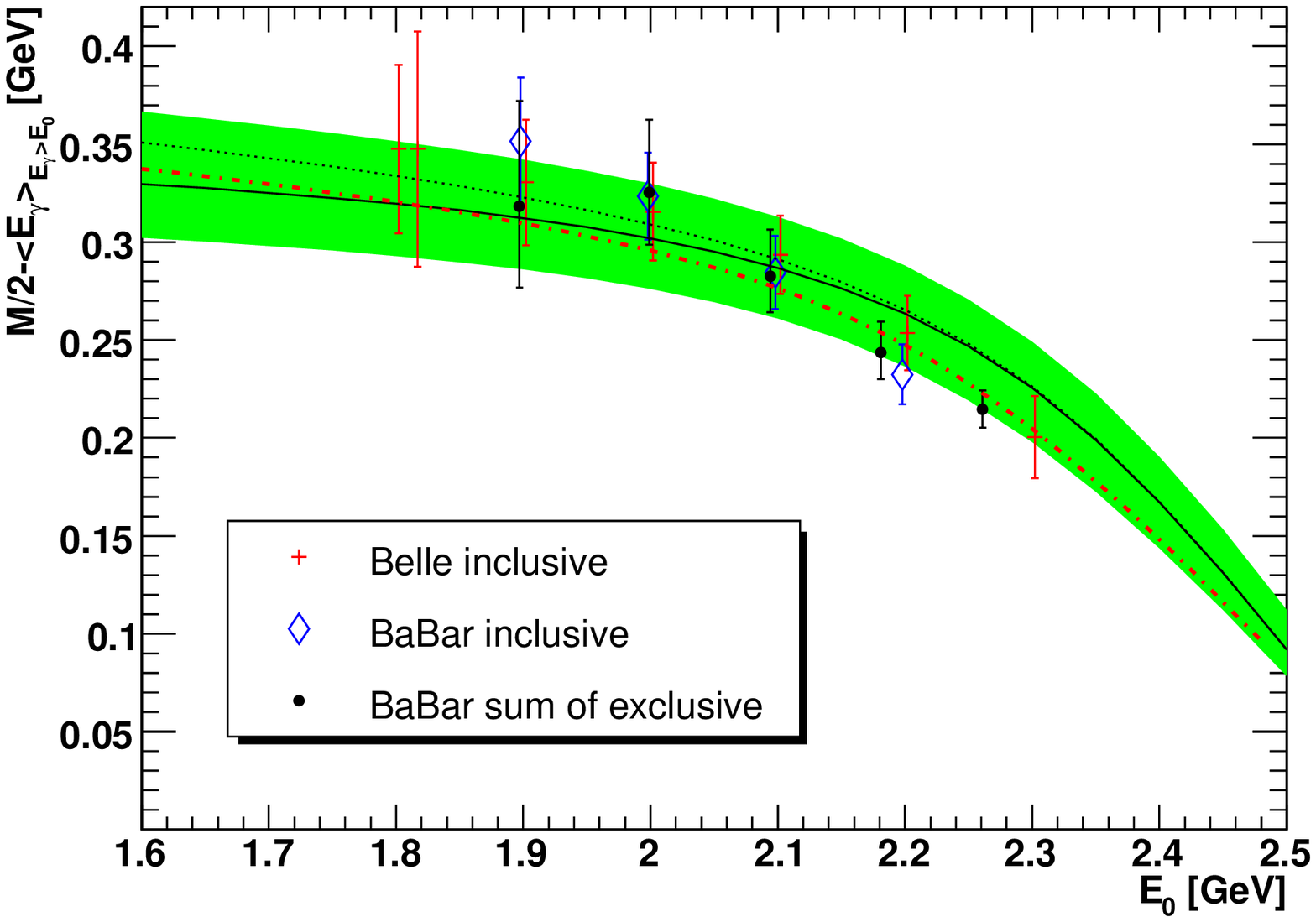,angle=0,width=.49\textwidth}\hfill
\epsfig{file=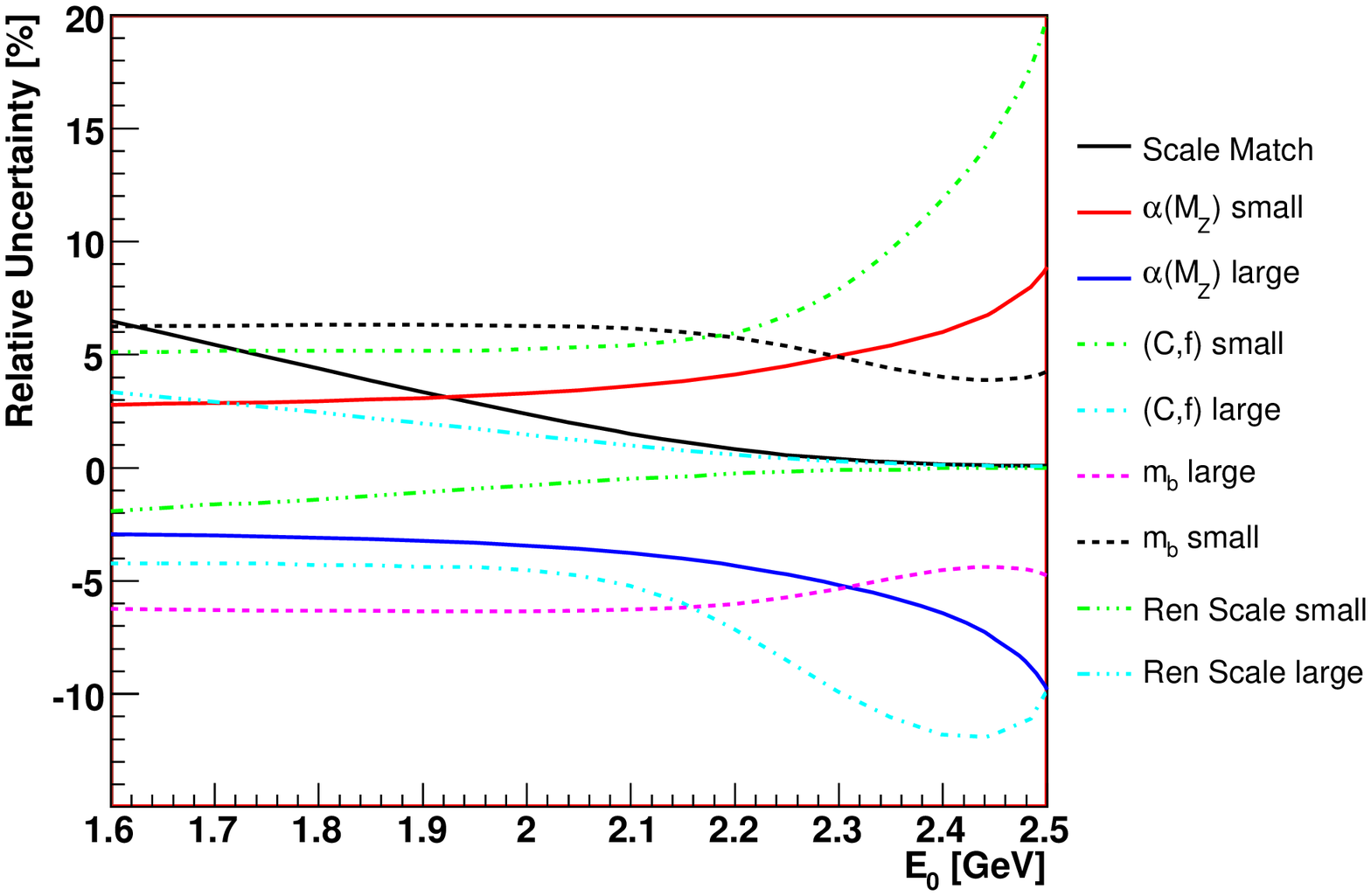,angle=0,width=.49\textwidth}
\caption{\label{fig:AvgEnergy} Left: The prediction for $M_B-\langle
  E_\gamma\rangle_{E_\gamma>E_0}$ as a function of $E_0$ compared with
  data from Belle and BaBar.
  The green band indicates the uncertainty, and is obtained by adding
  contributions in quadrature. The default choice of parameters, corresponding to
  $\mu=m_b^{\mathrm{PV}}$ in
  Eq.~(\ref{tilde_after_full_exponentiation_most_moment_space_mu}) and
  $(C_{3/2},f^{\mathrm{PV}})=(1,0)$ in Eqs.~(\ref{C32}) and
  (\ref{Sud_with_cubic_Renormalons_single_parameter}), is shown as a full (black)
  line. The dotted line is
  obtained by choosing $\mu=m_b^{\mathrm{PV}}/2$.
  A slightly better description of the data is obtained upon
  choosing $\mu=m_b^{\mathrm{PV}}/2$ and $(C_{3/2},f^{\mathrm{PV}})=(6.2,.3)$. This
  result is indicated by the (red) dot-dashed line.  These choices all have
  equal theoretical justification.  Right: The breakdown on sources for the uncertainty.}
\end{center}
\end{figure}

Spectral moments defined over a restricted energy range $E_{\gamma}>E_0$~\cite{Andersen:2005bj},
have proven to be a useful tool for comparison of the spectrum between
experimental data and theory~\cite{Abe:2005cv,Aubert:2005cb,Aubert:2005cu}.
This comparison is important for a few reasons. First, it allows to test the extrapolation
of ${\bar B}\longrightarrow X_s \gamma$ from the region of measurement.
Second, more generally, it allows to test our understanding of the Sudakov region, which is a critical
ingredient in extracting $|V_{\rm ub}|$ from semileptonic decays,
${\bar B}\longrightarrow X_u l {\bar\nu}$. It that case
the extrapolation from the region of measurement into the whole of phase space is very large, a
factor of $\sim 2-5$ depending on the specific kinematic cuts applied, and thus the sensitivity to
the details of the spectrum in the peak region is significant.

Following Ref.~\cite{Andersen:2005bj} we consider here the average
photon energy with a cut, namely
\begin{equation}
\left<E_{\gamma}\right>_{E_{\gamma}>E_0}\equiv\frac{\displaystyle
\int_{E_0} dE_{\gamma}\, \frac{d\Gamma(E_{\gamma})}{dE_{\gamma}}\,
E_{\gamma}}{\displaystyle \int_{E_0} dE_{\gamma}\,
\frac{d\Gamma(E_{\gamma})}{dE_{\gamma}}} \label{ave_E}
\end{equation}
and higher truncated moments, defined by:
\begin{equation}
\left<\left(\left<E_{\gamma}\right>_{E_{\gamma}>E_0}-
E_{\gamma}\right)^n\right>_{E_{\gamma}>E_0}\equiv\frac{\displaystyle\int_{E_0}
dE_{\gamma}\, \frac{d\Gamma(E_{\gamma})}{dE_{\gamma}}\,
\left(\left<E_{\gamma}\right>_{E_{\gamma}>E_0}-
E_{\gamma}\right)^n}{\displaystyle \int_{E_0} dE_{\gamma}\,
\frac{d\Gamma(E_{\gamma})}{dE_{\gamma}}}. \label{higher_cent_mom}
\end{equation}

It is obvious that the higher the cut, or the moment $n$ considered, the higher is the sensitivity
to the fine details of the peak and, eventually, the endpoint region, $E_{\gamma}\simeq M_B/2$.
In contrast to the BF with mild cuts, where one can obtain a purely perturbative prediction,
when considering the moments or high cuts one must take account of power corrections.
In DGE, this can be judged by the sensitivity of the Borel integral to values of $u$
away from the origin. In our parametrization of the soft function $B_{\cal S}(u)$ (see Eqs. (3.27)
to (3.29) in Ref.~\cite{Andersen:2005mj})
this directly depends on $C_{3/2}$. Since $B_{\cal S}(u)$ at $u\gsim 3/2$ is not known, and since
the significance of power corrections in (\ref{Sud_with_cubic_Renormalons}) directly depends on what
it is assumed to be, it becomes obvious that these two
aspects must be addressed together, and that
an additional constraint would be needed. Here we propose to use the support properties, namely
the vanishing of the spectrum for $E_{\gamma}>M_B/2$ for this purpose.

In taking power corrections in the peak region into account one has
to find a compromise that allows a sufficiently accurate description
of the spectrum and yet involves a sufficiently small number of
non-perturbative parameters. In
general, starting with perturbation theory, the closer the region of
interest to the endpoint, the harder it is to find such a
compromise. Here we wish to explore the relation between $C_{3/2}$
in the parametrization of $B_{\cal S}(u)$ entering Eq.~(\ref{Sud_with_cubic_small_x_general_color}),
the power corrections and the support properties.
As a first exploration of this issue, it is
reasonable to consider a single non-perturbative parameter. On the
other hand, since we would like to explore extreme cuts --- the support
properties --- there is no justification in using only the leading power
term, $k=3$ in Eq.~(\ref{Sud_with_cubic_Renormalons}). For $E_{\gamma}\simeq M_B/2$
such hierarchy does not exist.
In order to use
Eq.~(\ref{Sud_with_cubic_Renormalons}) we
therefore take a sum of \emph{all} renormalon ambiguities, all weighted by one non-perturbative
parameter, $f^{\PV}$:
\begin{align}
\label{Sud_with_cubic_Renormalons_single_parameter}
\begin{split}
\left.\widetilde{\rm Sud}^{(J)}(N,m_b )\right\vert_{\PV} &
\,\longrightarrow \,\left. \widetilde{\rm
Sud}^{(J)}(N,m_b)\right\vert_{\PV}\times\\
& \exp\Bigg\{\frac{C_F}{\beta_0}\,\pi\,f^{\PV} \,
\sum_{k=3}^{\infty} \,\frac{T(k/2)}{k/2}\, \,B_{\cal S}(k/2)
\left(\frac{\Lambda}{m_b}\right)^k \,R^{(J)}(N,k/2)\Bigg\},
\end{split}
\end{align}
where $f^{\PV}$ is expected to be of order $1$.
Eq.~(\ref{Sud_with_cubic_Renormalons_single_parameter}) includes a sum of
all powers of $(N\Lambda/m_b)^k$ with $k\geq 3$, which a priori (depending on $B_{\cal S}(k/2)$)
may all be relevant for $E_{\gamma}\simeq M_B/2$. Going over to milder
cuts, where only the leading power is relevant, $f^{\PV}$ can be
identified with $f^{\PV}_3$ in (\ref{Sud_with_cubic_Renormalons}).

\begin{figure}[t]
\begin{center}
\epsfig{file=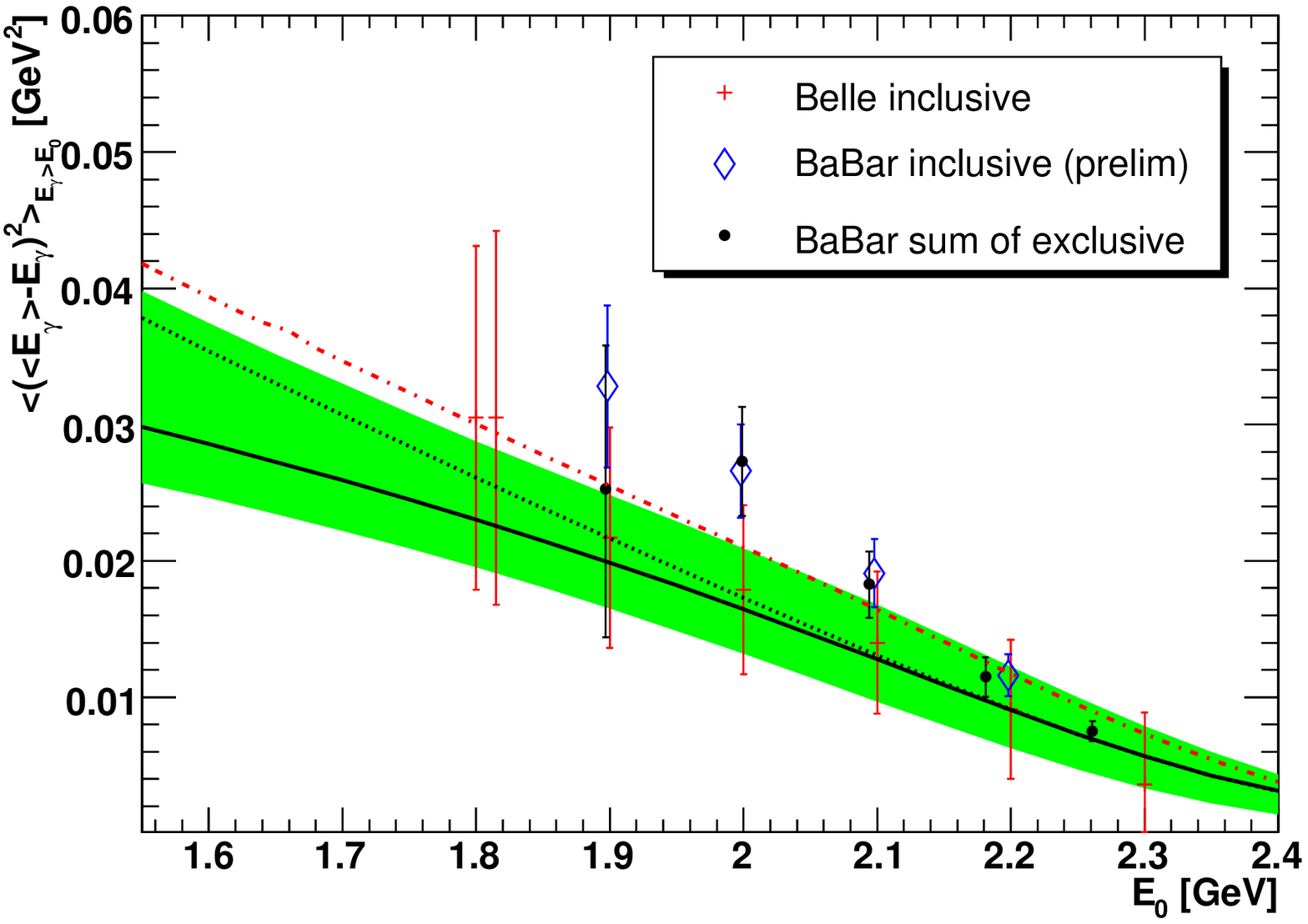,angle=0,width=.49\textwidth}\hfill
\epsfig{file=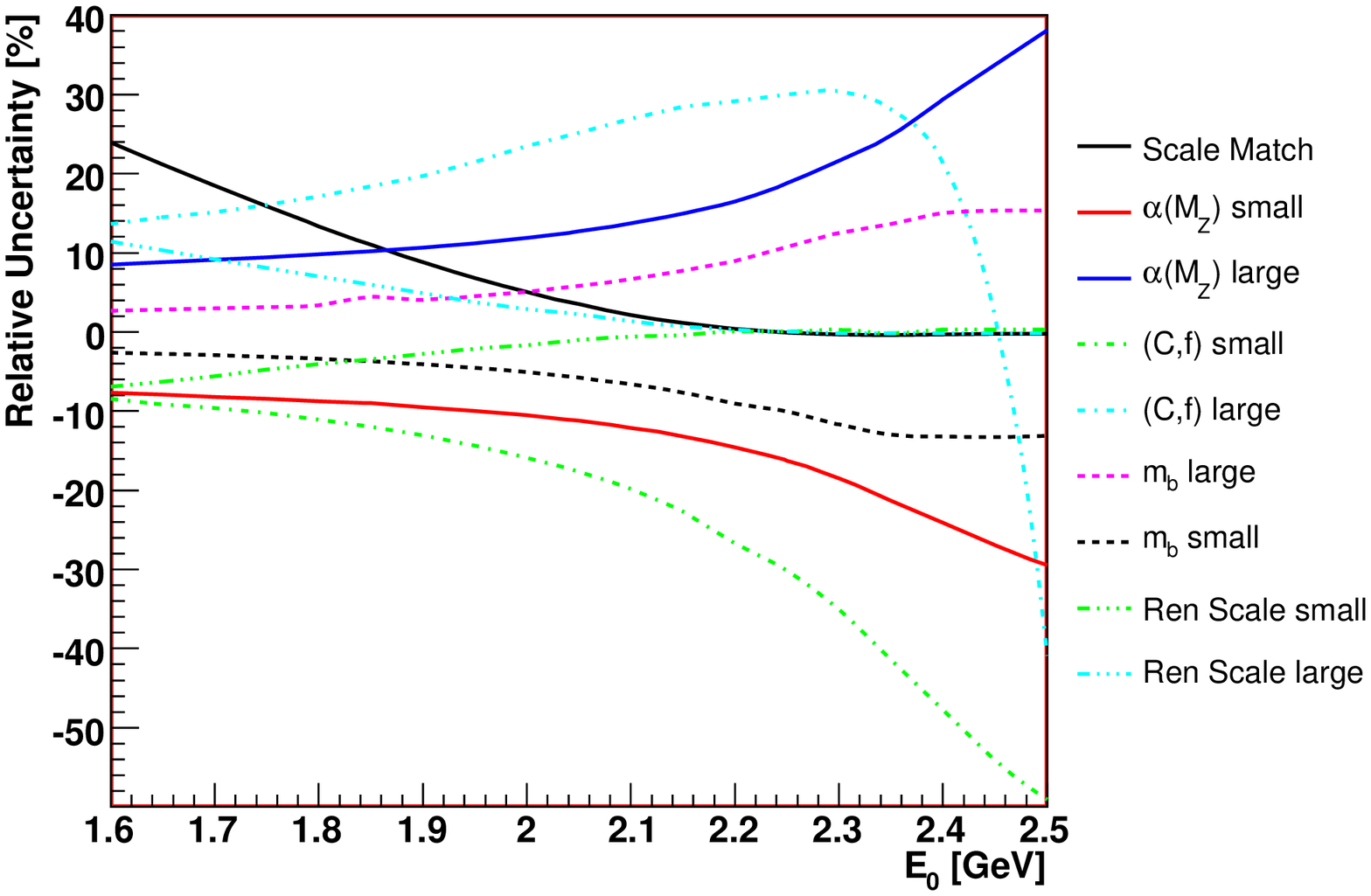,angle=0,width=.49\textwidth}
\caption{\label{fig:Variance} Left: The prediction for the second central
  moment $\left\langle\left(\langle
      E_\gamma\rangle_{E_\gamma>E_0}-E_\gamma\right)^2\right\rangle_{E_\gamma>E_0}$
  as a function of $E_0$ compared with data from Belle and BaBar. The band
  indicating the uncertainty and the various lines are obtained with the same
  choice of parameters as the corresponding ones in Fig.~\ref{fig:AvgEnergy}.
Right: The breakdown on sources for the
  uncertainty.}
\end{center}
\end{figure}

We already know from previous studies that very large values of $C_{3/2}$ or of power corrections
cannot be accommodated with the support properties, $E_{\gamma}<M_B/2$. Here we would like to
translate this into a concrete constraint. To this end we require
that neither the spectrum nor the first moment
extend too far beyond the physical endpoint.
Considering all possible values of $C_{3/2}$ and $f^{\mathrm{PV}}$, we implement
this requirement using the following cuts:
\begin{align}
  \begin{split}
    \label{eq:Cepscuts}
    \mathrm{CutA}(C_{3/2},f^{\mathrm{PV}})&=\left|\frac{\Gamma(\bar{B}\longrightarrow
        X_s\gamma,E_\gamma>M_B/2)}{\Gamma(\bar{B}\longrightarrow
        X_s\gamma,E_\gamma>M_B/20)}\right|\\
    \mathrm{CutB}(C_{3/2},f^{\mathrm{PV}})&=\left|1-\frac{\langle
        E_\gamma\rangle_{E_\gamma>M_B/2}}{M_B/2}\right|.
  \end{split}
\end{align}

In Fig.~\ref{fig:map} we have indicated by plus's the region in the plane
of $(C_{3/2},f^{\mathrm{PV}})$ for which the spectra conform to
$\mathrm{CutA}(C_{3/2},f^{\mathrm{PV}})<0.01$
with the remaining parameters at their default value. We have also indicated
by circles (asterisks) the region for which the spectra also obey
$\mathrm{CutB}(C_{3/2},f^{\mathrm{PV}})\,<\,0.01\, (0.02)$.

We observe that there is a
reasonably large range in the $(C_{3/2},f^{\mathrm{PV}})$ parameter space that conforms with the
physical support properties. The details depend on how stringent the constrains on
(\ref{eq:Cepscuts}) are. What is general, however, is that acceptable spectra have
power corrections that are typically of the order of the renormalon ambiguity: as shown in
Fig.~\ref{fig:map} $f^{\mathrm{PV}}\lsim 1$, except at \emph{very} small $C_{3/2}$ where the
power correction essentially has no effect.

Having excluded too large contributions to the Sudakov factor
from large $u$ values and too large power corrections
based on the support criterion, we still have a variety of spectra whose properties,
i.e. the first few cut moments, are different.
In the analysis of the theoretical uncertainty in this paper we
allow $(C_{3/2},f^{\mathrm{PV}})$ to vary within the
region of $\mathrm{CutA,\,CutB}<0.01$.
It is reassuring that the power
corrections satisfying these requirements are also of the
order of the renormalon ambiguity, or smaller. Note, on the other hand,
that for most of the range in $C_{3/2}$ vanishing power
corrections are excluded as well.

\begin{figure}[t]
\begin{center}
\epsfig{file=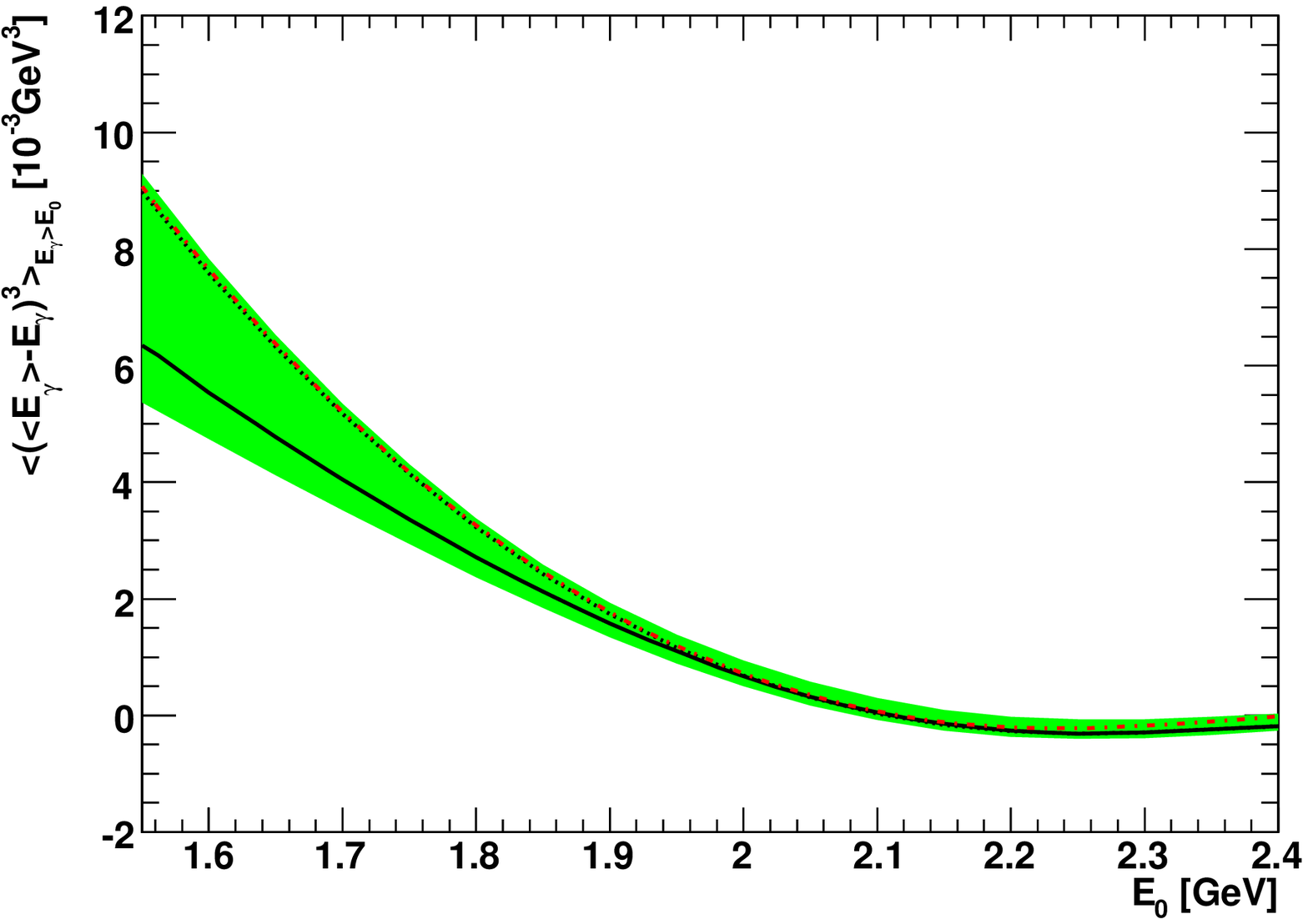,angle=0,width=.49\textwidth}\hfill
\epsfig{file=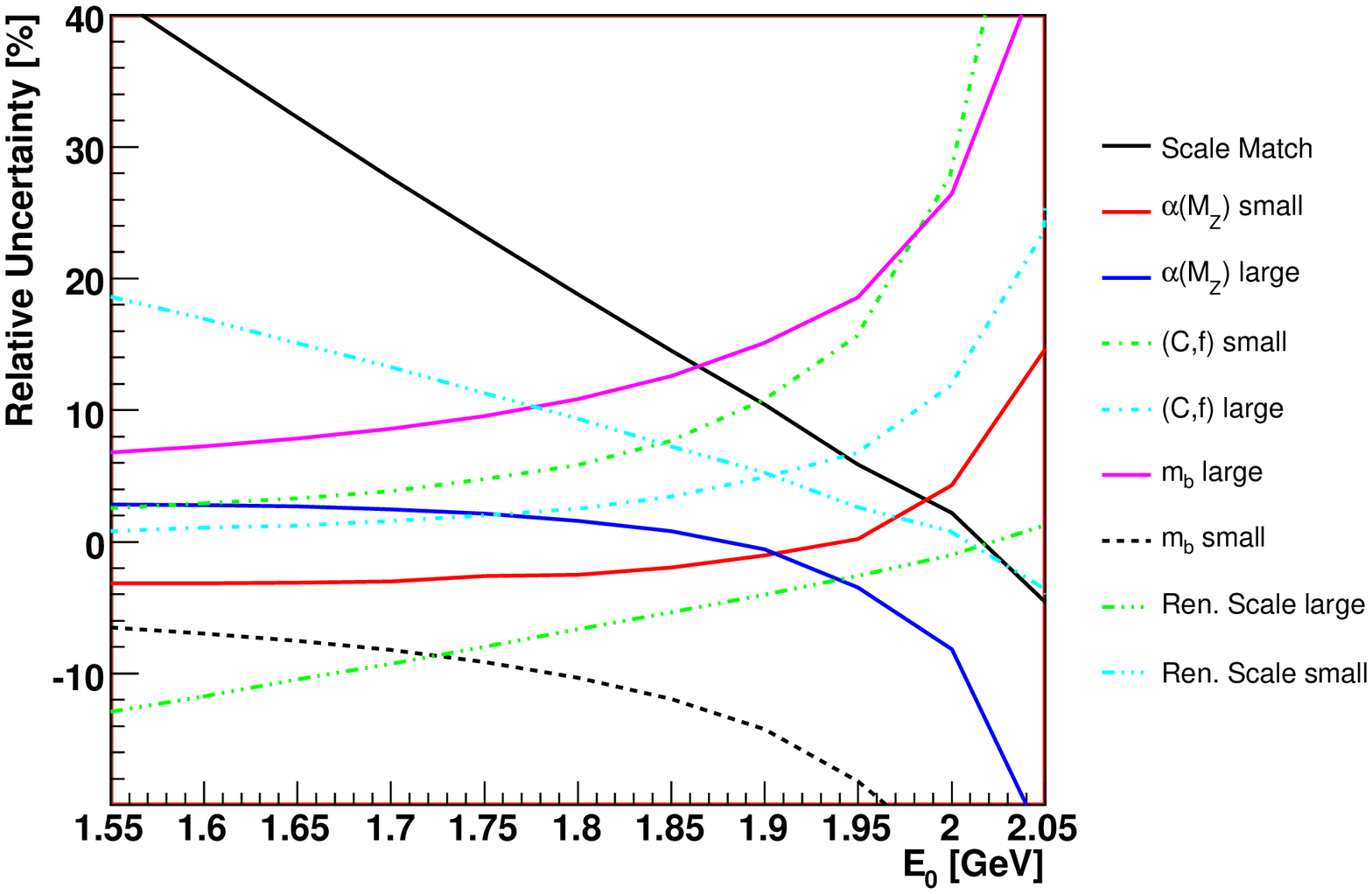,angle=0,width=.49\textwidth}
\caption{\label{fig:third_central_moment}  Left: The prediction for the third central
  moment $\left\langle\left(\langle
      E_\gamma\rangle_{E_\gamma>E_0}-E_\gamma\right)^3\right\rangle_{E_\gamma>E_0}$
  as a function of $E_0$. The band
  indicating the uncertainty and the various lines are obtained with the same
  choice of parameters as the corresponding ones in Fig.~\ref{fig:AvgEnergy}.
Right: The breakdown on sources for the
  uncertainty. This plot is terminated when the prediction for the third
  moment becomes consistent with zero.}
\end{center}
\end{figure}

In Figs.~\ref{fig:AvgEnergy} to~\ref{fig:third_central_moment} we show the cut dependence of the
first three central moments, $M_B-\left<E_{\gamma}\right>_{E_{\gamma}>E_0}$ where the average energy
is defined in (\ref{ave_E}), the variance $n=2$ in (\ref{higher_cent_mom}) and the third
moment $n=3$ in (\ref{higher_cent_mom}), respectively. The bands in the figures on the left hand side
represent
the estimated theoretical uncertainty obtained by the same procedure used in Sec.~\ref{sec:partial_BF}. The various contributions to the uncertainty are presented
in the plots on the right. As expected, the uncertainty associated with the behavior of the quark
distribution function, $B_{\cal S}(u)$ at $u\sim 3/2$, and the corresponding power corrections
increases as the cut is raised. It also gets more significant of course for higher moments.

In the average energy and variance figures, Figs.~\ref{fig:AvgEnergy} and~\ref{fig:Variance},
we also present experimental data~\cite{Abe:2005cv,Aubert:2005cb,Aubert:2005cu}
by Belle and two analysis by BaBar. Fits to data, with simultaneous
variation of $m_b$,  $C_{3/2}$ and $f^{\PV}$, can be very useful for the measurement of the bottom
mass and for testing the theoretical description of the peak region. We do not perform any such
fits\footnote{Fits should of course take into account the separation between statistical and systematic
experimental errors (which are summed here in quadrature)
as well as correlations between the data points.}
here, and the comparison with data is merely qualitative.

The default line (full line) in Figs.~\ref{fig:AvgEnergy} and~\ref{fig:Variance}, corresponding
to the $C_{3/2}=1$ choice made in Refs.~\cite{Andersen:2005bj,Andersen:2005mj} with no power
corrections $f^{\PV}=0$, seems to agree very well with the Belle data, and not as
good with the BaBar data that often has smaller errors.
In particular, in the average energy plot, the theoretical curve lies quite significantly
above the BaBar data point at $E_0=2.26$ GeV; in the variance plot it
is about two sigma below the BaBar data points at $E_0<2.1$ GeV.
When making this comparison one should take great care in interpreting theoretical uncertainties,
which do not reflect probability.
For example, there is absolutely no theoretical preference to the choice $C_{3/2}=1$ with $f^{\PV}=0$
as compared, for example to $C_{3/2}=6.2$ with $f^{\PV}=0.3$. Similarly, there is no preference for
choosing the renormalization scale in (\ref{tilde_after_full_exponentiation_most_moment_space_mu})
as $\mu=m_b$ (our default) as compared to $\mu=m_b/2$. Short of further calculations, these
choices remain arbitrary to a large extent. We note that with these different choices there is
good agreement with BaBar data. This is demonstrated by the dot-dashed lines in
Figs. \ref{fig:AvgEnergy} and \ref{fig:Variance}.

\section{Conclusions~\label{sec:conc}}

We presented here a calculation of the ${\bar B}\longrightarrow X_s \gamma$
branching fraction,
as well as the first few spectral moments as a function of a
cut on the photon energy, using the framework of
Dressed Gluon Exponentiation. Building on our previous work~\cite{Andersen:2005bj,Andersen:2005mj},
and on recent progress in fixed--order
calculations~\cite{Blokland:2005uk,Asatrian:2006ph,Melnikov:2005bx,Asatrian:2006sm,Bieri:2003ue},
we now have accurate predictions for these observables and good understanding of the theoretical
uncertainties and their dependence on the cut.

We made progress on several different aspects of the calculation:
\begin{itemize}
\item{} In matching the resummed spectrum into the fixed--order expansion
(Appendix~\ref{sec:matching}), we made full use of
the available NNLO results of the $G_{77}$ spectrum~\cite{Melnikov:2005bx,Asatrian:2006sm}.
In this sector we therefore have a complete NNLO with NNLL accuracy.
As shown in Fig.~\ref{fig:DGE_vs_FO} the DGE spectrum does not vary much in going from NLO to NNLO,
indicating that all important higher--order corrections are indeed resummed.
The largest relative variation is along the tail of the distribution where NNLO corrections
in the matching coefficient are important (see also
Fig.~\ref{fig:J_eq_3_BF_NLO_NNLO}).
\item{} We developed a method to perform Sudakov resummation without violating the analytic structure
of the perturbative series in moment space, and therefore, without generating artifacts
away from the Sudakov region. Using the Sudakov factor (\ref{Sud_with_cubic_small_x_general_color})
the resummed spectrum can therefore be used down to small $E_{\gamma}$ where it matches into
the fixed--order result (See Fig.~\ref{fig:J_dep}).
\item{} Matrix elements of other operators in the effective Weak Hamiltonian
and their interference with $O_7$ are known in full to ${\cal O}(\alpha_s)$ only.
However, we also know that independently of the nature of the short--distance interaction,
all important contributions in the peak region, i.e. ones that do not vanish as ${\cal O}(1/N)$,
necessarily involve the same Sudakov factor.
Here we computed the spectra of individual matrix elements $G_{ij}$ assuming the same Sudakov factor
multiplied by ${\cal O}(\alpha_s)$ hard coefficient functions that depend on the operators.
The resulting effect on the spectrum is shown in~Fig.~\ref{fig:spec_ind_contr}.
\item{} We performed a first numerical study of the relation between renormalon contributions
to the quark distribution function, power corrections, and \emph{support properties}.
To this end we used the parametrization of the soft anomalous dimension function $B_{\cal S}(u)$ proposed in
Ref.~\cite{Andersen:2005mj} (see Eqs.~(3.27) to (3.29) there), which is consistent with all
available constraints for $u\to 0$, and at $u=1/2$ and $u=1$; it includes a single parameter
$C_{3/2}$ that controls the $u\gsim 3/2$ region, which is not well constrained theoretically.
We also used a simple formula with a single parameter for the corresponding power corrections,
Eq.~(\ref{Sud_with_cubic_Renormalons_single_parameter}), which nevertheless takes into account all
powers of $(N\Lambda/m_b)^k$ with $k\geq 3$. Under these assumptions we showed that there
exists a range in the parameter space where the computed spectrum conforms with the physical
support properties. We observed that the size of the power corrections is of the order of
the renormalon ambiguity, as expected on general grounds.
\item{} We proposed a new method to evaluate the total ${\bar B}\longrightarrow X_s \gamma$
width that utilizes the semileptonic width, where the leading renormalon ambiguity has been explicitly
dealt with using Borel summation~\cite{Andersen:2005mj}. Our new method
avoids using any additional mass scheme.
The result for the total BF is summarized in \eq{total_BF_result}. It is
consistent with previous determinations~\cite{Gambino:2001ew,Buras:2002tp}.
Unfortunately, despite having partial NNLO
information on the matrix elements~\cite{Blokland:2005uk,Asatrian:2006ph,Bieri:2003ue},
the theoretical uncertainty is still large,~$\pm 13.7\%$. We observed large cancellations
between $G_{77}$ and $G_{27}$ contributions, which vary depending on the renormalization scale.
Therefore, significant improvement is expected upon completion of the NNLO calculation of $G_{27}$.
\item{} Finally, we devised a method to compute the BF with a cut on the photon energy, by combining
 the resummed calculation of the spectrum of individual matrix elements with the proper weight,
 given by the
 fixed--order calculation of the total BF, \eq{total_width_E0_using_normalized}.
 This framework facilitates the analysis of theoretical uncertainties
 associated with different ingredients, which are known at
 different orders, and have different r\^ole
  depending on how stringent the cut is.
\end{itemize}

It is of theoretical as well as practical interest to understand the behavior of
higher--order corrections. The overwhelming dominance of running--coupling corrections
at ${\cal O}({\alpha_s}^2)$, and the very late settling of the leading logarithmic behavior
at large $x$~\cite{Melnikov:2005bx}, might be interpreted as a signal that Sudakov resummation
is irrelevant. In order to address the relative significance of higher--order corrections
of different origin we presented in Sec.~\ref{sec:large_beta0_Borel} the all--order resummation of the
 $G_{77}$ spectrum in the large--$\beta_0$ limit, based on the calculation of Ref.~\cite{Gardi:2004ia}.
When working directly in momentum space ($x$ space) there are no renormalon ambiguities in the
real--emission result, but there are convergence constraints on the Borel integral. Consequently,
the Borel sum does not exist for $E_{\gamma}\gsim 2$ GeV. As shown in Fig.~\ref{fig:DGE_vs_FO}
and \ref{fig:Fixed_order_x_space},  the large--$\beta_0$ sum does not give
a viable description of the spectrum in the peak region. This stands in sharp contrast with DGE,
which through real--virtual cancellation and exponentiation in moments space,
accounts for multiple soft and collinear radiation. Thus, despite the
late settling of the leading logarithmic behavior at large $x$ up to ${\cal O}({\alpha_s}^2)$,
such logarithms \emph{are} important at higher orders. It is the combination of soft gluon
resummation with the resummation of running--coupling corrections that open the way for the
quantitative description of inclusive decay spectra.

Finally, the results of the present paper can help in various ways in understanding and using
the B factory ${\bar B}\longrightarrow X_s \gamma$ data.
In order to make comparison between the BF measurements with
a given kinematic cut, $E_{\gamma}> E_0$, and the theoretical result of~\eq{total_BF_result} we provided in
Table \ref{table:ext_factors} estimates of the extrapolation factor as a function of $E_0$.
This information, as well as all other details of the spectrum and the moments
computed here, can be obtained using a ${\tt c^{++}}$ program that is made
publicly available\footnote{\tt http://www.hep.phy.cam.ac.uk/$\sim$andersen/BDK/}.
Beyond the issue of the BF itself, combining the theoretical calculations presented here with
${\bar B}\longrightarrow X_s \gamma$ data is extremely valuable for other aspects of flavor
physics: it allows a precise determination of the the b--quark mass, as well as testing and improving
the description of inclusive decay spectra in the Sudakov region by quantifying Fermi--motion
corrections. This is particularly important for the determination of
$|V_{\rm ub}|$ from the B factory measurements of charmless semileptonic
decays\cite{HFAG,unknown:2006bi}.

\vskip 20pt

\noindent
{\bf{Note Added:}} Upon completion of this paper there appeared two new papers addressing
the calculation of the total BF \cite{Many_new,MS_new}. In particular, Ref.~\cite{MS_new} estimates the ${\cal O}(\alpha_s^2)$ contribution of the four--quark operator matrix elements using extrapolation from large $m_c$. This goes beyond
the large--$\beta_0$ results of Ref.~\cite{Bieri:2003ue} that we used here.
The final result for the total BF in \cite{Many_new,MS_new} is consistent with ours, but has a lower central value and a significantly smaller uncertainty from varying the renormalization scale. The reasons for that are not yet clear.

\appendix

\section{NNLO matching of the resummed $G_{77}$ spectrum\label{sec:matching}}

\subsection{Expansion of the Sudakov exponent and a basic NNLO matching formula~\label{sec:basic_matching}}

Let us begin by recalling the ${\cal O}({\alpha_s}^2)$ result for the
Sudakov exponent~\cite{Andersen:2005bj}. Expanding \eq{Sud_DGE} to
this order one obtains \eq{Sud_expand} with
\begin{eqnarray}
\label{E12} E_1(N) &=& -\frac{a_1}{2}\bigg(S_1^2(N)-S_2(N)\bigg)-
(b_1-d_1)S_1(N)\\\nonumber E_2(N) &=& \bigg[-\frac{a_1}{2}
S_1^3(N)+\left(d_1-\frac12 b_1-\frac12 a_2\right)S_1^2(N)+
\left(d_2-b_2+\frac{3}{2}a_1\,S_2(N)\right)S_1(N)\\\nonumber
&&+\left(-d_1+\frac{1}{2}b_1+\frac{1}{2}a_2\right)S_2(N)-\frac12
a_1\,S_3(N)\bigg] \beta_0
\end{eqnarray}
where we define\footnote{Note the different normalization compared
to the standard harmonic sum.}
\begin{eqnarray}
&S_k(N)\equiv&(-1)^{k-1}\,(k-1)!\sum_{l=1}^{N-1}\frac{1}{l^{k}}\,=\,\Psi_{k-1}(N)-\Psi_{k-1}(1),\\
&{\rm so}& S_1(N)=\Psi(N)+\gamma_E,\,\,\,
S_2(N)=\Psi_1(N)-\frac{\pi^2}{6},\nonumber\\\,\,\, &
&S_3(N)=\Psi_2(N)+2\zeta_3,\,\,\,
S_4(N)=\Psi_3(N)-\frac{\pi^4}{15}.\nonumber
\end{eqnarray}
and where the coefficients\footnote{For the precise relation with
$B_{\cal S}(u)$ and $B_{\cal J}(u)$ see Ref.~\cite{Andersen:2005bj}.
Note that the notation used there in Eqs. (2.5) through (2.7) is:
$A_n=C_F\beta_0^{n-1}a_n$ etc.} of the Sudakov anomalous dimensions,
defined in the $\overline {\rm MS}$ scheme,
are~\cite{Gardi:2005yi,Moch:2004pa,Andersen:2005bj,Melnikov:2005bx,Melnikov:2004bm,Korchemsky:1992xv}:
\begin{align}
\begin{split}
\label{a123_b12_d12} a_1&=1,\qquad
a_2= \frac{5}{3}+\left(\frac{1}{3}-\frac{1}{12}\pi^2\right)\frac{C_A}{\beta_0},\\
a_3&=-\frac{1}{3}+\frac{1}{\beta_0}
\left[\left(\frac{55}{16}-3\zeta_3\right)C_F+\left(\frac{253}{72}
-\frac{5}{18}\pi^2+\frac{7}{2}\zeta_3\right)
C_A\right]\\&+\frac{1}{\beta_0^2}
\left[\left(-\frac{7}{18}-\frac{1}{18}\pi^2-\frac{11}{4}\zeta_3+\frac{11}{720}\pi^4\right)C_A^2
+\left(-\frac{605}{192}+\frac{11}{4}\zeta_3\right)C_AC_F\right],\\
 b_1&=-\frac{3}{4},\qquad
b_2=-\frac{247}{72}+\frac{1}{6}\pi^2 +\frac{1}{\beta_0}
\left[\left(-\frac{3}{32}-\frac{3}{2}\zeta_3+\frac{1}{8}\pi^2\right)C_F+
\left(-\frac{73}{144}+\frac{5}{2}\zeta_3\right)C_A\right],\\
 d_1&=1,\qquad
d_2=\frac{1}{9}+\frac{C_A}{\beta_0}\left[\frac{9}{4}\zeta_3-\frac{\pi^2}{12}-\frac{11}{18}\right].
\end{split}
\end{align}
In the following we will compute the matching coefficient functions
$H(\alpha_s(m_b),N)$ and $\Delta R(\alpha_s(m_b),x)$ in
\eq{G_77_resummed} to ${\cal O}({\alpha_s}^2)$ based on the known NNLO
expansion~\cite{Melnikov:2005bx} and the expansion of the exponent
in \eq{Sud_expand} with (\ref{E12}).

To this end we will need the color decomposition of the NNLO result
in \eq{G77_diff_form} (note that an overall factor of $C_F$ was
extracted there). The NLO coefficients entering $R(\alpha_s(m_b),x)$
and $V(\alpha_s(m_b))$ are $r_1(x)$, which is is given explicitly
\eq{bar_r_reg_expression1} below, and $k_1=-31/12$. The NNLO ones
will be decomposed as follows:
\begin{align}
\begin{split}
\label{k2r2}
k_2&=N_f k_2^{N_f}+C_F k_2^{C_F}+C_A k_2^{C_A};\\
r_2^{\sing}(x)&= N_f r_2^{N_f,\sing}(x)+C_F r_2^{CF,\sing}(x)+C_A
r_2^{C_A,\sing}(x); \\
r_2^{\reg}(x)&=N_f r_2^{N_f,\reg}(x)+C_F r_2^{CF,\reg}(x)+C_A
r_2^{C_A,\reg}(x).
\end{split}
\end{align}
The explicit expression for
$r_2^{N_f,\reg}(x)=-\frac16r_2^{\beta_0,\reg}(x)$ is given in
\eq{bar_r_reg_expression2} while the expressions for the other
terms, corresponding to  the $C_F^2$ and $C_FC_A$ color factors, can
be found in Eqs. (7) and (8) in Ref.~\cite{Melnikov:2005bx},
respectively. Below we quote only their singular parts,
$r_2^{\sing}(x)$, which were derived already in
\cite{Andersen:2005bj}. The regular parts, $r_2^{\reg}(x)$, are used
here but owing to their length we do not write their explicit
expressions here; these can read off Eqs. (7) and (8) in
Ref.~\cite{Melnikov:2005bx} by removing the $\delta$ and the
plus--distribution terms. Using this color decomposition one finds
the constants $k_n$ by integrating $r_n^{\reg}(x)$ according
to~\eq{cn_def}~\cite{Melnikov:2005bx}:
\begin{equation}
\label{c2CACFNf}
k_2^{N_f}=\frac{49}{48}-\frac{\zeta_3}{3}+\frac{\pi^2}{16}=1.237;\qquad
k_2^{C_A}=-4.795;\qquad  k_2^{C_F}=1.216.
\end{equation}

Having set the notation let us consider now the
partially--integrated $G_{77}$ matrix element at ${\cal
O}({\alpha_s}^2)$, as obtained from \eq{integrated_G77} with
(\ref{G77_diff_form}):
\begin{eqnarray}
\label{first_integrated_G77}
\frac{G_{77}(E_0,m_b)}{G_{77}(0,m_b)}&=&1+
C_F\left[k_1\frac{\alpha_s(m_b)}{\pi}
+k_2\left(\frac{\alpha_s(m_b)}{\pi}\right)^2+\cdots\right]
\\ \nonumber
&+& C_F\int_{x_0=2E_0/m_b}^1\!\!\!\!\! dx
\left[r_1^{\sing}(x)\frac{\alpha_s(m_b)}{\pi}
+r_2^{\sing}(x)\left(\frac{\alpha_s(m_b)}{\pi}\right)^2+\cdots\right]_{+}
\\ \nonumber
&+&C_F \int_{x_0=2E_0/m_b}^1\!\!\!\!\! dx \bigg(
r_1^{\reg}(x)\frac{\alpha_s(m_b)}{\pi}+r_2^{\reg}(x)
\left(\frac{\alpha_s(m_b)}{\pi}\right)^2 +\cdots\bigg).
\end{eqnarray}
Defining the moments of $r_n^{\sing}(x)$ according to \eq{mom_def},
\begin{equation}
\label{Rn_sign_def} R_n^{\sing}(N) \equiv  \int_0^1 dx\,
x^{N-1}\,\Big[ r_n^{\sing}(x) \Big]_{+}=\int_0^1 dx
\left(x^{N-1}-1\right) r_n^{\sing}(x) ,
\end{equation}
we find:
\begin{equation}
\int_{x_0=2E_0/m_b}^1\!\!\!\!\! dx
\left[r_n^{\sing}(x)\right]_{+}=\frac{1}{2\pi i}
\int_{c-i\infty}^{c+i\infty} \frac{dN}{N-1}
\,\left(\frac{2E_0}{m_b}\right)^{1-N}\,R_n^{\sing}(N)
\end{equation}
Thus, we can express \eq{first_integrated_G77} as an inverse--Mellin
transform:
\begin{eqnarray}
\label{before_exponentiation}
\frac{G_{77}(E_0,m_b)}{G_{77}(0,m_b)}&=&\nonumber \frac{1}{2\pi i}
\int_{c-i\infty}^{c+i\infty} \frac{dN}{N-1}
\!\left(\frac{2E_0}{m_b}\right)^{1-N}\!\! \Bigg\{ 1+
C_F\bigg[k_1\frac{\alpha_s(m_b)}{\pi}
+\,k_2\left(\frac{\alpha_s(m_b)}{\pi}\right)^2+\cdots\bigg]\\ && +\,
C_F\,\left[R_1^{\sing}(N)\frac{\alpha_s(m_b)}{\pi}
+R_2^{\sing}(N)\left(\frac{\alpha_s(m_b)}{\pi}\right)^2+\cdots\right]\Bigg\}
\\
&&+\,C_F\,\int_{x_0=2E_0/m_b}^1\!\!\!\!\! dx \bigg(
r_1^{\reg}(x)\frac{\alpha_s(m_b)}{\pi}+r_2^{\reg}(x)
\left(\frac{\alpha_s(m_b)}{\pi}\right)^2+\cdots\bigg)\nonumber
\end{eqnarray}

Next, we can incorporate the resummation of Sudakov logarithms by
rewriting \eq{before_exponentiation} in terms of the Sudakov factor
of \eq{Sud_DGE} according to the general form in~\eq{G_77_resummed}:
\begin{eqnarray}
\label{after_exponentiation}
&&\hspace*{-20pt}
\left[\frac{G_{77}(E_0,m_b)}{G_{77}(0,m_b)}\right]_{\rm
Resummed} =\frac{1}{2\pi i} \int_{c-i\infty}^{c+i\infty}
\frac{dN}{N-1} \,\left(\frac{2E_0}{m_b}\right)^{1-N}\,{\rm
Sud}(N,m_b)
\\&&\hspace*{50pt} \nonumber \times \exp \left[C_F k_1\frac{\alpha_s(m_b)}{\pi}
+C_F\left(\bar{k}_2-C_F
k_1^2/2\right)\left(\frac{\alpha_s(m_b)}{\pi}\right)^2
+\cdots\right]\\\nonumber
&&\hspace*{50pt}+C_F\int_{x_0=2E_0/m_b}^1\!\!\!\!\! dx \bigg(
r_1^{\reg}(x)\frac{\alpha_s(m_b)}{\pi}+\bar{r}_2^{\reg}(x)
\left(\frac{\alpha_s(m_b)}{\pi}\right)^2+\cdots\bigg).
\end{eqnarray}
Note that $\bar{r}_2^{\reg}(x)$ introduced here is different from
$r_2^{\reg}(x)$ used in \eq{before_exponentiation} since the
exponentiation generates some additional integrable terms (see
below). Such terms need to be subtracted\footnote{This is done in
\eq{r2_bar_def_integrable} below.} of $r_2^{\reg}(x)$ to avoid
double counting. Of course, $\bar{k}_2$ is defined accordingly,
$\bar{k}_n\equiv -\int_0^1 dx\, \bar{r}_n^{\reg}(x)$.

Upon expansion, this formula together with the coefficients of the
Sudakov exponent was used to predict~\cite{Andersen:2005bj} the
terms that diverge at $N\longrightarrow \infty$ at ${\cal
O}(\alpha_s^2)$, see Appendix A there. Explicit calculations later
confirmed these results~\cite{Melnikov:2005bx,Asatrian:2006sm}.
Expanding~\eq{after_exponentiation} using~\eq{Sud_expand} one gets,
under the inverse--Mellin integral:
\begin{equation}
\label{expansion_of_after_exponentiation}
1+C_F\left[
\left(R_1^{\sing}(N)+ k_1\right)\frac{\alpha_s(m_b)}{\pi}+
\left(\bar{R}_2^{\sing}(N)+\bar{k}_2\right)\left(\frac{\alpha_s(m_b)}{\pi}\right)^2+
\cdots\right],
\end{equation}
where
\begin{equation}
\label{R12_sing} R_1^{\sing}(N)= E_1(N),\,\,\,\qquad
\bar{R}_2^{\sing}(N)=\left(\frac{1}{2}E_1^2(N)+k_1E_1(N)\right)C_F
+E_2(N).
\end{equation}
\eq{expansion_of_after_exponentiation} should be compared with the
contents of the curly brackets in \eq{before_exponentiation}. The
${\cal O}(\alpha_s)$ terms are clearly identical, but the ${\cal
O}(\alpha_s^2)$ terms are not. They are equal as far as the terms
that diverge at $N\longrightarrow \infty$ are concerned, yet they
differ by ${\cal O}(1/N)$ corrections proportional to $C_F^2$ that
are generated by exponentiating $E_1(N)$. To account for this
difference we have introduced the bar notation. Specifically
${\bar{R}}_2^{\sing}(N)$ is different from the originally defined
$R_2^{\sing}(N)$ of \eq{Rn_sign_def} by some additional terms that
are finite at $N\longrightarrow \infty$:
\begin{align}
\begin{split}
\bar{R}_2^{\sing}(N)&=R_2^{\sing}(N)+\int_0^1 dx
\left(x^{N-1}-1\right) r_2^{\sing,\,\integrable} (x).
\end{split}
\end{align}
The constant at $N\longrightarrow \infty$ is retained at its
computed value by requiring
$\bar{R}_2^{\sing}(N)+\bar{k}_2=R_2^{\sing}(N)+k_2+{\cal O}(1/N)$,
see \eq{k2_bar_k2_relation} below. In the remainder of this section
we derive explicit expressions for the barred matching coefficients
in \eq{after_exponentiation}.

Starting with \eq{R12_sing} and using \eq{E12} we get the following
explicit expressions:
\begin{align}
\begin{split}
\label{R12_bar_explicit}
R_1^{\sing} (N)&=
-\frac{1}{2}\bigg(S_1^2(N)-S_2(N)\bigg)- (b_1-d_1)S_1(N)\\
\begin{split}
\bar{R}_2^{\sing} (N)&= C_F\Bigg[\frac{1}{8}S_1^4(N)+\frac12
\left(b_1-d_1\right)S_1^3(N)+ \left(-\frac12 k_1-\frac14
S_2(N)+\frac{1}{2}(d_1-b_1)^2\right)S_1^2(N)\\&+
(d_1-b_1)\left(\frac{1}{2}S_2(N)+k_1\right)S_1(N)
+\frac{1}{8}S_2^2(N)+\frac{1}{2}k_1S_2\Bigg]\\
&+\beta_0\Bigg[-\frac{1}{2}S_1^3(N)+\left(d_1-\frac12 b_1-\frac12
a_2\right)S_1^2(N)
+\left(\frac{3}{2}S_2(N)+d_2-b_2\right)S_1(N)\\
&+\left(-d_1+\frac{1}{2}b_1+\frac{1}{2}a_2\right)S_2(N)
-\frac{1}{2}S_3(N) \Bigg]
\end{split}
\end{split}
\end{align}
where we substituted $a_1=1$; other coefficients can be read
off~\eq{a123_b12_d12}. The additional terms contained in
$\bar{R}_2^{\sing} (N)$ but not in $R_2^{\sing} (N)$, (finite terms
at $N\longrightarrow \infty$) are:
\begin{align}
\begin{split}
&\int_0^1 dx \left(x^{N-1}-1\right) r_2^{\sing,\,\integrable}
(x)=C_F \,
\Bigg[-S_2(N)\left(\frac{\pi^2}{12}-\frac12(d_1-b_1)^2\right)
\\
&+(b_1-d_1) \left( \left(S_2(N)+\frac{\pi^2}{6}\right)S_1(N)-\frac12
S_3(N)\right)
\\
&-\frac12\left(\frac{\pi^2}{6}\left(S_2(N)-S_1^2(N)\right)-S_1^2(N)S_2(N)+\left(S_3(N)-2\zeta_3\right)S_1(N)-\frac13S_4(N)+S_2^2(N)
\right)\\
&+\frac12 \left(
\frac12S_2^2(N)-\frac{1}{12}S_4(N)+\frac{\pi^2}{6}S_2(N) \right)
\Bigg]
\end{split}
\end{align}
In $x$ space this corresponds to
\begin{align}
\begin{split}
\label{r12_bar_x}
r_1^{\sing}(x)&=-\frac{\ln(1-x)}{1-x}+(b_1-d_1)\frac{1}{1-x}
\\
\bar{r}_2^{\sing}(x)&=r_2^{\sing,\,\integrable}(x)+r_2^{\sing}(x)
\end{split}
\end{align}
where the original terms, appearing under the plus prescription in
\eq{G77_diff_form}, are\footnote{These terms are the same as in Eq.
(A.4) in \cite{Andersen:2005bj}, as they must be.}
\begin{align}
\begin{split}
\label{r2sing} r_2^{\sing}(x)&=C_F \Bigg[
\left(\zeta_3-k_1(d_1-b_1)+\frac{\pi^2}{6}(b_1-d_1)\right)\frac{1}{1-x}
\\
&+\left((d_1-b_1)^2-\frac{\pi^2}{6}-k_1\right)\frac{\ln(1-x)}{1-x}+\frac{3}{2}\left(d_1-b_1\right)\frac{\ln^2(1-x)}{1-x}
+\frac12 \frac{\ln^3(1-x)}{1-x}
\Bigg]\\
&+\beta_0
\Bigg[(b_2-d_2)\frac{1}{1-x}+(2d_1-b_1-a_2)\frac{\ln(1-x)}{1-x}
+\frac32\frac{\ln^2(1-x)}{1-x}\Bigg]
\end{split}
\end{align}
while the additional, integrable terms that arise from $C_F^2
E_1^2(N)$ are:
\begin{align}
\begin{split}
\label{r2_integrable}
&r_2^{\sing,\,\integrable}(x)=C_F\Bigg[\left(\frac{\pi^2}{12}-\frac{1}{2}(d_1-b_1)^2\right)\frac{\ln
(x)}{1-x}
+(b_1-d_1)\frac{\ln (x)\ln(1-x)}{1-x}\\
&\qquad-\frac12\frac{\ln
(x)\ln^2(1-x)}{1-x}+\frac{1}{2}\frac{1}{1-x}\,
\left(\displaystyle-\ln(x)\left({\rm
Li}_2(x)+\frac{\pi^2}{6}\right)+2\,{\rm Li}_3(x)-2\zeta_3
\right)\Bigg].
\end{split}
\end{align}

It is now straightforward to obtain the modified (barred)
$r_2^{\reg}(x)$ that enters the matching formula of
\eq{after_exponentiation}:
\begin{equation}
\label{r2_bar_def_integrable}
\bar{r}_2^{\reg}(x)=r_2^{\reg}(x)-r_2^{\sing,\,\integrable}(x).
\end{equation}
Note that only the $C_F^2$ term is modified. Finally, the
corresponding change in $k_2$ can be easily determined by computing
\begin{align}
\label{k2_bar_k2_relation}
\begin{split}
\bar{k}_2-k_2=\int_0^1dx
\,r_2^{\sing,\,\integrable}(x)&=-\lim_{N\longrightarrow
\infty}\left[\bar{R}_2^{\sing}(N)-R_2^{\sing}(N)\right]
\\&=\left(\frac{\pi^4}{720}+\frac{49}{192}\pi^2-\frac{7}{4}\zeta_3
\right) C_F = 0.550496\, C_F.
\end{split}
\end{align}
Using \eq{c2CACFNf} we therefore find $k_2^{C_F}=1.7662$, so
\begin{equation}
\label{k2_bar} \bar{k}_2\,=\,-4.795\,C_A  \,+\, 1.237 \,N_f \,+\,
1.7662 \,C_F.
\end{equation}

\subsection{Including more in moment space~\label{sec:more_in_mom}}

One advantage of the DGE, which is tightly connected to the fact
that the calculation is done in moment space, is that the resummed
spectrum smoothly extends beyond the perturbative endpoint $x=1$ and
tends to zero at $x=1+{\cal O}(\Lambda/m_b)$, even in the absence of
power corrections.

For the differential spectrum to be \emph{smooth} at $x=1$, not just
integrable, one must take into account \emph{in moment space} not
only terms that diverge (or are finite) for $N\longrightarrow
\infty$, but also ${\cal O}(1/N)$ corrections, corresponding in
particular to powers of $\ln(1-x)$. In the basic matching formula of
\eq{after_exponentiation} these correction are still part of
$r_n^{\reg}(x)$. In the following we rearrange the split between
moment space and $x$ space to incorporate all the terms that are
\emph{finite} for $x \longrightarrow 1$ in moment space, leaving
only subleading terms, that \emph{vanish} at $x \longrightarrow 1$
in $x$ space. In addition we include in moments space \emph{all}
running--coupling, ${\cal O}(\beta_0\alpha_s^2)$ effects.

One way to ensure this is, of course, full moment--space matching.
One defines:
\begin{equation}
\bar{R}_n^{\reg}(N) \equiv \int_0^1 dx \left(x^{N-1}-1\right)
\bar{r}_n^{\reg}(x)= \int_0^1 dx x^{N-1}
\bar{r}_n^{\reg}(x)+\bar{k}_n.
\end{equation}
and then
\begin{align}
\begin{split}
\label{after_full_exponentiation_all_moment_space}
&\hspace*{-5pt}\frac{G_{77}(E_0,m_b)}{G_{77}(0,m_b)} =\frac{1}{2\pi
i} \int_{c-i\infty}^{c+i\infty} \frac{dN}{N-1}
\,\left(\frac{2E_0}{m_b}\right)^{1-N}\,{\rm Sud}(N,m_b)
 \times \,\exp \left\{C_F R_1^{\reg}(N)\frac{\alpha_s(m_b)}{\pi}
\right.   \\ & \left. \hspace*{2pt}+\,C_F\left[ \bar{R}_2^{\reg}(N)-
C_F\left( \frac12\left(   R_1^{\reg}(N)\right)^2+
\left(R_1^{\reg}(N)-k_1\right) E_1(N)\right)\right]
\left(\frac{\alpha_s(m_b)}{\pi}\right)^2 +\cdots\right\}.
\end{split}
\end{align}

Here we choose to implement full moment--space matching at NLO while
splitting the NNLO corrections as follows:
\begin{equation}
\label{reg_leading}
\bar{r}_n^{\reg}(x)=\bar{r}_n^{\reg,\,\leading}(x) +
\bar{r}_n^{\reg,\,\subleading}(x)
\end{equation}
where $\bar{r}_n^{\reg,\,\subleading}(x)$ vanishes as $(1-x)$  at
$x\longrightarrow 1$ (up to logarithms) and
\begin{eqnarray}
\label{N_space_reg_leading}
\bar{R}_n^{\reg,\,\leading}(N)&=&\int_0^1dx \,
\bar{r}_n^{\reg,\,\leading}(x) \left(x^{N-1}-1\right) -
 \int_0^1dx\, \bar{r}_n^{\reg,\,\subleading}(x)\nonumber \\
 &=&\,\bar{k}_n\,+\,\int_0^1dx \, \bar{r}_n^{\reg,\,\leading}(x) x^{N-1}
\end{eqnarray}
such that at $N\longrightarrow \infty$:
$\bar{R}_n^{\reg,\,\leading}(N)\longrightarrow \bar{k}_n$, while at
finite $N$ $\bar{R}_n^{\reg,\,\leading}(N)$ contains all the terms
that fall as $1/N$ (including $1/N$ times a power of $\ln N$). We
shall not make this split for $r_1^{\reg}$ nor for the $\beta_0$
contribution to $\bar{r}_2^{\reg}$ (the so-called BLM term) ---
these will be fully contained in the moment--space expression.

For the leading terms in $\bar{r}_2^{\reg}$ at $x\longrightarrow 1$,
\eq{reg_leading} we find:
\begin{eqnarray}
\label{r2_reg_leading_x} \bar{r}_2^{N_f,\,\reg,\,\leading}(x) &=&
-\frac{7}{24}-\frac{\pi^2}{36}+\frac{13}{36}\ln(1-x)-\frac{1}{4}\ln^2(1-x)\\
\nonumber \bar{r}_2^{C_F,\,\reg,\,\leading}(x)  &=&
\frac{17}{72}\pi^2-\frac{9}{4}\zeta_3-\frac{379}{96}
+\left(-\frac{5}{12}-\frac{\pi^2}{4}
\right)\ln(1-x)+\ln^2(1-x)+\frac{1}{2}\ln^3(1-x)
\\ \nonumber
\bar{r}_2^{C_A,\,\reg,\,\leading}(x)  &=&
\frac{9}{8}\zeta_3+\frac{3}{4}-\frac{\pi^2}{36}
+\left(-\frac{22}{9}+\frac{\pi^2}{8}\right) \ln(1-x)+
\frac{11}{8}\ln^2(1-x).
\end{eqnarray}
In moment space, the coefficients of the different color factors of
$\bar{R}_n^{\reg,\,\leading}(N)$~read:
\begin{eqnarray}
\label{R2_reg_leading} \bar{R}_2^{N_f,\,\reg,\,\leading}(N) &=&
\bar{k}_2^{N_f} +\left(-\frac{7}{24}-\frac{\pi^2}{36}\right)
\mu_0(N) +\frac{13}{36}    \mu_1(N) -\frac{1}{4}      \mu_2(N)
\\ \nonumber
\bar{R}_2^{C_F,\,\reg,\,\leading}(N)  &=& \bar{k}_2^{C_F}
+\left(\frac{17}{72}\pi^2-\frac{9}{4}\zeta_3-\frac{379}{96}\right)
\mu_0(N) +\left(-\frac{5}{12}-\frac{\pi^2}{4}\right)
\mu_1(N)\\\nonumber &&\hspace*{200pt} +\, \mu_2(N) +\frac{1}{2}
\mu_3(N)
\\ \nonumber
\bar{R}_2^{C_A,\,\reg,\,\leading}(N)  &=& \bar{k}_2^{C_A}
+\left(\frac{9}{8}\zeta_3+\frac{3}{4}-\frac{\pi^2}{36}\right)
\mu_0(N) +\left(-\frac{22}{9}+\frac{\pi^2}{8}\right) \mu_1(N)
+\frac{11}{8} \mu_2(N),
\end{eqnarray}
$\mu_j(N)\equiv \int_0^1dx\,x^{N-1}\ln^j(1-x)$, e.g. $\mu_0(N)=1/N$,
$\mu_1(N)=-(\Psi(N)+\gamma_E)/N-1/N^2$, etc.

Making this split, and shifting the renormalization scale of the
coupling in the matching coefficient to an arbitrary scale $\mu$, we
obtain:
\begin{eqnarray}
\label{after_full_exponentiation_most_moment_space_mu}
&&\hspace*{-5pt}\left[\frac{G_{77}(E_0,m_b)}{G_{77}(0,m_b)}\right]_{\rm
Resummed} =\frac{1}{2\pi i} \int_{c-i\infty}^{c+i\infty}
\frac{dN}{N-1} \,\left(\frac{2E_0}{m_b}\right)^{1-N}\,{\rm
Sud}(N,m_b)
 \times
 \nonumber  \\ &&  \,\exp \Bigg\{C_F R_1^{\reg}(N)\frac{\alpha_s(\mu)}{\pi}
 +\,C_F\bigg[
\beta_0 \,\ln \left(\frac{\mu^2}{m_b^2}\right)\,R_1^{\reg}(N)+
 \bar{R}_2^{\reg, \,\leading}(N)\nonumber \\ &&\hspace*{30pt}-\,\, C_F\left(
\frac12\left(   R_1^{\reg}(N)\right)^2+
\left(R_1^{\reg}(N)-k_1\right) E_1(N)\right)
\bigg]\left(\frac{\alpha_s(\mu)}{\pi}\right)^2
\!+\cdots\!\Bigg\}\nonumber \\
&&\hspace*{50pt}+\,C_F\int_{x_0=2E_0/m_b}^1\!\!\!\!\! dx\,\,
\bar{r}_2^{\reg,\,\subleading}(x)
\left(\frac{\alpha_s(\mu)}{\pi}\right)^2+\cdots.
\end{eqnarray}
where the explicit expression for $R_1^{\reg}(N)$ is given in
\eq{R1_reg_N}, $\beta_0=\frac{11}{12}C_A-\frac{1}{6}N_f$ and
\begin{eqnarray}
\label{R2_reg_mom} \bar{R}_2^{\reg\,\leading}(N)&\equiv &
\underbrace{-6\beta_0 \bar{R}_2^{N_f,\reg}(N)}_{\rm BLM} \\
\nonumber &+& \underbrace{C_F \bar{R}_2^{C_F,\,\reg,\, \leading}(N)
+\,\,C_A \left(\bar{R}_2^{C_A,\,\reg, \,\leading}(N)
+\frac{11}{2}\bar{R}_2^{N_f,\,\reg,\,\leading}(N)\right)}_{\rm
non-BLM},
\end{eqnarray}
where $\bar{R}_2^{N_f,\reg}(N)$ is given explicitly in
\eq{R2_reg_N_Nf}, while the leading terms of the non-BLM color
factors are given in \eq{R2_reg_leading}. As announced above, terms
that are excluded from \eq{R2_reg_mom} are ${\cal O}(1/N^2)$ and do
not involve running--coupling effects. These residual terms are
included in $x$ space through
\begin{eqnarray}
\label{R2_reg_x} \bar{r}_2^{\reg,\,\subleading}(x)&\equiv& C_F
\bar{r}_2^{C_F,\,\reg,\,\subleading}(x)\nonumber \\&+&C_A
\left(\bar{r}_2^{C_A,\,\reg,\,\subleading}(x)
+\frac{11}{2}\bar{r}_2^{N_f,\,\reg,\,\subleading}(x)\right)\!\!.
\end{eqnarray}
As usual, the renormalization--scale dependence in
\eq{after_full_exponentiation_most_moment_space_mu} can serve as a
measure of subleading perturbative corrections at ${\cal
O}(\alpha_s^3)$ and beyond. In the numerical analysis we will vary
$\mu$ between $m_b/2$ and $m_b$ as one of several means to estimate
the theoretical uncertainty.

\subsection{NNLO matching under constraints on the analytic structure ($J\neq 0$)\label{sec:mathcing_Jneq0}}

Let us describe now the matching procedure with the Sudakov factor of
\eq{Sud_with_cubic_small_x_general_color}. The matching coefficients, just like the exponent, are
constructed under a constraint on the analytic structure in moment space:
no poles should appear for $N>-J$, and so the small--$x$ asymptotic behavior
would coincide with that of the fixed--order result, $d\Gamma/dx\sim x^J$.

The expansion of the $J$--modified Sudakov exponent,
\eq{Sud_with_cubic_small_x_general_color}, takes the form:
\begin{equation}
\label{tilde_Sud_expand} \widetilde{{\rm
Sud}}^{(J)}(N,m_b)=\exp\left\{
C_F\left[\widetilde{E}_1(N)\frac{\alpha_s(m_b)}{\pi}
+\widetilde{E}_2(N)\left(\frac{\alpha_s(m_b)}{\pi}\right)^2+\cdots\right]\right\}.
\end{equation}
with
\begin{eqnarray}
\label{tildeEn}
\widetilde{E}_n(N) &=& E_n(N+J)-E_n(J+1)
\end{eqnarray}
where $E_{n}$ for $n=1,2$ are given explicitly in \eq{E12}.

We now proceed to match \eq{Sud_with_cubic_small_x_general_color} into the NNLO
result. First, let us write a basic matching formula, in analogy with
\eq{after_exponentiation},
\begin{eqnarray}
\label{tilde_after_exponentiation}
&&\hspace*{-20pt}\left.\frac{G_{77}(E_0,m_b)}{G_{77}(0,m_b)}\right
\vert_{\rm Resummed} =\frac{1}{2\pi i} \int_{c-i\infty}^{c+i\infty}
\frac{dN}{N-1}
\,\left(\frac{2E_0}{m_b}\right)^{1-N}\,\widetilde{{\rm
Sud}}^{(J)}(N,m_b)
\\&&\hspace*{50pt} \nonumber \times \exp \left\{C_F \tilde{k}_1\frac{\alpha_s(m_b)}{\pi}
+C_F\left(\tilde{k}_2-C_F \tilde{k}_1^2/2\right)\left(\frac{\alpha_s(m_b)}{\pi}\right)^2
+\cdots\right\}\\\nonumber
&&\hspace*{50pt}+C_F\int_{x_0=2E_0/m_b}^1\!\!\!\!\! dx \bigg( \tilde{r}_1^{\reg}(x)\frac{\alpha_s(m_b)}{\pi}+
\tilde{r}_2^{\reg}(x)
\left(\frac{\alpha_s(m_b)}{\pi}\right)^2+\cdots\bigg).
\end{eqnarray}
Defining  $\widetilde{R}_n^{\sing}(N)$ as the coefficients arising from the expansion of the Sudakov factor times
the matching coefficient in \eq{tilde_after_exponentiation}:
\begin{equation}
\label{tilde_R12_sing}
\widetilde{R}_1^{\sing}(N)=\widetilde{E}_1(N),\,\,\,\qquad
\widetilde{R}_2^{\sing}(N)=\left(\frac{1}{2}\widetilde{E}_1^2(N)+\tilde{k}_1\widetilde{E}_1(N)\right)C_F+
\widetilde{E}_2(N)
\end{equation}
we can compute the modified regular parts as follows:
\begin{equation}
\label{rn_reg_tilde}
\tilde{r}_n^{\reg}(x)=r_n^{\reg}(x)-\Delta r_n^{\sing}(x)
\end{equation}
where $\Delta r_n^{\sing}(x)\equiv \tilde{r}_n^{\sing}(x)-r_n^{\sing}(x)$
and where
\begin{equation}
\label{tilde_Rn_sign_def}
\widetilde{R}_n^{\sing}(N)=\int_0^1dx \left( x^{N-1}-1\right)\tilde{r}_n^{\sing}(x),
\end{equation}
similarly to \eq{Rn_sign_def}. It is straightforward to compute $\tilde{k}_1$:
\begin{align}
\begin{split}
\label{tilde_k1}
\tilde{k}_1-k_1=\int_0^1dx \,\Delta r_1^{\sing}(x)&=-\lim_{N\longrightarrow \infty}\left[
\widetilde{R}_1^{\sing}(N)-R_1^{\sing}(N)\right]
\\&=-\left(\frac12 {S_1}^2 (1+J)+(b_1-d_1)S_1 (1+J)-\frac12S_2 (1+J)\right)C_F
\end{split}
\end{align}
Because of the structure of $\widetilde{E}_n$ in \eq{tildeEn} one can express $\widetilde{R}_n^{\sing}(N)$
as\footnote{As we show below $f_{n,J}(x)$ are in fact $J$ independent upon substituting for $\tilde{k}_1$
in terms of $k_1$. They are given by $f_{1,J}(x)=r_1^{\sing}(x)$ and $f_{2,J}(x)=\bar{r}_2^{\sing}(x)$.}
\begin{align}
\begin{split}
\label{tildeRn}
\widetilde{R}_n^{\sing}(N)&=F_n(N+J,J)-F_n(1+J,J)\equiv\int_0^1dx\,
x^{N+J-1}\left[f_{n,J}(x)\right]_{+} -\int_0^1dx\,
 x^{J}\left[f_{n,J}(x)\right]_{+}\\
&=
\int_0^1dx\,
\left( x^{N-1}-1\right)x^Jf_{n,J}(x).
\end{split}
\end{align}
This means the inverse Mellin transform of $\widetilde{R}_n^{\sing}(N)$ is readily obtained by multiplying
the inverse Mellin transform of $F_n(N,J)$ by $x^J$ \emph{under the plus prescription}.
Moreover, since for $J=0$ \eq{tildeRn} must coincide with \eq{R12_sing} we have:
\begin{equation}
F_1(N,0)=R_1^{\sing}(N);\qquad
F_2(N,0)=\bar{R}_2^{\sing}(N),
\end{equation}
and therefore \eq{tildeRn} implies\footnote{One can explicitly compute $\widetilde{R}_{1,2}^{\sing}(N)$ and therefore $F_n(N+J,J)$
using \eq{tilde_R12_sing}.
It is straightforward to verify that upon substituting for $\tilde{k}_1$
in terms of $k_1$ using \eq{tilde_k1}, the dependence of $F_n(N+J,J)$ on $J$ appears \emph{only} through $N+J$ and
the result coincides with \eq{R12_bar_explicit}.}
\begin{equation}
\label{tilde_Rn_sing_in_terms_of_non_tilde_ones}
\widetilde{R}_1^{\sing}(N)=R_1^{\sing}(N+J)-R_1^{\sing}(1+J);\qquad
\widetilde{R}_2^{\sing}(N)=\bar{R}_2^{\sing}(N+J)-\bar{R}_2^{\sing}(1+J)
\end{equation}
where the explicit expressions of the functions on the
r.h.s. are given in \eq{R12_bar_explicit} and
\begin{align}
\begin{split}
\label{Delta_r12_reg}
\Delta r_1^{\sing}(x)&=(x^J-1)\,r_1^{\sing}(x)\\
\Delta r_2^{\sing}(x)&=x^J\,\bar{r}_2^{\sing}(x)-r_2^{\sing}(x)
=x^J\,r_2^{\sing,\,\integrable}(x)\,+\,(x^J-1)\,r_2^{\sing}(x),
\end{split}
\end{align}
where $r_1^{\sing}(x)$ and $r_2^{\sing}(x)$ and $r_2^{\sing,\,\integrable}(x)$
are given in Eqs (\ref{r12_bar_x}), (\ref{r2sing}) and  (\ref{r2_integrable}), respectively.
Finally we find $\tilde{k}_2$ by
\begin{align}
\begin{split}
\tilde{k}_2 -\bar{k}_2 &=\int_0^1dx \,\Delta r_2^{\sing}(x)=-\lim_{N\longrightarrow \infty}\left[
\widetilde{R}_2^{\sing}(N)-\bar{R}_2^{\sing}(N)\right]\\
&=
C_F \Bigg[{\displaystyle \frac {1}{8}} \,{S_{1}}^{4}(1+ J)
- {\displaystyle \frac {1}{2}} \,( {d_1} -  {b_1})\,{S_{1}}^{3}(1 + J) \\
& + \left( \frac12(d_1-b_1)^2- {\displaystyle \frac {1}{4}} \,{S_{2}}(1 + J) -
{\displaystyle \frac { {k_1}}{2}}   \right)\,{S_{1}}^{2}(1 + J)
 \\
& + \left({\displaystyle \frac {1}{2}} \,(  {d_1}-  {b_1}
)\,{S_{2}}(1 + J) + ( {d_1} -  {b_1} )\,
 {k_1}\right)\,{S_{1}}(1 + J) + {\displaystyle \frac {1}{8}} \,{S
_{2}}^{2}(1 + J) + {\displaystyle \frac {1}{2}} \, {k_1}\,{S
_{2}}(1 + J)\Bigg] \\
&+{\beta _{0}}\Bigg[ - {\displaystyle \frac {1}{2}} \,{S_{1}}(1
 + J)^{3} + \left( {d_1} - {\displaystyle \frac { {b_1}}{2}
}  - {\displaystyle \frac { {a_2}}{2}} \right)\,{S_{1}}^{2}(1 + J)
  \\
& + \left({\displaystyle \frac {3}{2}} \,{S_{2}}(1 + J) +  {d_2}
 -  {b_2}\right)\,{S_{1}}(1 + J)+ \left( -  {d_1} + {\displaystyle \frac { {b_1}}{2}
}  + {\displaystyle \frac { {a_2}}{2}} \right)\,{S_{2}}(1 + J) -
{\displaystyle \frac {1}{2}} \,{S_{3}}(1 + J)\Bigg],
\end{split}
 \end{align}
where we used \eq{tilde_k1} and wrote $\tilde{k}_1$ in terms of $k_1$. $\bar{k}_2$ on the l.h.s is given in
\eq{k2_bar}.

Having determined all the ingredients in the matching formula \eq{tilde_after_exponentiation},
we can easily convert it to the preferred form where all the NLO terms and the leading NNLO terms are
evaluated in moment space, in analogy with
\eq{after_full_exponentiation_most_moment_space_mu}:
\begin{eqnarray}
\label{tilde_after_full_exponentiation_most_moment_space_mu}
&&\hspace*{-5pt}\left.\frac{G_{77}(E_0,m_b)}{G_{77}(0,m_b)}\right
\vert_{\rm Resummed} =\frac{1}{2\pi i} \int_{c-i\infty}^{c+i\infty}
\frac{dN}{N-1} \,
\left(\frac{2E_0}{m_b}\right)^{1-N}\,\widetilde{{\rm
Sud}}^{(J)}(N,m_b)
 \times
 \nonumber  \\ && \,\exp \Bigg\{C_F \widetilde{R}_1^{\reg}(N)\frac{\alpha_s(\mu)}{\pi}
 +\,C_F\bigg[
\beta_0 \,\ln \left(\frac{\mu^2}{m_b^2}\right)\,\widetilde{R}_1^{\reg}(N)+
\widetilde{R}_2^{\reg, \,\leading}(N)\nonumber \\
&& \hspace*{30pt}-\,\, C_F\left(
\frac12\left(   \widetilde{R}_1^{\reg}(N)\right)^2+
\left(\widetilde{R}_1^{\reg}(N)-\tilde{k}_1\right)
\widetilde{E}_1(N)\right)
\bigg]\left(\frac{\alpha_s(\mu)}{\pi}\right)^2
 +\cdots \Bigg\}\nonumber \\
&&\hspace*{50pt}+\,C_F\int_{x_0=2E_0/m_b}^1\!\!\!\!\! dx\,\,
\tilde{r}_2^{\reg,\,\subleading}(x)
\left(\frac{\alpha_s(\mu)}{\pi}\right)^2+\cdots,
\end{eqnarray}
where we used \eq{tilde_R12_sing}. Here $\widetilde{R}_1^{\reg}(N)$ is
\begin{align}
\label{tildeR1reg}
\begin{split}
\widetilde{R}_1^{\reg}(N)&\equiv \int_0^1dx (x^{N-1}-1)
\tilde{r}_1^{\reg}(x) = R_1^{\reg}(N)+R_1^{\sing}(N)
-\widetilde{R}_1^{\sing}(N)\\
&= R_1^{\reg}(N)-\bigg[R_1^{\sing}(N+J)-R_1^{\sing}(1+J)
-R_1^{\sing}(N)\bigg]
\end{split}
\end{align}
where the second line in based on \eq{Delta_r12_reg},
and $\widetilde{R}_1^{\sing}(N)$ is given in \eq{tilde_Rn_sing_in_terms_of_non_tilde_ones}.
An explicit expression for the regular part in $x$ space can be obtained
using Eqs. (\ref{rn_reg_tilde}) and (\ref{Delta_r12_reg}):
\begin{equation}
\label{tilde_r2_reg}
\tilde{r}_2^{\reg}(x)=r_2^{\reg}(x)-
\Bigg(x^J\,r_2^{\sing,\,\integrable}(x)\,+\,(x^J-1)\,r_2^{\sing}(x)\Bigg).
\end{equation}
Finally, $\widetilde{R}_2^{\reg, \,\leading}(N)$ is defined in analogy with $R_2^{\reg, \,\leading}(N)$ in Eqs.
(\ref{reg_leading}) and (\ref{N_space_reg_leading}). To this end we decompose the regular part:
\begin{equation}
\label{tilde_reg_leading}
\tilde{r}_2^{\reg}(x)=\tilde{r}_2^{\reg,\,\leading}(x) + \tilde{r}_2^{\reg,\,\subleading}(x)
\end{equation}
such that $\tilde{r}_2^{\reg,\,\subleading}(x)$ vanishes as $(1-x)$ for $x\longrightarrow 1$ and
\begin{eqnarray}
\label{N_space_tilde_reg_leading}
\widetilde{R}_2^{\reg,\,\leading}(N)&=&\int_0^1dx \, \tilde{r}_2^{\reg,\,\leading}(x) \left(x^{N-1}-1\right) -
 \int_0^1dx\, \tilde{r}_2^{\reg,\,\subleading}(x)\nonumber \\
 &=&\,\tilde{k}_2\,+\,\int_0^1dx \, \tilde{r}_2^{\reg,\,\leading}(x) x^{N-1}.
\end{eqnarray}
In contrast with $\bar{r}_2^{\reg,\,\leading}(x)$, however, we require that
$\tilde{r}_2^{\reg,\,\leading}(x)$ --- and therefore also $\tilde{r}_2^{\reg,\,\subleading}(x)$ (!) ---
will behave as $x^J$ at $x\longrightarrow 0$.
To compute $\tilde{r}_2^{\reg,\,\leading}(x)$  we therefore first extract $x^J$ out of
$\tilde{r}_2^{\reg}(x)$ of \eq{tilde_r2_reg} before expanding near $x\longrightarrow 1$.
Using $x^J=1-J\,(1-x)+{\cal O}((1-x)^2)$ and \eq{r2_bar_def_integrable} we obtain:
\begin{equation}
\tilde{r}_2^{\reg,\,\leading}(x)=x^J\left[\bar{r}_2^{\reg,\,\leading}(x)\,
+\,J\,(1-x)\,r_2^{\sing}(x)\right].
\end{equation}
Finally, returning to~\eq{N_space_tilde_reg_leading} and using \eq{N_space_reg_leading} we obtain:
\begin{eqnarray}
\label{N_space_tilde_reg_leading_explicit}
\widetilde{R}_2^{\reg,\,\leading}(N)
 &=&\,\tilde{k}_2\,+\,\int_0^1dx \, \bigg(\bar{r}_2^{\reg,\,\leading}(x)\,+\,J\,(1-x)\,r_2^{\sing}
 (x)\bigg) x^{N+J-1}\\
 &=&\,\tilde{k}_2\,-\bar{k}_2+ \bar{R}_2^{\reg,\,\leading}(N+J)\,+
 \,J\,\int_0^1dx \, \,(1-x)\,r_2^{\sing}(x)\, x^{N+J-1}.\nonumber
\end{eqnarray}
where the explicit expression for $\bar{R}_2^{\reg,\,\leading}(N)$ is
 given in \eq{R2_reg_mom} with \eq{R2_reg_leading}.

\section{The normalized
$G_{77}$ spectrum in the large--$\beta_0$ limit: results
\label{sec:bar_r_reg_expression}}

\subsection{Expansion coefficients in $x$ space}

The coefficients $r^{\beta_0}_n$ of \eq{large_beta_0_partial_sum}
are obtained upon expanding the Borel function in \eq{B_BXSG}:
\begin{equation}
\label{Bx_sum_and_separation}
 \frac12 B(x,u)=\sum_{n=0}^{\infty} r^{\beta_0}_n(x)\,
u^n/n!;\qquad \quad
 r^{\beta_0}_n(x)=r^{\beta_0,\,\sing}_n(x)+r^{\beta_0,\,\reg}_n(x),
\end{equation}
To this end we need the expansion of the hypergeometric function,
which is (see~\cite{Kalmykov:2006pu}):
\begin{eqnarray}
_2F_1\Big([1, 1],[2-u],x\Big)
&=&\frac{1-u}{x}\Bigg\{-\ln(1-x)+u\bigg[\frac{1}{2}\ln^2(1-x)+{\rm
Li}_2(x)\bigg]\\ \nonumber
&&\hspace*{-30pt}+u^2\left[-S_{1,2}(x)-\ln(1-x){\rm Li}_2(x) +{\rm
Li}_3(x)-\frac{1}{6}\ln^3(1-x)\right]\\ \nonumber
&&\hspace*{-30pt}+u^3\bigg[-S_{2,2}(x)-\ln(1-x){\rm Li}_{3}(x)
+\ln(1-x)S_{1,2}(x)\\ \nonumber
&&\hspace*{-10pt}+\frac{1}{2}\ln^2(1-x) {\rm
Li}_2(x)+\frac{1}{24}\ln^4(1-x) +S_{1,3}(x)+{\rm
Li}_4(x)\bigg]+\cdots\Bigg\},
\end{eqnarray}
where Nielsen integrals are defined by
\begin{equation}
S_{a,b}(x)\equiv \frac{(-1)^{a+b-1}}{(a-1)!b!}
\int_0^1\frac{d\xi}{\xi}\ln^{a-1}(\xi)\ln^b(1-\xi x).
\end{equation}

The resulting coefficients are as follows:
\begin{align}
\begin{split}
\label{bar_r_reg_expression1} r^{\sing}_1(x)&= - {\displaystyle
\frac {\mathrm{ln}(1 - x)}{1 - x}}  - {\displaystyle \frac {7}{4\,(1
- x)}}
\\
r^{\reg}_1(x)&=-\frac{1+x}{2}\,\mathrm{ln}(1 - x) - {\displaystyle
\frac {x^{2}}{2}}  + {\displaystyle \frac {x}{4}}  + {\displaystyle
\frac {7}{4}}
\end{split}
\end{align}
\begin{align}
\begin{split}
\label{bar_r_reg_expression2}
r^{\beta_0,\,\sing}_2(x)&={\displaystyle \frac {3}{2}}
\,{\displaystyle \frac {\mathrm{ln}^{2}( 1 - x)}{1 - x}}  +
{\displaystyle \frac {13}{12}} \, {\displaystyle \frac
{\mathrm{ln}(1 - x)}{1 - x}}  + {\displaystyle \left({{\displaystyle
\frac {\pi ^{2}}{6}}  - {\displaystyle \frac {85}{24}}
}\right)\frac{1}{1 - x}}
\\
r^{\beta_0\,\reg}_2(x)&= \left({\displaystyle \frac {x}{2}}  +
{\displaystyle \frac {1}{2}}  + {\displaystyle \frac {1}{  1 - x}}
\right)\,{\rm Li}_2(x) + \left( {\displaystyle \frac {3\,x}{4}}  +
{\displaystyle \frac {3}{4}} \right)\,\mathrm{ln}^{2}(1 - x)\\   &
 + \left( - {\displaystyle \frac {25\,x}{12}}
 - {\displaystyle \frac {3}{2\,x}}  - {\displaystyle \frac {1}{12
}}  + {\displaystyle \frac {x^{2}}{2}} \right)\,\mathrm{ln}(1 - x) -
{\displaystyle \frac {19\,x^{2}}{12}}  
+ {\displaystyle \frac {49}{24}}  + {\displaystyle \frac
{55\,x}{24}}  - {\displaystyle \frac {\pi ^{2}}{6\,(  1 - x)}}
\end{split}
\end{align}
\begin{align}
\begin{split}
\label{bar_r_reg_expression3} r^{\beta_0\,\sing}_3(x)&= -
{\displaystyle \frac {7}{3}} \,{\displaystyle \frac {\mathrm{
ln}^{3}(1 - x)}{1 - x}}  + {\displaystyle \frac {1}{4}} \,
{\displaystyle \frac {\mathrm{ln}^{2}(1 - x)}{1 - x}}  +\left(
{\displaystyle \frac {275}{36}} - {\displaystyle \frac {\pi
^{2}}{3}}
 \right)
{\displaystyle \frac {\,\mathrm{ln}(1 - x)}{1 - x}}
 + {\displaystyle \left( {{\displaystyle \frac {29\,\pi ^{2}}{36}
}  - {\displaystyle \frac {581}{72}} }\right)\frac{1}{1 - x}}
\\
r^{\beta_0\,\reg}_3(x)&= \left(x + 1 + {\displaystyle \frac {2}{ 1 -
x}} \right)\,{\rm Li}_3(x) + \bigg[\left(
 - 2 - 2\,x - {\displaystyle \frac {4}{  1 - x}} \right)\,\mathrm{ln}(
1 - x) + {\displaystyle \frac {11\,x}{3}}  + {\displaystyle \frac
{4}{3\,(  1 - x)}}  \\   &+ {\displaystyle \frac {3}{x}}  -
{\displaystyle \frac {10}{3}} \bigg]\,{\rm Li}_2(x) \mbox{} + \left(
- x - 1 - {\displaystyle \frac {2}{  1 - x}} \right)\,{S_{ 1,
\,2}}(x) + \left( - {\displaystyle \frac {7}{6}}  - {\displaystyle
\frac {7\,x}{6}} \right)\,\mathrm{ln}^{3}(1 - x) \\
&\mbox{} + \left({\displaystyle \frac {23\,x}{4}}  + { \displaystyle
\frac {9}{2\,x}}  - {\displaystyle \frac {13}{4}}  -
{\displaystyle \frac {x^{2}}{2}} \right)\,\mathrm{ln}^{2}(1 - x) \\
&\mbox{} + \left({\displaystyle \frac {19\,x^{2}}{6}}  -
{\displaystyle \frac {7}{x}}  + {\displaystyle \frac {\pi ^{2}}{6 }}
+ {\displaystyle \frac {247}{36}}  + \left( - {\displaystyle \frac
{407}{36}}  + {\displaystyle \frac {\pi ^{2}}{6}} \right)\,x +
{\displaystyle \frac {2\,\pi ^{2}}{3\,(  1 - x)}}
\right)\,\mathrm{ln}
(1 - x) \\
&\mbox{}  + \left({\displaystyle \frac {\pi ^{2}}{6}}  -
{\displaystyle \frac {203}{36}} \right)\,x^{2}- {\displaystyle \frac
{2\,\pi ^{2}}{9\,(  1 - x)}}  + \left( {\displaystyle \frac
{923}{72}}  - {\displaystyle \frac {\pi ^{2} }{12}} \right)\,x -
{\displaystyle \frac {139}{72}}  - {\displaystyle \frac {7\,\pi
^{2}}{12}}
\end{split}
\end{align}
\begin{align}
\begin{split}
\label{bar_r_reg_expression4} r^{\beta_0\,\sing}_4(x)&=
{\displaystyle \frac {15}{4}} \,{\displaystyle \frac
{\mathrm{ln}^{4} (1 - x)}{1 - x}}  - {\displaystyle \frac {35}{12}}
\, {\displaystyle \frac {\mathrm{ln}^{3}(1 - x)}{1 - x}}
+\left({\displaystyle \frac {\pi ^{2}}{2}}  - {\displaystyle \frac
{105}{8}} \right) {\displaystyle \frac {\,\mathrm{ln}^{2}(1 - x)}{1
- x}
}  \\
&+\left({\displaystyle \frac {6029}{216}} - {\displaystyle \frac
{29\,\pi ^{2} }{12}}    \right) {\displaystyle \frac
{\,\mathrm{ln}(1 - x) }{1 - x}}  + {\displaystyle
\left(-\frac{\pi^4}{30}+\frac{235}{72}\pi^2-\frac{9557}{432}+3\zeta_3
\right)\frac {1}{1 - x}}
\\
r^{\beta_0,\,\reg}_4(x)&= 3\left(1+ x + {\displaystyle \frac {2}{  1
- x}} \right)\,{\rm Li}_4(x) + \bigg[-6 \left( 1+ x  +
{\displaystyle \frac {2}{  1 - x}} \right)\,\mathrm{
ln}(1 - x)+{\displaystyle\frac {9}{x}}  + 11\,x - 10 \\
&   + {\displaystyle \frac {4}{  1 - x}} \bigg]\,{\rm Li}_3(x) +
\bigg[ 6 \left(1+ x + {\displaystyle \frac {2}{  1 - x}}
\right)\,\mathrm{ln}^{2}(1
 - x) \\
&+ \left(20 - {\displaystyle \frac {8}{  1 - x}}  - {\displaystyle
\frac {18}{x}}  - 22\,x\right)\,\mathrm{ln}(1 - x) -
{\displaystyle \left( {\pi ^{2} + \frac53}\right)\frac{1}{  1 - x}}  \\
&\mbox{} - {\displaystyle \frac {197}{6}}  - {\displaystyle \frac
{\pi ^{2}}{2}}  + \left({\displaystyle \frac {121}{6}}  -
{\displaystyle \frac {\pi ^{2}}{2}} \right)\,x + {\displaystyle
\frac {
21}{x}} \bigg]{\rm Li}_2(x) \\
&\mbox{} + \bigg[6\left(1+\,x + {\displaystyle \frac {2}{  1 - x}}
\right)\, \mathrm{ln}(1 - x) - {\displaystyle \frac {4}{  1 - x}}  +
10 -
11\,x - {\displaystyle \frac {9}{x}} \bigg]\,{S_{1, \,2}}(x) \\
&\mbox{} + 3 \left(1+ \,x  + {\displaystyle \frac {2}{  1 - x}}
\right)\,\left({S_{1, \,3}}(x)-{S_{2, \,2}}(x)\right)     \\
&\mbox{} + \bigg({\displaystyle \frac {15}{8}} +{\displaystyle \frac
{15\,x}{8}} \bigg)\,\mathrm{ln}^{4}(1 - x) + \left({\displaystyle
\frac {x^{2}}{2}}  + {\displaystyle \frac {119}{12}}  -
{\displaystyle \frac {157\,x}{12}}  -
{\displaystyle \frac {21}{2\,x}} \right)\,\mathrm{ln}^{3}(1 - x) \\
&\mbox{} + \left( - {\displaystyle \frac {3\,\pi ^{2}}{4}}  -
{\displaystyle \frac {2\,\pi ^{2}}{  1 - x}}  + {\displaystyle \frac
{63}{2\,x}}  - {\displaystyle \frac {19\,x^{2}}{4}}  -
{\displaystyle \frac {345}{8}}  + \left({\displaystyle \frac
{297}{8} }  - {\displaystyle \frac {3\,\pi ^{2}}{4}}
\right)\,x\right)\,\mathrm{ln}^{2}(1
 - x) +  \\
&\left[{\displaystyle \frac {12163}{216}}  + {\displaystyle \frac
{3\, \pi ^{2} - 57}{2\,x}}  + {\displaystyle \frac {4\,\pi
^{2}}{3\,(
  1 - x)}}  + \left( {\displaystyle \frac {203}{12}}- {\displaystyle \frac {\pi ^{2}}{2}}
 \right)\,x^{2} + \left(  {\displaystyle \frac {25\,\pi ^{2}}{12}}- {\displaystyle
\frac {14063}{216}}   \right)
\,x + {\displaystyle \frac {\pi ^{2}}{12}} \right] \\
&\mathrm{ln}(1 - x)\mbox{} + \left( - {\displaystyle \frac
{4955}{216} }  + {\displaystyle \frac {19\,\pi ^{2}}{12}}
\right)\,x^{2} - {\displaystyle \frac {15715}{432}}  + \left( -
{\displaystyle \frac {
55\,\pi ^{2}}{24}}  + {\displaystyle \frac {30227}{432}} \right)\,x \\
&\mbox{} - {\displaystyle \frac {1}{90}} \,{\displaystyle \frac {
 - 25\,\pi ^{2} + 270\,\zeta_3 - 3\,\pi ^{4}}{  1 - x}}  -
{\displaystyle \frac {49\,\pi ^{2}}{24}}
\end{split}
\end{align}

\subsection{Expansion coefficients in moment space}

Defining the moments, as in (\ref{mom_def}), by
\begin{equation}
\bar{M}_N^{\PT,\,O_7 }=\int_0^1 dx
\frac{1}{\Gamma_{O_7}}\frac{d\Gamma_{O_7}}{dx} x^{N-1}
\end{equation}
and using \eq{BXg_full} with the final expression in~\eq{B_BXSG} we
get:
\begin{equation}
\bar{M}_N^{\PT,\,O_7 }=1+ \frac{C_F}{2\beta_0}
\int_0^{\infty}du\,T(u)\,\left(\frac{\Lambda^2}{m^2}\right)^u\,\tilde{B}(N,u)
\end{equation}
with
\begin{eqnarray}
\tilde{B}(N,u)&\equiv& \int_0^1 dx \Big(x^{N-1}-1\Big) B(x,u)\\
\nonumber &=& {\rm e}^{\frac53 u}\,\frac{\sin\pi u}{\pi u}
\,\int_0^1dx
\,\Big(x^{N-1}-1\Big)\,(1-x)^{-u}\Bigg\{\frac{1}{1-x}\frac{1}{(1-u)(2-u)}+\\
\nonumber&&
\left[-(1-4x+x^2)\left(\frac{1}{1-x}+\frac{1}{1-u}\right)+\frac{2(1-x)^2}{(1-u)^2}\right]
\,\, _2F_1\Big([1, 1],[2-u],x\Big)\\ \nonumber&&
+(1-4x+x^2)\frac{1}{1-x}+\frac{(x+1)(x^2-3x+1)}{(1-u)(1-x)}-\frac{(2-x)x}{(2-u)}-
\frac{2(1-x)}{(1-u)^2} \Bigg\}
\end{eqnarray}

The perturbative series to leading order in the flavor expansion
takes the form:
\begin{eqnarray}
\label{N_space} &&\hspace*{-15pt} \bar{M}_N^{\PT,\,O_7 }=1   + C_F
R_1(N) \frac{\alpha_s(m_b^2)}{\pi} +C_F N_f R_2^{N_f}(N)
\left(\frac{\alpha_s(m_b^2)}{\pi}\right)^2+\cdots
\end{eqnarray}
where at each order $k$ we separate the coefficients into singular
and regular parts, as in \eq{Bx_sum_and_separation}, namely
\begin{equation}
R_k(N)=R_k^{\sing}(N)+ R_k^{\reg}(N).
\end{equation}
Here \emph{both} parts of the coefficients are defined with
vanishing first moments at each order: $R_k^{\sing}(N=1)=0$ and
$R_k^{\reg}(N=1)=0$.

At NLO the coefficients are:
\begin{equation}
\label{R1_sing_N} R_1^{\sing}(N)=\underbrace{\frac12\Psi_1(N) -
{\displaystyle \frac {\pi ^{2}}{12}}- \frac12\left(\Psi (N) +
\gamma_E \right)^{2}}_{\rm LL}+\underbrace{{\displaystyle \frac
{7}{4}} \,\left(\Psi (N) + \gamma_E \right)}_{\rm NLL}
\end{equation}
\begin{eqnarray}
\label{R1_reg_N} R_1^{\reg}(N)&=&\left( - {\displaystyle \frac
{1}{2N\,(N + 1)}}  + {\displaystyle \frac {1}{N}}
\right)\,\left(\Psi (N) + \gamma_E \right)- {\displaystyle \frac
{31}{12}} \\ \nonumber && \hspace*{40pt}
  + {\displaystyle \frac {9}{4\,N}}
 + {\displaystyle \frac {1}{2(N + 1)^{2}}}  -
{\displaystyle \frac {1}{2(N + 2)}}  - {\displaystyle \frac
{1}{4\,(N
 + 1)}}  + {\displaystyle \frac {1}{2N^{2}}}
\end{eqnarray}
At NNLO we have:
\begin{eqnarray}
R_2^{N_f, \sing}(N)&=&\underbrace{-\frac14\left[
-\frac{1}{3}(\Psi(N)+\gamma_E)^3+\left(\Psi_1(N)-\frac{\pi^2}{6}\right)(\Psi(N)+\gamma_E)
-\frac{1}{3}(\Psi_2(N)+2\zeta_3)\right]}_{\rm LL}\nonumber \\
&&\hspace*{-30pt}+\underbrace{\frac{13}{72}\left[\frac12\Psi_1(N) -
{\displaystyle \frac {\pi ^{2}}{12}}- \frac12\left(\Psi (N) +
\gamma_E \right)^{2}\right]}_{\rm NLL}
+\underbrace{\left(\frac{\pi^2}{36}-\frac{85}{144}\right)\left(\Psi
(N) + \gamma_E \right)}_{\rm NNLL}
\end{eqnarray}
and
\begin{eqnarray}
\label{R2_reg_N_Nf} R_2^{N_f, \reg}(N)&=&\frac{1}{12}
\sum_{k=1}^{\infty}\frac{\Gamma(k)\Gamma(N)}{\Gamma(N+k+2)}\left(3+3\frac{1+2N}{k}+2\frac{N(N+1)}{k^2}\right)
+\frac{49}{48}\\\nonumber &&\hspace*{-30pt}
-\frac{1}{12N^3}-\frac{13}{72
N^2}-\frac{31}{48}\frac{1}{N}-\frac{1}{8}\left(\frac{1}{N}+\frac{1}{N+1}\right)
(\Psi(N)+\gamma_E)^2+\bigg(-\frac{19}{72}\frac{1}{N}-\frac{1}{6}\frac{1}{N^2}\\\nonumber
&&\hspace*{-30pt}
+\frac{1}{12}\frac{1}{N+2}-\frac{7}{72}\frac{1}{N+1}-\frac{1}{6}\frac{1}{(N+1)^2}+\frac{1}{6}\Psi_1(N)
-\frac{1}{4}\frac{1}{N-1}\bigg)\left(\Psi(N)+\gamma_E\right)\\\nonumber
&&\hspace*{-30pt}
+\frac{1}{24}\left(\frac{1}{N+1}+\frac{1}{N}\right)\Psi_1(N)+\left(\frac{1}{16}
-\frac{5}{144}\frac{1}{N}-\frac{5}{144}\frac{1}{N+1}\right)\pi^2-\frac{1}{12}\Psi_2(N)-\frac{1}{3}\zeta_3
\\\nonumber
&&\hspace*{-30pt}-\frac{13}{72}\frac{1}{(N+1)^2}+\frac{1}{12(N+2)^2}+\frac{5}{36}\frac{1}{N+2}
-\frac{1}{12(N+1)^3}+\frac{7}{144(N+1)}.
\end{eqnarray}

\subsection{The small--$x$ limit~\label{sec:large_beta0_small_x}}

Upon expanding Eq.~(2.3) in Ref.~\cite{Gardi:2004ia}, or \eq{B_BXSG}
above, in powers of $x$ one finds:
\begin{eqnarray}
\label{BXg_full_small_x}
\hspace*{-3pt}\left.\frac{1}{\Gamma_{O_7}}\frac{d\Gamma_{O_7}}{dx}\right\vert_{{\rm
large}\,\,\beta_0}\hspace*{-10pt} &=& \frac{C_F}{2\beta_0}
\int_0^{\infty}du\,T(u)\,\left(\frac{\Lambda^2}{m_b^2}\right)^u
\left\{\,B^{(x^3)}(u)\, x^3+B^{(x^4)}(u)\,x^4+{\cal O}(x^5)\right\}
\\ \nonumber&=&
\frac{C_F\alpha_s}{2\pi}\left\{\bigg(1+\frac{35}{12}\frac{\alpha_s\beta_0}{\pi}
+\cdots\bigg)\,x^3\,+\,
\bigg(\frac54+\frac{1099}{240}\frac{\alpha_s\beta_0}{\pi}
+\cdots\bigg) \,x^4+{\cal O}(x^5)\right\}
\end{eqnarray}
with
\begin{align}
\begin{split}
B^{(x^3)}(u)&={\rm e}^{\frac{5}{3}u}\,\frac{\sin \pi u}{\pi
u}\frac43\left[\frac{1}{1-u}-\frac{1}{4-u}\right]\\
B^{(x^4)}(u)&={\rm e}^{\frac{5}{3}u}\,\frac{\sin \pi u}{\pi
u}\left[\frac{5}{2}\frac{1}{1-u}+\frac{4}{2-u}
-\frac{6}{3-u}-\frac{7}{4-u}+\frac{5}{2}\frac{1}{5-u} \right]
\end{split}
\end{align}
Performing the Borel integration in \eq{BXg_full_small_x} (with the
default values $m_b=4.875$ GeV $\Lambda=0.332$ GeV and $N_f=4$) we
obtain the following numerical coefficients of the small $x$
expansion:
\begin{align}
\begin{split}
\label{BXg_full_small_x_numerically} \hspace*{-20pt}\left.
\frac{1}{\Gamma_{O_7}}\frac{d\Gamma_{O_7}}{dx} \right\vert_{{\rm
large}\,\,\beta_0} =0.0729\,x^3+0.105 \,x^4+{\cal O}(x^5).
\end{split}
\end{align}

\bibliographystyle{JHEP}
\bibliography{database}

\end{document}